\documentclass{article}
\usepackage{amsmath}
\usepackage{amssymb}

\input{tcilatex}

\begin{document}

\begin{center}
{\LARGE Phase Space Theory of\bigskip }

{\LARGE \ Bose-Einstein Condensates and\bigskip\ }

{\LARGE Time-Dependent\ Modes}
\end{center}

\bigskip

\begin{center}
B J Dalton$^{\ast }$

\bigskip

ARC Centre for Quantum-Atom Optics

and

Centre for Atom Optics and Ultrafast Spectroscopy

Swinburne University of Technology

Melbourne, Victoria 3122, Australia

\bigskip

\bigskip

\bigskip
\end{center}

\pagebreak

\subsection{Abstract}

A phase space theory approach for treating dynamical behaviour of
Bose-Einstein condensates applicable to situations such as interferometry
with BEC in time-dependent double well potentials is presented.
Time-dependent mode functions are used, chosen so that one, two,.. highly
occupied modes describe well the physics of interacting condensate bosons in
time dependent potentials at well below the transition temperature. Time
dependent mode annihilation, creation operators are represented by time
dependent phase variables, but time independent total field annihilation,
creation operators are represented by time independent field functions. Two
situations are treated, one (mode theory) is where specific mode
annihilation, creation operators and their related phase variables and
distribution functions are dealt with, the other (field theory) is where
only field creation, annihilation operators and their related field
functions and distribution functionals are involved. The field theory
treatment is more suitable when large boson numbers are involved. The paper
focuses on the hybrid approach, where the modes are divided up between
condensate (highly occupied) modes and non-condensate (sparsely occupied)
modes. It is found that there are extra terms in the Ito stochastic
equations both for the stochastic phases and stochastic fields, involving
coupling coefficients defined via overlap integrals between mode functions
and their time derivatives. For the hybrid approach both the Fokker-Planck
and functional Fokker-Planck equations differ from those derived via the
correspondence rules, the drift vectors are unchanged but the diffusion
matrices contain additional terms involving the coupling coefficients.

Results are also presented for the combined approach where all the modes are
treated as one set. Here both the Fokker-Planck and functional Fokker-Planck
equations are exactly the same as those derived via the correspondence
rules. However, although the Ito stochastic field equations are also
unchanged, the Ito equations for the stochastic phases contain an extra
classical term involving the coupling coefficients. \bigskip

\bigskip

* Tel: +61 0 3 9214 8187; fax: +61 0 3 9214 5160

\textit{Email address: }bdalton@swin.edu.au\pagebreak

\section{Introduction}

\label{Section 1 Introduction}

Bose-Einstein condensates (BEC) in cold atomic gases are a quantum system on
a macroscopic scale. At temperatures well below the critical temperature
almost all the bosonic atoms occupy the same single particle state or mode 
\cite{Leggett01a}. Consequently, BEC exhibit coherence properties analogous
to those for an optical laser where macroscopic occupancy of photons in a
single mode can occur, and therefore their application for interferometry
was a logical outcome following early experiments demonstrating that BEC
could show interference effects \cite{Andrews97a}, \cite{Hall98a}. However,
unlike the photons in a laser, the bosonic atoms in a BEC interact with each
other and this can cause decoherence effects that degrade interference
patterns. Ideally, the BEC should be at a temperature well below the
transition temperature, so that coherence is enhanced due to all the bosons
occupying only one or two modes, with decoherence effects due to thermally
occupied excited modes becoming small. However such low temperatures are not
always realised. The interactions between the bosons could in principle be
made zero via Feshbach resonance methods \cite{Cornish00a}, but this is hard
to accomplish over the range of trapping magnetic fields involved during an
interferometric process and requires a broad Feshbach resonance.
Furthermore, it is bosonic interactions that provide for the possibility of
realising Heisenberg limited interferometry via spin squeezing effects \cite%
{Orzel01a}, \cite{Esteve08a}, and proposals for such interferometry have
been made \cite{Bouyer97a}, \cite{Dunningham02a} in which the bosonic
interactions are central. Recent reviews of BEC interferometry include \cite%
{Bongs04a}, \cite{Dunningham05a}, \cite{Cronin09a}, \cite{Schumm11a}.

The theoretical treatment of BEC interferometry must take into account the
many body nature of the BEC including the presence of boson-boson
interactions. The interferometry process is time dependent and may involve
non-static trapping potentials, such as when the trap changes from a single
well to a possibly asymmetric double well and back again. The possibility of
fragmentation of the BEC into more than a single mode \cite{Leggett01a} must
be allowed for, since it cannot just be assumed that all bosons stay in one
single condensate mode throughout the process even if this was initially the
case. Decoherence effects associated with bosons being lost from condensate
modes need to be treated, including the creation of Bogoliubov excitations 
\cite{Pitaevskii03a}. Unless the BEC is prepared well below the transition
temperature thermal effects would also need to be taken into account. Also,
the experiments may not control the total number of bosons in the BEC. These
considerations suggest that the quantum state of the BEC should be described
via a density operator rather than a pure state, and the many body nature of
the BEC would best be discussed in terms of field operators and second
quantisation rather than via standard first quantisation methods involving
symmetrised products of wave functions.

A number of different theoretical approaches have been used for studying
BECs, including treatments using variational methods, Heisenberg equations,
master equations and phase space distributions. General descriptions of and
references to these methods are set out in \cite{Dalton10a}. The present
paper focuses on phase space methods which involve the overall bosonic field
rather than just the separate modes - these were originally developed for
applications in quantum optics \cite{Graham70a}, \cite{Graham70b}, \cite%
{Drummond87a}, \cite{Kennedy88a}, \cite{Gatti97a}, but are now applied in
quantum-atom optics \cite{Steel98a}, \cite{Molmer03a}. The quantum density
operator is represented either by a quasi-distribution function in a phase
space involving variables that replace the mode annihilation, creation
operators for treatments where it is practical to consider separate modes,
or by a quasi-distribution functional in a phase space involving field
functions which replace the bosonic field annihilation, creation operators.
As these two situations are interconvertible it is convenient to consider
both though ultimately the paper is focused on the latter. Also, it is often
convenient to define field operators for condensate and non-condensate modes
separately \cite{Gardiner98a}, \cite{Gardiner00a}, and for situations well
below the transition temperature often only one or two condensate modes are
involved. Quantum correlation functions involving products of annihilation,
creation operators either for the separate modes or for the bosonic field
are given by phase space ordinary or functional integrals of the equivalent
products of phase space variables or field functions, weighted by the
quasi-distribution function or functional. Furthermore, there are several
possible choices for the type of quasi-distribution function or functional
that may be used. The positive P type is suitable for consideration of
quantum correlation functions involving normally ordered products of
annihilation, creation operators, the Wigner type is better suited to treat
those that involve symmetrically ordered products. Indeed, when the BEC is
well below the transition temperature a hybrid approach, such as where the
highly occupied condensate modes are described via a Wigner
quasi-distribution function or functional and the positive P type is used
for the mainly unoccupied non-condensate modes \cite{Steel98a}, \cite%
{Dalton07a}, \cite{Hoffmann08a}, \cite{Dalton10a}, might best describe the
physical situation where the condensate bosons behave essentially like a
classical mean field and the non-condensate bosons exhibit mainly quantum
features . In the phase space approaches the Liouville-von Neumann or master
equation for the density operator is replaced by either Fokker-Planck or
functional Fokker-Planck equations for the quasi-distribution function or
functional. Finally, the Fokker-Planck equations are replaced by equivalent
Ito stochastic equations for either stochastic phase space variables or
stochastic fields, and the phase space ordinary or functional integrals for
the quantuim correlation functions given by stochastic averages.

Phase space methods first involve a consideration of what are suitable
single particle states or modes that could be used to describe states of the
bosonic system. In treating both the dynamics of bosonic systems and their
static behaviour at non-zero temperatures it is usually convenient to use
Fock states as a basis. In first quantization the Fock states are
symmetrised products where the identical bosons occupy single particle
states or modes, and for bosons there are Fock states in which more than one
particle can occupy any particular mode. In an idealised Bose-Einstein
condensate of non-interacting particles at zero temperature all $N$ bosons
occupy the mode with lowest single particle energy. Finite temperature
effects for non-interacting bosons allow for excitations to higher energy
modes. Below a critical temperature $T_{c}$ Bose-Einstein condensation
occurs as a phase transition. For non-interacting bosons in a static
potential $V$ the usual choice for the single particle states is the time
independent energy eigenstates for a single boson in the static potential.
However, for interacting bosons in cases where the trap potential $V$
changes with time - such as in interferometry experiments involving
Bose-Einstein condensates - there is no obvious set of time independent
modes that could be used. Of course theoretical treatments using time
independent modes can still be carried out, often these are chosen for
mathematical convenience such as in facilitating numerical calculations. For
example Blakie et al \cite{Blakie08a} use single particle states for
non-interacting bosons in a static trap potential, Egorov et al \cite%
{Egorov11a} use plane wave states. It is well known however that for
interacting bosons at temperatures close to zero, the solution to the time
dependent Gross-Pitaevskii equation \cite{Gross61a}, \cite{Pitaevskii61a}
usually provides a good physical choice for the single particle state that
all the bosons occupy. This mode function is obtained from the Dirac-Frenkel
variational principle \cite{Dirac30a}, \cite{Frenkel34a} in which the state
vector is a product with all bosons occupying the same single particle
state. However, as Leggett has pointed out \cite{Leggett01a} Bose-Einstein
condensates may sometimes exist in fragmented states, in which there is
macroscopic occupancy of more than one mode. Such a situation could occur in
double-well interferometry with Bose-Einstein condensates, where two modes
may be involved. These may be two localised modes in the separate potential
wells, or they may be delocalised modes spread over both wells, in a
symmetric double well one could be symmetric and the other antisymmetric.
Generalised coupled Gross-Pitaevskii equations \cite{Spekkens99a}, \cite%
{Menotti01a}, \cite{Dalton07b}, \cite{Streltsov06a}, \cite{Streltsov07a}, 
\cite{Alon08a}, \cite{Sakmann09a}, \cite{Dalton10a}, \cite{Dalton12a} can be
obtained via the Dirac-Frenkel variational principle for the actual pair of
mode functions. Theories in which all the bosons occupy only or a few single
particle states where the mode functions satisfy Gross-Pitaevskii type
equations are referred to as mean field theories, since the Gross-Pitaevskii
type equations contain non-linear terms which can be interpreted as due to
mean fields. However, the actual dynamics is more complicated than that can
be treated via mean field theories, processes associated with unoccupied
modes such as Bogoliubov excitations \cite{Bogoliubov47a}, \cite{Leggett01a}%
, \cite{Pitaevskii03a} need to be considered. Phase space theories are a
standard approach to moving beyond mean field theories. Since such time
dependent modes or single particle states provide a useful first
approximation to treating the dynamics of interacting bosons in time
dependent potentials at temperatures well below $T_{c}$, it is natural to
base the phase space theory on \emph{time dependent modes} as these more
closely describe the BEC before interactions involving both occupied and
non-occupied modes are included. It is also possible that choosing
time-dependent modes could facilitate numerical calculations. A
redevelopment of phase space theory to allow for this modal time dependence
is warranted.

However, if the theory is to be based on time dependent modes it follows
that as the \emph{total} bosonic \emph{field} annihilation, creation
operators are not time dependent, then the \emph{mode} annihilation,
creation operators must be. Since in phase space theories these mode
annihilation, creation operators are represented via the phase space
variables, the question arises as to whether these should be chosen to be 
\emph{time dependent} to reflect the time dependence of the mode functions.
It turns out that although the phase space theory can also be developed in
terms of using \emph{time independent} phase space variables to represent
the time dependent mode annihilation, creation operators, the field
functions that represent the \emph{total} bosonic\emph{\ field}
annihilation, creation operators via mode expansions with the phase space
variables as coefficients then become time dependent. This is rather
unsatisfactory as it would be preferable to represent time independent
bosonic total field operators by time independent field functions. There are
other complications with using time independent phase space variables, so
the present paper is framed around using phase space variables whose time
dependence is related to the time dependence of the mode functions. The time
dependence of the phase space variables is actually chosen to be identical
to that for the mode annihilation, creation operators. Both the Wigner and
positive P distribution functions or functionals are treated, both in a
double phase space. The theory presented here treats two situations, one (%
\emph{mode theory}) being where mode annihilation, creation operators and
their related phase variables and distribution functions are dealt with
specifically, the other (\emph{field theory}) being where field creation,
annihilation operators and their related field functions and distribution
functionals involve a description where individual modes are not
distinguished.

A so-called \emph{combined approach} could be adopted, in which the modes
are not divided into condensate and non-condensate types, either when
separate modes are considered or in the field theory treatment involving the
total fields. However, even in the combined approach there are some
unexpected consequences. For the phase space distribution \emph{functions}
involved in the mode theory situation,\emph{\ }although the Fokker-Planck
equations and the results for\emph{\ }the quantum correlation functions as
averages via the phase space \emph{integrals} are unchanged from those for
the case of conventional time independent mode functions, the relationship
between the Ito stochastic equations and the Fokker-Planck equations \emph{%
now includes} an additional \emph{classical term} that depends on the time
dependence of the mode functions via the so-called \emph{coupling
coefficients}. These are defined via integrals of products of mode functions
with the mode function time derivatives. On the other hand, when the field
theory results are derived, not only are the \emph{functional} Fokker-Planck
equations for the phase space distribution \emph{functional} and the results
for the quantum correlation functions as averages involving phase space 
\emph{functional integrals} unchanged from those for the situation of
conventional time independent mode functions, but the relationship between
the Ito stochastic field equations and the functional Fokker-Planck
equations \emph{no longer includes} any additional term that depends on the
time dependence of the mode functions. These results confirm the validity of
phase space distribution functional results for functional Fokker-Planck
equations and Ito stochastic field equations presented elsewhere \cite%
{Steel98a}\textbf{, }\cite{Blakie08a} in the situation where the \emph{total}
bosonic fields are considered.

The situation changes for the so-called \emph{hybrid approach} \cite%
{Steel98a}, \cite{Dalton07a}, \cite{Hoffmann08a}, \cite{Dalton10a}, which is
the approach treated in the present paper. In the hybrid approach, the modes
are divided into the highly occupied condensate modes and the sparsely
occupied non-condensate modes, these being treated in terms of a Wigner or
positive P distribution functions or functionals respectively. The total
bosonic field operators are the sum of condensate and non-condensate field
operators. As in Refs \cite{Dalton07a}, \cite{Dalton10a} time dependent mode
functions are involved in both cases, and the condensate and non-condensate 
\emph{field operators} now also become \emph{time dependent}. Hence the
field functions that represent them in the phase space theory are \emph{no
longer} time independent, as was the case for the total fields. The phase
space theory is therefore more complex than that outlined previously in \cite%
{Dalton10a}, both when separate modes are considered and when condensate and
non-condensate fields are involved. However, in both the mode and field
theory situations, ordinary or functional Fokker-Planck equations and Ito
stochastic equations for the phase space variables or fields can still be
derived, and relationships between the drift, diffusion terms in the
Fokker-Planck equations and the classical and noise terms in the Ito
stochastic equations established both for separate phase variables and for
the fields. Hybrid Wigner and positive P distribution functions or
functionals are treated, both in a double phase space. Both the ordinary and
functional Fokker-Planck equations now have \emph{additional diffusion terms}
involving coupling coefficients between condensate and non-condensate modes.
The \emph{drift terms} are unchanged. Also, \emph{additional classical terms}
involving the coupling coefficients need to be included in the relationship
between the Ito stochastic (field) equations and the (functional)
Fokker-Planck equation, and the \emph{noise terms} are now related to the 
\emph{new diffusion term}. These additional terms modify the functional
Fokker-Planck and Ito stochastic field equations presented previously in
Ref. \cite{Dalton10a}. We show that the condensate and non-condensate
stochastic fields involve the \emph{same} expansions in terms of time
dependent stochastic phases and mode functions as do the condensate and
non-condensate field functions when expanded in terms of non-stochastic
phase variables.

The theory for the hybrid approach is presented in two parts, the first
(mode theory) dealing with separate time dependent modes, their annihilation
and creation operators, phase variables, the mode quantum correlation
functions, characteristic and distribution functions, Fokker-Planck and Ito
stochastic equations for the phase variables associated with separate mode
annihilation, creation operators. This more familiar separate mode theory
treatment will be covered in Section \ref{Section 2 Separate Modes Case} for
the hybrid approach. In the second part (field theory) we consider field
annihilation and creation operators, field functions, field correlation
functions, characteristic and distribution functionals, functional
Fokker-Planck and Ito stochastic equations for the fields associated with
condensate and non-condensate field annihilation, creation operators. This
less familiar quantum field theory treatment will be covered in Section \ref%
{Section 3 - Quantum Field Case} for the hybrid approach where the total
fields are divided into condensate and non-condensate fields. However,
although the quantum field theory treatment is more useful for dealing with
systems containing a large number of bosons, it is less familiar than the
separate mode approach, involving as it does concepts based on functional
calculus. Consequently the key steps for the quantum field theory treatment
will be set out as a separate entity, rather than just deriving the
functional Fokker-Planck and Ito stochastic field equations from the
previous separate mode results. Connections between the separate mode and
quantum field theory treatments will be frequently emphasised however. Both
in the separate mode and the field theory treatments the stochastic
equations are the sum of a classical contribution related to the drift
vector and a noise contribution involving Gaussian-Markov random noise
terms. The stochastic properties of the noise contributions are derived and
related to the diffusion matrices. The conclusions are presented in Section %
\ref{Section 5 - Conclusions}, along with a comparison to phase space
theories based on time independent modes. Key results are summarised in
Section \ref{Section 4 - Summary of Key Results}, both for the hybrid
approach and for the combined approach. In Appendix \ref{Appendix -
Functional Calculus} a brief outline of functional calculus is presented,
and the equivalence of the separate mode and field theory treatments is
shown in Appendix \ref{Appendix Equivalence of Separate Mode and Field
Theory} for the hybrid approach. The paper is also accompanied by several
further Appendices, available as On-line Supplementary Material. Appendices %
\ref{Appendix - Derivation of Langevin Equations}, \ref{Appendix Derivation
of Functional Fokker-Planck Equation} and\ref{Appendix Derivation of Ito
Stochastic Field Equations} include details of derivations for Sections \ref%
{Section 2 Separate Modes Case} and \ref{Section 3 - Quantum Field Case} to
avoid overloading the main body of the paper. Many of the technical details
on deriving Fokker-Planck equations can also be found in Ref. \cite%
{Dalton10a}. Time derivatives will be written as $\frac{{\LARGE \partial }}{%
{\LARGE \partial t}}$ though perhaps $\frac{{\LARGE d}}{{\LARGE dt}}$ should
sometimes be used to emphasise that all time dependences are differentiated.

\pagebreak

\section{Theory - Separate Modes Treatment}

\label{Section 2 Separate Modes Case}

Phase space theory is based on the idea of representing the \emph{quantum
density operator} $\widehat{\rho }$ by a \emph{quasi-distribution function} $%
P(\alpha ,\alpha ^{+},\alpha ^{\ast },\alpha ^{+\ast })$ which depends on 
\emph{phase variables} $\alpha =\{\alpha _{1},\alpha _{2},..,\alpha
_{k},..,\alpha _{n}\},\alpha ^{+}=\{\alpha _{1}^{+},\alpha
_{2}^{+},..,\alpha _{k}^{+},..,\alpha _{n}^{+}\}$. Here $\alpha _{k}$ is
associated with the \emph{annihilation} operator $\widehat{a}_{k}$ and $%
\alpha _{k}^{+}$ is associated with the \emph{creation }operator $\widehat{a}%
_{k}^{\dag }$ for the $k$th mode involved in an expansion of the field
operators $\hat{\Psi}(\mathbf{r})$ and $\hat{\Psi}^{\dag }(\mathbf{r})$ in
terms of \emph{orthonormal mode functions} $\phi _{k}(\mathbf{r},t\mathbf{)}$
or their complex conjugates $\phi _{k}^{\ast }(\mathbf{r},t\mathbf{)}$.
Quantum correlation functions can be expressed in terms of \emph{phase space
averages} involving the quasi-distribution function. The time dependence of
the distribution function is given by a \emph{Fokker-Planck equation} which
includes \emph{drift} and \emph{diffusion} terms associated with first and
second order derivatives of the distribution function with respect to the
phase variables. Ultimately the phase space variables will be replaced by 
\emph{stochastic} variables satisfying \emph{Ito stochastic equations} that
are equivalent to the Fokker-Planck equation, and the quantum correlation
determined via \emph{stochastic averages}. In the present paper dealing with
time dependent mode functions, a \emph{hybrid} distribution function is used
where the condensate modes are treated via a \emph{Wigner} distribution
function and the non-condensate modes via a \emph{positive P} distribution
function. Thus the phase variables for the condensate modes will be treated 
\emph{differently} to those for the non-condensate modes. The treatment
involves using \emph{time dependent} phase variables, and leads to \emph{%
non-standard} Fokker-Planck and Ito equations. Details of derivations are
covered in Appendix \ref{Appendix - Derivation of Langevin Equations}.

\subsection{Field Operators, Hamiltonians, Quantum Correlation Functions}

\label{SubSection 2. FieldOprs}

In the Schrodinger picture the \emph{field operators} associated with
annihilating or creating a boson at position $\mathbf{r}$ are time
independent and satisfy the standard bosonic commutation rules, for which
the non-zero expressions are%
\begin{equation}
\lbrack \hat{\Psi}(\mathbf{r}),\hat{\Psi}^{\dagger }(\mathbf{s})]=\delta 
\mathbf{(r-s})  \label{Eq.BoseCommRules}
\end{equation}%
For simplicity we consider only single component bosonic systems, however
cases where the bosons may involve differing internal (hyperfine) states can
be treated via appropriate generalisations of the present approach.

The field operators are fundamental, appearing in expressions for the system 
\emph{Hamiltonian} 
\begin{equation}
\hat{H}=\int d\mathbf{r(}\frac{\hbar ^{2}}{2m}\nabla \hat{\Psi}(\mathbf{r}%
)^{\dag }\cdot \nabla \hat{\Psi}(\mathbf{r})+\hat{\Psi}(\mathbf{r})^{\dag }V%
\hat{\Psi}(\mathbf{r})+\frac{g}{2}\hat{\Psi}(\mathbf{r})^{\dag }\hat{\Psi}(%
\mathbf{r})^{\dag }\hat{\Psi}(\mathbf{r})\hat{\Psi}(\mathbf{r}))
\label{Eq.HamBosonFields}
\end{equation}%
where $m$ is the mass, $V$ is the trapping potential and the zero range
approximation is used for interactions between the particles, with $g=4\pi
\hbar ^{2}a_{s}/m$ involving the s-wave scattering length $a_{s}$.

In addition, measurement results are often expressed in terms of \emph{%
quantum correlation functions}, which involve expectation values of products
of field operators. A typical quantum correlation function is the \emph{%
normally ordered} case 
\begin{eqnarray}
&&G^{n}(\mathbf{r}_{1}\cdots \mathbf{r}_{p};\mathbf{s}_{q}\cdots \mathbf{\ s}%
_{1})=\langle \hat{\Psi}(\mathbf{r}_{1})^{\dag }\cdots \hat{\Psi}(\mathbf{r}%
_{p})^{\dag }\hat{\Psi}(\mathbf{s}_{q})\cdots \hat{\Psi}(\mathbf{s}%
_{1})\rangle  \notag \\
&=&\mbox{Tr}(\hat{\Psi}(\mathbf{s}_{q})\cdots \hat{\Psi}(\mathbf{s}_{1})\hat{%
\rho}(t)\hat{\Psi}(\mathbf{r}_{1})^{\dag }\cdots \hat{\Psi}(\mathbf{r}%
_{p})^{\dag })  \label{Eq.QuantCorrFns}
\end{eqnarray}%
where for an $N$ particle system we require $p,q\leq N$ to give a non-zero
result. Various spatial interference and coherence effects in Bose Einstein
condensates can be described via such quantum correlation functions. The
case for $p=q$ where $\mathbf{r}_{i}=\mathbf{s}_{i}$ for all $i$ is
proportional to the probability of simultaneously detecting bosons at $%
\mathbf{r}_{1},\mathbf{r}_{2}\cdots \mathbf{r}_{p}$ \cite{Bach04a}, \cite%
{Dalton12a}.

\subsection{Time Dependent Modes}

\label{SubSection 2. TimeDepModes}

For reasons set out in the Introduction, the phase space theory presented
here is based on using \emph{time dependent mode functions} to describe
single particle states of the bosonic system. Such mode functions are
required to be \emph{orthonormal} and to form a \emph{complete} set in the
function space of interest. Thus the mode functions $\phi _{k}(x,t\mathbf{)}$
satisfy 
\begin{eqnarray}
\int dx\phi _{k}^{\ast }(x,t\mathbf{)}\phi _{l}(x,t\mathbf{)} &=&\delta _{kl}
\label{Eq.Orthonorm} \\
\sum_{k}\phi _{k}(x,t\mathbf{)}\phi _{k}^{\ast }(y,t\mathbf{)} &\mathbf{=}%
&\delta (x-y)  \label{Eq.Completeness1}
\end{eqnarray}%
Henceforth $x$ refers to position in a 3D space unless otherwise stated.

For the case where the set of modes is not complete and restricted to a set $%
S$ the completeness relation is replaced by 
\begin{equation}
\sum_{k\epsilon S}\phi _{k}(x,t\mathbf{)}\phi _{k}^{\ast }(y,t\mathbf{)=\;}%
\delta ^{S}(x,y)  \label{Eq.CompletenessRestrict}
\end{equation}%
where $\delta ^{S}(x,y)$ is a restricted delta function \cite{Blakie08a}, 
\cite{Dalton10a}. The restricted delta function can be used to define a
projector $\mathcal{P}_{x}^{S}$ that turns any function $F(y)$ into a linear
combination of the $\phi _{k}(x,t\mathbf{)}$ within the restricted function
space. Thus%
\begin{eqnarray}
\mathcal{P}_{x}^{S}[F(x)] &=&\int dy\,\delta ^{S}(x,y)F(y)  \notag \\
&=&\sum_{k\epsilon S}\phi _{k}(x,t\mathbf{)}\int dy\,\phi _{k}^{\ast }(y,t%
\mathbf{)}F(y)  \label{Eq.Projector}
\end{eqnarray}%
If the set of modes is complete then $\delta ^{S}(x,y)=\delta (x-y)$, the
usual Dirac delta function. Unless otherwise stated we will consider sets of
modes $\phi _{k}(x,t\mathbf{)}$ or $\phi _{k}^{\ast }(x,t\mathbf{)}$ that
are complete.

\subsection{Mode Creation and Annihilation Operators}

\label{SubSection 2. ModeCreationOprs}

The field operators may be expanded in terms of a complete set of time
dependent orthonormal single mode functions as%
\begin{equation}
\hat{\Psi}(x\mathbf{)=}\sum_{k}\widehat{a}_{k}(t)\phi _{k}(x,t\mathbf{)}%
,\qquad \hat{\Psi}^{\dag }(x\mathbf{)=}\sum_{k}\widehat{a}_{_{k}}^{\dag
}(t)\phi _{k}^{\ast }(x,t\mathbf{)}  \label{Eq.FieldOprExpn}
\end{equation}%
where $\widehat{a}_{k}(t),\widehat{a}_{k}^{\dag }(t)$ are time dependent 
\emph{mode annihilation, creation operators}. These operators satisfy
standard bosonic \emph{commutation rules} that are equivalent to (\ref%
{Eq.BoseCommRules}) for which the non-zero expressions are%
\begin{equation}
\lbrack \widehat{a}_{k}(t),\widehat{a}_{_{l}}^{\dag }(t)]=\delta _{kl}
\label{Eq.BoseModeCommRules}
\end{equation}%
The annihilation, creation operators can be expressed in terms of the field
operators and mode functions as 
\begin{equation}
\widehat{a}_{k}(t)=\dint dx\,\phi _{k}^{\ast }(x,t\mathbf{)}\hat{\Psi}(x%
\mathbf{)\qquad }\widehat{a}_{k}^{\dag }(t)=\dint dx\,\phi _{k}(x,t\mathbf{)}%
\hat{\Psi}^{\dag }(x\mathbf{)}  \label{Eq.ModeOprs}
\end{equation}%
so they can be considered as functionals of the mode function or its complex
conjugate.

A necessary and sufficient condition for the total field operators to be
time independent is that the mode annihilation, creation operators evolve as%
\begin{eqnarray}
\frac{\partial \widehat{a}_{k}(t)}{\partial t} &=&\dsum\limits_{l}C_{kl}(t)%
\widehat{a}_{l}(t)\qquad \frac{\partial \widehat{a}_{k}^{\dag }(t)}{\partial
t}=\dsum\limits_{l}C_{kl}^{\ast }(t)\widehat{a}_{l}^{\dag }(t)
\label{Eq.TimeDerivAnnihCreatOprs} \\
C_{kl}(t) &=&\dint dx\frac{\partial \phi _{k}^{\ast }(x,t\mathbf{)}}{%
\partial t}\phi _{l}(x,t\mathbf{)}=iD_{kl}(t)  \label{Eq.Coupling Coefts}
\end{eqnarray}%
showing that the time derivatives of the annihilation, creation operators
are just linear combinations of these operators. Note that the sum is over 
\emph{all} modes. By differentiating the orthonormality condition (\ref%
{Eq.Orthonorm}) we see that 
\begin{equation}
C_{kl}+C_{lk}^{\ast }=0  \label{Eq.RelnCouplingCoefts}
\end{equation}%
so it follows that the related matrix $D$ is \emph{hermitian}, $%
D_{kl}=D_{lk}^{\ast }$. We will refer to the $C_{kl}(t)$ as the \emph{%
coupling coefficients}. As will be seen later the coupling coefficients play
a key role in the functional Fokker-Planck and Ito stochastic field
equations.

Assuming that the mode functions $\phi _{k}(x,t\mathbf{)}$ or $\phi
_{k}^{\ast }(x,t\mathbf{)}$ form a complete set for expansions of $\frac{%
{\LARGE \partial }}{{\LARGE \partial t}}\phi _{k}(x,t)$ or $\frac{{\LARGE %
\partial }}{{\LARGE \partial t}}\phi _{k}^{\ast }(x,t)$ respectively, we can
also expand the mode derivatives in terms of the modes with the coupling
coefficients as expansion coefficients 
\begin{eqnarray}
\frac{\partial }{\partial t}\phi _{k}(x,t) &=&\dsum\limits_{l}\beta
_{kl}\phi _{l}(x,t)=\dsum\limits_{l}C_{kl}^{\ast }\phi _{l}(x,t)  \notag \\
\frac{\partial }{\partial t}\phi _{k}^{\ast }(x,t)
&=&\dsum\limits_{l}C_{kl}\phi _{l}^{\ast }(x,t)  \label{Eq.TimeDerivModes1}
\end{eqnarray}

We can use these results to confirm that the mode expansions for the field
operators are time independent. We have using (\ref{Eq.FieldOprExpn}), (\ref%
{Eq.TimeDerivAnnihCreatOprs}) and (\ref{Eq.TimeDerivModes1})%
\begin{eqnarray}
\frac{\partial }{\partial t}\widehat{\psi }(x\mathbf{)} &=&0  \notag \\
\frac{\partial }{\partial t}\widehat{\psi }^{\dag }(x\mathbf{)} &\mathbf{=}&0
\label{Eq.TimeDerivTotalFldOprs}
\end{eqnarray}%
after applying (\ref{Eq.RelnCouplingCoefts}).

\subsection{Determination of Mode Functions}

\label{SubSection2 - Determinination of Modes}

There are several possibilities for determining the time dependent modes.
One possibility is to use the complex conjugates of the eigenfunctions for
the first order quantum correlation function $G^{1}(\mathbf{r}_{1};\mathbf{s}%
_{1})=\langle \hat{\Psi}(\mathbf{r}_{1})^{\dag }\hat{\Psi}(\mathbf{s}%
_{1})\rangle $ - these are the so-called \emph{natural orbitals}. The
corresponding eigenvalues give the occupancies of these single boson states,
and the condensate modes could be chosen as those with macroscopic
occupancy, with the remainder specifying non-condensate modes. However this
requires \emph{knowing} the quantum correlation function, and it is hard to
see how this could be done without \emph{first} determining the dynamical
behaviour. Another possibility would be to use the time dependent single
particle states for a \emph{single boson} in the trap potential, but this
ignores the effect of boson-boson interactions. Perhaps the best alternative
is to apply the Dirac-Frenkel \emph{variational principle} \cite{Dirac30a}, 
\cite{Frenkel34a} to a state in which the bosons are restricted to only
occupying as few modes as possible, and where the mode functions along with
amplitudes for the allowed Fock states are treated as variational functions.
Such an approach leads to the Gross-Pitaevskii equation \cite{Gross61a}, 
\cite{Pitaevskii61a} in the case where only one mode is involved or
generalised Gross-Pitaevskii equations where two or more modes are
considered.

The Dirac-Frenkel principle involves minimising the \emph{dynamical action}
given by 
\begin{equation}
S_{dyn}=\tint dt\,\left( 
\begin{array}{c}
\{\left\langle \partial _{t}\Phi \mathbf{|\,}\Phi \right\rangle
-\left\langle \Phi \mathbf{|\,}\partial _{t}\Phi \right\rangle \}/\mathbf{\,}%
2i \\ 
-\left\langle \Phi \mathbf{|\,}\widehat{H}\,\mathbf{|}\Phi \right\rangle /%
\mathbf{\,}\hbar%
\end{array}%
\right)  \label{Eq.Action}
\end{equation}%
where $\left\vert \,\Phi (t)\right\rangle $ is the quantum state. For
unrestricted states this minimisation leads to the time-dependent
Schrodinger equation, so the Dirac-Frenkel principle enables the fundamental
equation for quantum dynamics to be expressed as a Principle of Least
Action. In the present application however, the quantum state is restricted
to a \emph{specific form} involving variational quantities, so applying the
Dirac-Frenkel principle then does not result in a quantum state that
satisfies the time-dependent Schrodnger equation, but nevertheless is as
close to being a solution as that the specific form allows.

For the \emph{two mode case} that is relevant to double well BEC
interferometry the two condensate mode functions are determined as in \cite%
{Dalton10a}, \cite{Dalton12a}. The \emph{quantum state} $\left\vert \,\Phi
(t)\right\rangle $ of the $N$ boson system as a superposition of the $N+1$ 
\emph{basis states} $\left\vert \,\frac{{\small N}}{{\small 2}}%
,k\right\rangle $, where there are $\frac{{\small N}}{{\small 2}}-k$ and $%
\frac{{\small N}}{{\small 2}}+k$ bosons (respectively) occupying the two
modes with (time dependent) \emph{mode functions} $\phi _{1}(x,t)$ and $\phi
_{2}(x,t)$. The \emph{amplitude} for this basis state is $b_{k}(t)$. 
\begin{equation}
\left\vert \,\Phi (t)\right\rangle =\tsum\limits_{k=-\frac{\mathbf{N}}{%
\mathbf{2}}}^{\frac{\mathbf{N}}{\mathbf{2}}}\,b_{k}(t)\,\left\vert \,\frac{%
{\small N}}{{\small 2}},k\right\rangle .  \label{Eq.TwoModeQState1}
\end{equation}%
and the basis states are \emph{Fock states} given by 
\begin{equation}
\left\vert \,\frac{{\small N}}{{\small 2}},k\right\rangle =\frac{\left( 
\widehat{a}_{1}(t)^{\dag }\right) ^{(\frac{{\LARGE N}}{{\LARGE 2}}-k)}}{[(%
\frac{N}{2}-k)!]^{\frac{\mathbf{1}}{\mathbf{2}}}}\frac{\left( \widehat{a}%
_{2}(t)^{\dag }\right) ^{(\frac{{\LARGE N}}{{\LARGE 2}}+k)}}{[(\frac{N}{2}%
+k)!]^{\frac{\mathbf{1}}{\mathbf{2}}}}\left\vert \,0\right\rangle \qquad
(k=-N/2,-N/2+1,..,+N/2)  \label{Eq.OrthonormalBasisStates}
\end{equation}%
These basis states are \emph{fragmented} or \emph{number squeezed} states,
allowing for both modes to have \emph{macroscopic occupancy }when $|k|\ll N/2
$. The notation $\left\vert \,\frac{{\small N}}{{\small 2}},k\right\rangle $
for the basis states reflects the feature that the two mode Bose condensate
behaves like a \emph{giant spin system} - details are given in \cite%
{Dalton10a}, \cite{Dalton12a}. The \emph{total angular momentum} quantum
number $j=\frac{{\small N}}{{\small 2}}$ is \emph{macroscopic}, and $k=-%
\frac{{\small N}}{{\small 2}},-\frac{{\small N}}{{\small 2}}{\small +1,..,}%
\frac{{\small N}}{{\small 2}}{\small -1,+}\frac{{\small N}}{{\small 2}}$
specifies the \emph{magnetic} quantum number as well as $2k$ determining the
difference in \emph{mode occupancy}.

The quantum state (\ref{Eq.TwoModeQState1})\ is a functional of the
amplitudes $b_{k}$ and (via (\ref{Eq.ModeOprs})) the mode functions $\phi
_{1}(x,t)$ and $\phi _{2}(x,t)$. In \cite{Dalton10a}, \cite{Dalton12a} a 
\emph{self-consistent} set of equations for the amplitudes and mode
functions has been determined by applying the Dirac-Frenkel principle \cite%
{Dirac30a}, \cite{Frenkel34a} and minimising the \emph{dynamic action} with
respect to these variational functions using the state vector given by (\ref%
{Eq.TwoModeQState1}). Here the mode functions and amplitudes act as
variational quantities.

The mode functions satisfy the \emph{coupled generalised} \emph{%
Gross-Pitaevskii }equations 
\begin{eqnarray}
i\hbar \tsum\limits_{j}X_{ij}\,\frac{\partial }{\partial t}\phi _{j}
&=&\tsum\limits_{j}X_{ij}(-\frac{\hbar ^{2}}{2m}\nabla ^{2}+V)\,\phi _{j} 
\notag \\
&&+\tsum\limits_{j}(g\tsum\limits_{mn}Y_{im\,jn}\,\phi _{m}^{\ast }\,\phi
_{n})\,\phi _{j}\qquad \qquad (i=1,2).  \label{Eq.GenGrossPitEqns}
\end{eqnarray}%
and allow for boson-boson interactions and are time-dependent. They follow
the changes in the time dependent potential $V(\mathbf{r},t)$. This is a
generalised \emph{mean field theory} - the quantity in brackets in the final
term of the last equation being a mean field. The quantities $X_{ij}$ and $%
Y_{im\,jn}$ are \emph{one-body} and \emph{two-body} \emph{correlation
functions} 
\begin{eqnarray}
X_{ij} &=&\left\langle \Phi \right\vert \widehat{a}_{i}^{\dag }\widehat{a}%
_{j}\left\vert \Phi \right\rangle   \label{Eq.OneBodyCorrFn} \\
Y_{im\,jn} &=&\left\langle \Phi \right\vert \widehat{a}_{i}^{\dag }\widehat{a%
}_{m}^{\dag }\widehat{a}_{j}\widehat{a}_{n}\left\vert \Phi \right\rangle 
\label{Eq.TwoBodyCorrFn}
\end{eqnarray}%
Detailed expressions given in the Appendix for \cite{Dalton10a}, \cite%
{Dalton12a}, showing that $X_{ij}$ and $Y_{im\,jn}$ are quadratic forms of
the amplitudes $b_{k}$. They are of order $N$ and $N^{2}$ respectively.

The \emph{amplitudes} satisfy coupled \emph{matrix mechanics} equations%
\begin{equation}
i\hbar \frac{\partial b_{k}}{\partial t}=\tsum\limits_{l}(H_{kl}-\hbar
U_{kl})b_{l}\qquad (k=-N/2,..,N/2).  \label{Eq.AmpEqns}
\end{equation}%
These $N+1$ equations (\ref{Eq.AmpEqns}) describe the system dynamics as it
evolves amongst the possible fragmented states. In these equations the
matrix elements $H_{kl}$, $U_{kl}\ $depend on the mode functions $\phi
_{i}(x,t)$. Detailed expressions for $H_{kl}$, $U_{kl}$ are given in
Appendix for \cite{Dalton10a}, \cite{Dalton12a}. The quantities $H_{kl}$ are
matrix elements of the \emph{Hamiltonian} $\widehat{H}$ in equation (\ref%
{Eq.HamBosonFields}) between the fragmented states $\left\vert \,\frac{%
{\small N}}{{\small 2}},k\right\rangle $, $\left\vert \,\frac{{\small N}}{%
{\small 2}},l\right\rangle $. The quantities $U_{kl}$ are elements of the
so-called \emph{rotation matrix}, and allow for the time dependence of the
mode functions. 
\begin{eqnarray}
H_{kl} &=&\left\langle \frac{{\small N}}{{\small 2}},k\right\vert \widehat{H}%
\left\vert \frac{{\small N}}{{\small 2}},l\right\rangle  \label{Eq.HamMatrix}
\\
U_{kl} &=&\frac{1}{2i}\left( \left\{ \partial _{t}\left\langle \frac{{\small %
N}}{{\small 2}},k\right\vert \right\} \left\vert \frac{{\small N}}{{\small 2}%
},l\right\rangle -\left\langle \frac{{\small N}}{{\small 2}},k\right\vert
\left\{ \partial _{t}\left\vert \frac{{\small N}}{{\small 2}},l\right\rangle
\right\} \right)  \label{Eq.RotationMatrix}
\end{eqnarray}%
The specific forms of the $X_{ij}$, $Y_{im\,jn}$, $H_{kl}$, $U_{kl}$ are not
important in what follows, all that is required is that they have been
determined. Equations for the mode functions and amplitudes similar to (\ref%
{Eq.GenGrossPitEqns}) and (\ref{Eq.AmpEqns}) have been obtained by Alon et
al \cite{Alon08a} for single component BECs. The key feature is the
self-consistent nature of the equations - the amplitudes are required to
determine the form of the mode equations and the mode functions are required
to determine the matrix elements in the amplitude equations.

From the amplitude and mode equations it can be shown that%
\begin{eqnarray}
\frac{\partial }{\partial t}\tsum\limits_{k=-\frac{\mathbf{N}}{\mathbf{2}}}^{%
\frac{\mathbf{N}}{\mathbf{2}}}\left\vert b_{k}(t)\right\vert ^{2} &=&0
\label{Eq.AmplitudeNorm} \\
i\hbar \tsum\limits_{ij}X_{ij}\,\frac{\partial }{\partial t}\tint dx\mathbf{%
\,}\phi _{i}^{\ast }(x,t)\,\phi _{j}(x,t) &=&0  \label{Eq.ModeOrthonorm}
\end{eqnarray}%
The first result shows that the amplitudes would remain normalised to unity
and the second result is consistent with the modes remaining orthogonal and
normalised, assuming they were so chosen at $t=0$. The second result
involves the trace of the product of a positive definite invertible matrix $%
X $ with a matrix which is the time derivative of the mode orthogonality
matrix.

For the non-condensate modes a different approach is required since physical
considerations suggest the mode occupancy would be small. The variational
approach might be extended to consider quantum states for an $N+1$ boson
system where all except a single boson occupy one or two condensate modes.
For the case of two condensate modes such a quantum state would be of the
form 
\begin{equation}
\left\vert \,\Phi (t)\right\rangle =\tsum\limits_{k=-\frac{\mathbf{N}}{%
\mathbf{2}}}^{\frac{\mathbf{N}}{\mathbf{2}}}\,\dsum\limits_{l}b_{k,l}(t)\,%
\frac{\left( \widehat{a}_{1}(t)^{\dag }\right) ^{(\frac{{\LARGE N}}{{\LARGE 2%
}}-k)}}{[(\frac{N}{2}-k)!]^{\frac{\mathbf{1}}{\mathbf{2}}}}\frac{\left( 
\widehat{a}_{2}(t)^{\dag }\right) ^{(\frac{{\LARGE N}}{{\LARGE 2}}+k)}}{[(%
\frac{N}{2}+k)!]^{\frac{\mathbf{1}}{\mathbf{2}}}}\left( \widehat{a}%
_{l}(t)^{\dag }\right) \left\vert \,0\right\rangle \,
\label{Eq.QuantumStateTwoCondOneNonCondModes}
\end{equation}%
where the variational functions are now the two condensate mode functions $%
\phi _{1}(x,t)$, $\phi _{2}(x,t)$, the non-condensate mode functions $\phi
_{l}(x,t)$ and the state amplitudes $b_{k,l}(t)$. Presumably the mode
equations for the condensate modes would be similar to those discussed
above, whilst the mode equation for the singly occupied non-condensate modes
would contain mean field terms associated with the highly occupied
condensate modes - rather like what happens for the highest energy orbital
for electrons in an alkali metal atom, the closed shell orbitals being
analogous to the condensate modes. There would be a multitude of time
dependent solutions for the singly occupied mode, and these could be used as
the non-condensate mode functions. Alternatively, the non-condensate modes
might be just chosen via a Schmidt orthogonality process to be orthogonal to
the previous condensate modes, starting with a set of plane wave modes or
single atom trapped modes.

\subsection{Characteristic and Distribution Functions}

\label{SubSection2. CharDistnFns}

Phase space theory involves first introducing characteristic functions that
can be used to specify all the quantum correlation functions for a given
density operator. We wish to divide the modes up into two sets - condensate
and non-condensate modes. The former are to be treated via the Wigner
distribution, the latter via the positive P distribution. The definition of
the \emph{characteristic function} $\chi (\xi ,\xi ^{+})$ for this hybrid
approach is%
\begin{eqnarray}
\chi (\xi ,\xi ^{+}) &=&Tr(\hat{\Omega}_{C}^{W}(\xi _{C},\xi _{C}^{+})\,\hat{%
\Omega}_{NC}^{+}(\xi _{NC}^{+})\,\hat{\rho}\,\hat{\Omega}_{NC}^{-}(\xi
_{NC}))  \notag \\
\hat{\Omega}_{NC}^{+}(\xi _{NC}^{+}) &=&\exp i\sum\limits_{k\epsilon NC}\hat{%
a}_{k}\xi _{k}^{+}\qquad \hat{\Omega}_{NC}^{-}(\xi _{NC})=\exp
i\sum\limits_{k\epsilon NC}\xi _{k}\hat{a}_{k}^{\dag }\,  \notag \\
\hat{\Omega}_{C}^{W}(\xi _{C},\xi _{C}^{+}) &=&\exp i\sum\limits_{k\epsilon
C}(\hat{a}_{k}\xi _{k}^{+}+\xi _{k}\hat{a}_{k}^{\dag })  \label{Eq.CharFn}
\end{eqnarray}%
where the \emph{characteristic variables} are $\xi =\{\xi _{1},\xi
_{2},..,\xi _{k},..\xi _{n}\}=\{\xi _{C},\xi _{NC}\},\xi ^{+}=\{\xi
_{1}^{+},\xi _{2}^{+},..,\xi _{k}^{+},..\xi _{n}^{+}\}=\{\xi _{C}^{+},\xi
_{NC}^{+}$ $\}$. There are two sub-sets - condensate $\xi _{C}$ or $\xi
_{C}^{+}$ and non-condensate $\xi _{NC}$ or $\xi _{NC}^{+}$

Using the \emph{Baker-Hausdorff theorem }%
\begin{equation}
\exp i\sum\limits_{k}(\hat{a}_{k}\xi _{k}^{+}+\xi _{k}\hat{a}_{k}^{\dag
})=\left( \exp i\sum\limits_{k}\xi _{k}\hat{a}_{k}^{\dag }\right) \left(
\exp i\sum\limits_{k}\hat{a}_{k}\xi _{k}^{+}\right) \left( \exp (-\frac{1}{2}%
\sum_{k}\xi _{k}\xi _{k}^{+})\right)   \label{Eq.BakerHauss}
\end{equation}%
and the cyclic properties of the trace, we may combine exponential operators
to find that 
\begin{eqnarray}
\chi (\xi ,\xi ^{+}) &=&\exp (-\frac{1}{2}\sum_{k\epsilon C}\xi _{k}\xi
_{k}^{+})\,\chi _{P+}(\xi ,\xi ^{+})  \notag \\
\chi _{P+}(\xi ,\xi ^{+}) &=&Tr(\,\hat{\Omega}^{+}(\xi ^{+})\,\hat{\rho}\,%
\hat{\Omega}^{-}(\xi ))  \notag \\
\hat{\Omega}^{+}(\xi ^{+}) &=&\exp i\sum\limits_{k\epsilon C,NC}\hat{a}%
_{k}\xi _{k}^{+}\qquad \hat{\Omega}^{-}(\xi )=\exp i\sum\limits_{k\epsilon
C,NC}\xi _{k}\hat{a}_{k}^{\dag }  \label{Eq.CharFnReln}
\end{eqnarray}%
which relates the \emph{hybrid} characteristic function to that for the case
where all modes are treated via a \emph{normally ordered} characteristic
function.

The characteristic function is related to the \emph{quasi-distribution}
function $P(\alpha ,\alpha ^{+},\alpha ^{\ast },\alpha ^{+\ast })$ via%
\begin{eqnarray}
&&\chi (\xi ,\xi ^{+})  \notag \\
&=&\int \int d^{2}\alpha \,d^{2}\alpha ^{+}\,  \notag \\
&&\times \exp (i\sum\limits_{k\epsilon C}(\alpha _{k}\xi _{k}^{+}+\xi
_{k}\alpha _{k}^{+}))\exp (i\sum\limits_{k\epsilon NC}\alpha _{k}\xi
_{k}^{+})\,P(\alpha ,\alpha ^{+},\alpha ^{\ast },\alpha ^{+\ast })\,\exp
(i\sum\limits_{k\epsilon NC}\xi _{k}\alpha _{k}^{+})  \notag \\
&=&\int \int d^{2}\alpha \,d^{2}\alpha ^{+}\,\exp (i\sum\limits_{k\epsilon
C,NC}(\alpha _{k}\xi _{k}^{+}+\xi _{k}\alpha _{k}^{+}))\,P(\alpha ,\alpha
^{+},\alpha ^{\ast },\alpha ^{+\ast })\,  \label{Eq.DistnFn}
\end{eqnarray}%
where in the second line all the exponentials have been combined. As in \cite%
{Dalton10a}, \cite{Hoffmann08a}, the \emph{hybrid} quasi-distribution
function is of the Wigner type for the condensate modes and of the positive
P type for the non-condensate modes. The \emph{phase space variables} are $%
\alpha =\{\alpha _{1},\alpha _{2},..,\alpha _{k},..,\alpha _{n}\}=\{\alpha
_{C},\alpha _{NC}\},\alpha ^{+}=\{\alpha _{1}^{+},\alpha _{2}^{+},..,\alpha
_{k}^{+},..,\alpha _{n}^{+}\}=\{\alpha _{C}^{+},\alpha _{NC}^{+}\}$, where
the set of phase space variables for the condensate modes is $\alpha _{C}$
or $\alpha _{C}^{+}$ and that for the non-condensate modes is $\alpha _{NC}$
or $\alpha _{NC}^{+}$. With $\alpha _{k}=\alpha _{kx}+i\alpha _{ky},\alpha
_{k}^{+}=\alpha _{kx}^{+}+i\alpha _{ky}^{+}$ the phase space integration is 
\begin{equation}
\int \int d^{2}\alpha \,d^{2}\alpha ^{+}\equiv \int \int
\dprod\limits_{k}d\alpha _{kx}d\alpha _{ky}\,\dprod\limits_{k}d\alpha
_{kx}^{+}d\alpha _{ky}^{+}  \label{Eq.PhaseSpaceIntn}
\end{equation}

It needs to be emphasised for this double phase space approach that although
the characteristic function is uniquely determined from the density operator
and vice versa, the same is not true for the distribution function. As
emphasised in several references \cite{Drummond80a}, \cite{Gardiner91a}
dealing with the positive P representation, the distribution is \emph{%
non-unique} - many equivalent distribution functions determine the same
characteristic function and density operator, and these may satisfy
different Fokker-Planck equations based on different choices of the
correspondence rules used to convert the dynamical equation for the density
operator into an equivalent Fokker-Planck equation for the distribution
function (see Section \ref{SubSection2. FPE}). Although this non-uniqueness
precludes interpreting the distribution function as a probability, it is
seen as having the advantage of allowing flexibility in the Fokker-Planck
equations in terms of obtaining related Ito stochastic equations that are
suitable for numerical work.

Although the distribution function is not unique, it is known to exist - at
least for the positive P case. In that case \cite{Drummond80a} proved the
existence of a so-called \emph{canonical form} of the positive P
distribution function in the form 
\begin{eqnarray}
&&P_{canon}(\alpha ,\alpha ^{+},\alpha ^{\ast },\alpha ^{+\ast })  \notag \\
&=&\left( \frac{1}{4\pi ^{2}}\right) ^{n}\mathbf{\exp (-}\frac{1}{2}%
\sum_{k}(\alpha _{k}\alpha _{k}^{\ast }+\alpha _{k}^{+\ast }\alpha _{k}^{+}%
\mathbf{))}\times \left\langle \frac{\alpha +\alpha ^{+\ast }}{2}\right\vert
_{B}\widehat{\rho }\left\vert \frac{\alpha +\alpha ^{+\ast }}{2}%
\right\rangle _{B}  \notag \\
&&  \label{Eq.CanonicalRepnFnBoseFermiDensityOpr}
\end{eqnarray}%
where 
\begin{equation*}
\left\vert \alpha \right\rangle _{B}=\exp (\sum_{k}\alpha _{k}\widehat{a}%
_{k}^{\dag })\,\left\vert 0\right\rangle
\end{equation*}%
are \emph{Bargmann coherent states} and where the density operator may be
written as 
\begin{eqnarray}
\hat{\rho} &=&\int \int d^{2}\alpha ^{+}d^{2}\alpha \,\,P_{canon}(\alpha
,\alpha ^{+},\alpha ^{\ast },\alpha ^{+\ast })\,\frac{\left\vert \alpha
\right\rangle _{B}\left\langle \alpha ^{+\ast }\right\vert _{B}}{%
Tr(\left\vert \alpha \right\rangle _{B}\left\langle \alpha ^{+\ast
}\right\vert _{B})}\,  \notag \\
&&  \label{Eq.CanonicalRepnBoseFermiDensityOpr}
\end{eqnarray}

\subsection{Quantum Correlation Functions}

The standard approach can be applied to relate the quantum correlation
functions to derivatives of the characteristic functions with respect to $%
i\xi _{k}$ or $i\xi _{k}^{+}$ followed by letting the $\xi _{k},\xi
_{k}^{+}\rightarrow 0$. This enables the quantum correlation functions to be
expressed as phase space integrals. The \emph{quantum correlation functions}
are for \emph{symmetrically ordered} products of \emph{condensate mode }%
operators and \emph{normally ordered} products for \emph{non-condensate mode}
operators. We have%
\begin{eqnarray}
G(l_{1},l_{2},..l_{p};m_{q},..,m_{2},m_{1} &\mathbf{:}%
&j_{1},j_{2},..j_{r};k_{s},..,k_{2},k_{1})  \notag \\
&=&\left\langle \{\hat{a}_{l_{1}}^{\dag }\,\hat{a}_{l_{2}}^{\dag }..\hat{a}%
_{l_{n}}^{\dag }\hat{a}_{m_{q}}..\,\hat{a}_{m_{2}}\hat{a}_{m_{1}}\}\hat{a}%
_{j_{1}}^{\dag }\,\hat{a}_{j_{2}}^{\dag }..\hat{a}_{j_{r}}^{\dag }\hat{a}%
_{k_{s}}..\,\hat{a}_{k_{2}}\hat{a}_{k_{1}}\right\rangle  \notag \\
&=&Tr(\hat{\rho}\,\{\hat{a}_{l_{1}}^{\dag }\,\hat{a}_{l_{2}}^{\dag }..\hat{a}%
_{l_{p}}^{\dag }\,\hat{a}_{m_{q}}..\,\hat{a}_{m_{2}}\hat{a}_{m_{1}}\}\,\hat{a%
}_{j_{1}}^{\dag }\,\hat{a}_{j_{2}}^{\dag }..\hat{a}_{j_{r}}^{\dag }\hat{a}%
_{k_{s}}..\,\hat{a}_{k_{2}}\hat{a}_{k_{1}})  \notag \\
&=&\int \int d^{2}\alpha \,d^{2}\alpha ^{+}\,P(\alpha ,\alpha ^{+},\alpha
^{\ast },\alpha ^{+\ast })  \label{Eq.QCFHybridCase} \\
&&\times \lbrack \alpha _{l_{1}}^{+}\,\alpha _{l_{2}}^{+}..\alpha
_{l_{p}}^{+}\,\alpha _{m_{q}}..\,\alpha _{m_{2}}\alpha _{m_{1}}][\alpha
_{j_{1}}^{+}\,\alpha _{j_{2}}^{+}..\alpha _{j_{r}}^{+}\,\alpha
_{k_{s}}..\,\alpha _{k_{2}}\alpha _{k_{1}}]\,  \notag
\end{eqnarray}%
where $l_{1},l_{2},..l_{p};m_{q},..,m_{2},m_{1}$ refer to condensate modes
and $j_{1},j_{2},..j_{r};k_{s},..,k_{2},k_{1}$ to non-condensate modes.
Symmetrically ordered means the average of the product of the operators
taken in any order. Thus%
\begin{eqnarray}
&&\{\hat{a}_{l_{1}}^{\dag }\,\hat{a}_{l_{2}}^{\dag }..\hat{a}_{l_{p}}^{\dag
}\,\hat{a}_{m_{q}}..\,\hat{a}_{m_{2}}\hat{a}_{m_{1}}\}  \notag \\
&=&\frac{1}{(p+q)!}\dsum\limits_{R}\Re (\hat{a}_{l_{1}}^{\dag }\,\hat{a}%
_{l_{2}}^{\dag }..\hat{a}_{l_{p}}^{\dag }\,\hat{a}_{m_{q}}..\,\hat{a}_{m_{2}}%
\hat{a}_{m_{1}}).  \label{Eq.SymmOrder}
\end{eqnarray}%
In Eq.(\ref{Eq.SymmOrder}) the sum over $R$ is over all $(p+q)!$ orderings $%
\Re $ of the factors $\hat{a}_{l_{1}}^{\dag }\,\hat{a}_{l_{2}}^{\dag }..\hat{%
a}_{l_{p}}^{\dag }\,\hat{a}_{m_{q}}..\,\hat{a}_{m_{2}}\hat{a}_{m_{1}}$.

\subsection{Time Dependent Phase Variables}

\label{SubSection2. PhaseVariables}

As outlined in the Introduction, the complex phase variables $\alpha
_{k}(t),\alpha _{k}^{+}(t)$ associated with the time dependent mode
annihilation, creation operators $\widehat{a}_{k}(t),\widehat{a}_{k}^{\dag
}(t)$ will be chosen to be time dependent. Hybrid quasi-distribution
functions $P(\alpha ,\alpha ^{+},\alpha ^{\ast },\alpha ^{+\ast })$ in
double phase spaces will be considered in which the condensate modes will be
treated in terms of Wigner and non-condensate modes in terms of positive P
quasi-distribution functions. Note that the distribution function $P(\alpha
,\alpha ^{+},\alpha ^{\ast },\alpha ^{+\ast })$ is not an analytic function
of the $\alpha _{k}(t),\alpha _{k}^{+}(t)$ but also depends on the complex
conjugates $\alpha _{k}^{\ast }(t),\alpha _{k}^{+\ast }(t)$. The time
dependence of the phase variables is \emph{arbitrary}, but since they are
associated with the mode annihilation, creation operators one logical choice
would be to require the phase variables to have the \emph{same} time
dependence. As we will see, this leads to theoretical results that allow for
the consequences of the basic time dependence of the mode functions is a
reasonably \emph{simple} form. In particular, the field functions that are
associated with the time independent total field operators are also time 
\emph{independent}, as might be hoped for. An alternative choice of
conventional time independent phase variables would lead to an
unsatisfactory result that time \emph{dependent} field functions are
associated with the total field operators. The theory can of course also be
developed via such other choices of time dependence for the phase variables,
we believe however that the present choice is the most satisfactory.

The time dependences \emph{chosen} for the phase variables are thus as for
the mode annihilation, creation operators%
\begin{equation}
\frac{\partial \alpha _{k}(t)}{\partial t}=\dsum\limits_{l}C_{kl}(t)\alpha
_{l}(t)\qquad \frac{\partial \alpha _{k}^{+}(t)}{\partial t}%
=\dsum\limits_{l}C_{kl}^{\ast }(t)\alpha _{l}^{+}(t)
\label{Eq.TimeDerivModePhases}
\end{equation}%
and since $C_{kl}(t)=iD_{kl}(t)$ with $D$ hermitian, it is possible to find
a \emph{unitary} matrix $T$ given by%
\begin{equation}
\frac{\partial T_{kl}(t)}{\partial t}=i\dsum\limits_{m}D_{km}(t)T_{ml}(t)
\label{Eq.UnitaryMatrix}
\end{equation}%
such that the formal solution for the phase variables is 
\begin{equation}
\alpha _{k}(t)=\dsum\limits_{l}T_{kl}(t)\alpha _{l}(0)\qquad \alpha
_{k}^{+}(t)=\dsum\limits_{l}T_{kl}^{\ast }(t)\alpha _{l}^{+}(0)
\label{Eq.TimeDepPhaseVar}
\end{equation}%
This choice of the time dependence for the phase variables is a key element
in the derivations.

By writing $\alpha _{k}(t)=\alpha _{kx}(t)+i\alpha _{ky}(t),\alpha
_{k}^{+}(t)=\alpha _{kx}^{+}(t)+i\alpha
_{ky}^{+}(t),T_{kl}(t)=T_{kl}^{x}(t)+iT_{kl}^{y}(t)$ in terms of real,
imaginary parts, it is then possible to relate phase space integrals over
the $\alpha (t),\alpha ^{+}(t)$ to phase space integrals over the $\alpha
(0),\alpha ^{+}(0)$. Thus%
\begin{eqnarray}
\int \int d^{2}\alpha (t)d^{2}\alpha ^{+}(t) &\equiv &\int \int
\dprod\limits_{k}d\alpha _{kx}(t)d\alpha _{ky}(t)\,\dprod\limits_{k}d\alpha
_{kx}^{+}(t)d\alpha _{ky}^{+}(t)  \notag \\
&=&\int \int \dprod\limits_{k}d\alpha _{kx}(0)d\alpha
_{ky}(0)\,\dprod\limits_{k}d\alpha _{kx}^{+}(0)d\alpha _{ky}^{+}(0)  \notag
\\
&=&\int \int d^{2}\alpha (0)d^{2}\alpha ^{+}(0)  \label{Eq.PhaseSpaceIntegn}
\end{eqnarray}%
The Jacobian involves the $T_{kl}(t),T_{kl}^{\ast }(t)$ and the unitarity of 
$T$ leads to the Jacobian being unity. Differentiation under the phase space
integral can therefore be carried out without concern about time dependent
differentials - a key simplification for the derivations.

The time dependence of the characteristic function variables $\xi =\{\xi
_{1},\xi _{2},..,\xi _{k},..\xi _{n}\},\xi ^{+}=\{\xi _{1}^{+},\xi
_{2}^{+},..,\xi _{k}^{+},..\xi _{n}^{+}\}$ is also arbitrary. However, if we 
\emph{choose} the $\xi _{k}(t),\xi _{k}^{+}(t)$ to evolve as for the mode
annihilation, creation operators%
\begin{equation}
\frac{\partial \xi _{k}(t)}{\partial t}=\dsum\limits_{l}C_{kl}(t)\xi
_{l}(t)\qquad \frac{\partial \xi _{k}^{+}(t)}{\partial t}=\dsum%
\limits_{l}C_{kl}^{\ast }(t)\xi _{l}^{+}(t)  \label{Eq.TimeDepCharFnVar}
\end{equation}%
- which is the\emph{\ same} as for the phase space variables, then this
leads to many key expressions involving the characteristic variables being
time independent, thereby simplifying the theory.

\subsection{Fokker-Planck Equations}

\label{SubSection2. FPE}

The \emph{Fokker-Planck equation} gives the time dependence of the
distribution function $P(\alpha ,\alpha ^{+},\alpha ^{\ast },\alpha ^{+\ast
})$. The derivation of it here requires special treatment due to the time
dependent modes and phase space variables used in the theory.

\subsubsection{Derivation}

We first need to consider the time derivative of the characteristic function
(\ref{Eq.CharFn})\ in the form given by Eq. (\ref{Eq.CharFnReln}) \ It
follows using Eqs.(\ref{Eq.TimeDerivAnnihCreatOprs}), (\ref%
{Eq.TimeDepCharFnVar}) together with $C_{kl}+C_{lk}^{\ast }=0$ that%
\begin{eqnarray}
\frac{\partial }{\partial t}\sum\limits_{k}\hat{a}_{k}\xi _{k}^{+} &=&0 
\notag \\
\frac{\partial }{\partial t}\sum\limits_{k}\xi _{k}\hat{a}_{k}^{+} &=&0
\label{Eq.TimeIndepResults1}
\end{eqnarray}%
This means that the time derivative of the normally ordered characteristic
function $\chi _{P+}(\xi ,\xi ^{+})$ is only dependent on the time
derivative of the density operator, with the time dependences of the mode
annihilation, creation operators $\widehat{a}_{k}(t),\widehat{a}_{k}^{\dag
}(t)$ and characteristic function variables $\xi _{k}(t),\xi _{k}^{+}(t)$
cancelling out. However, although the quantities $\sum\limits_{k\epsilon
C,NC}\hat{a}_{k}\xi _{k}^{+}$ and $\sum\limits_{k\epsilon C,NC}\xi _{k}\hat{a%
}_{k}^{\dag }$ are time independent, the same is not true for $%
\sum_{k\epsilon C}\xi _{k}\xi _{k}^{+}$. We find that 
\begin{equation}
\frac{\partial }{\partial t}\sum_{k\epsilon C}\xi _{k}\xi
_{k}^{+}=\sum\limits_{k\epsilon C}\dsum\limits_{l\epsilon NC}(C_{kl}\xi
_{l}\xi _{k}^{+}+C_{kl}^{\ast }\xi _{k}\xi _{l}^{+})
\label{Eq.TimeDepResults2}
\end{equation}%
where the $k\epsilon C,l\epsilon C$ terms cancel out. This means that there
are additional terms in the equation of motion for the characteristic
function. Thus%
\begin{eqnarray}
\frac{\partial }{\partial t}\chi (\xi ,\xi ^{+}) &=&(-\frac{1}{2}%
\sum_{k\epsilon C}\dsum\limits_{l\epsilon NC}(C_{kl}\xi _{l}\xi
_{k}^{+}+C_{kl}^{\ast }\xi _{k}\xi _{l}^{+}))\exp (-\frac{1}{2}%
\sum_{k\epsilon C}\xi _{k}\xi _{k}^{+})\,\chi _{P+}(\xi ,\xi ^{+})  \notag \\
&&+\exp (-\frac{1}{2}\sum_{k\epsilon C}\xi _{k}\xi _{k}^{+})\,Tr(\,\hat{%
\Omega}^{+}(\xi ^{+})\,\frac{\partial }{\partial t}\hat{\rho}\,\hat{\Omega}%
^{-}(\xi ))  \notag \\
&=&(\frac{1}{2}\sum_{k\epsilon C}\dsum\limits_{l\epsilon NC}(C_{kl}\,i\xi
_{l}\,i\xi _{k}^{+}+C_{kl}^{\ast }\,i\xi _{k}\,i\xi _{l}^{+}))\chi (\xi ,\xi
^{+})+\chi (\xi ,\xi ^{+};\frac{\partial }{\partial t}\hat{\rho}\,)  \notag
\\
&&  \label{Eq.TimeDerivCharFn}
\end{eqnarray}%
The first term in (\ref{Eq.TimeDerivCharFn}) then produces extra terms in
the Fokker-Planck equation. The last term in (\ref{Eq.TimeDerivCharFn}) is
the characteristic function that would occur if $\hat{\rho}$ is replaced by $%
\frac{{\LARGE \partial }}{{\LARGE \partial t}}\hat{\rho}\,$and this term
just gives the standard result for the Fokker-Planck equation.

The first term on the right side of (\ref{Eq.TimeDerivCharFn}) is given by%
\begin{eqnarray}
&&(\frac{1}{2}\sum_{k\epsilon C}\dsum\limits_{l\epsilon NC}(C_{kl}\,i\xi
_{l}\,i\xi _{k}^{+}+C_{kl}^{\ast }\,i\xi _{k}\,i\xi _{l}^{+}))\chi (\xi ,\xi
^{+})  \notag \\
&=&\int \int d^{2}\alpha \,d^{2}\alpha ^{+}\,\exp (i\sum\limits_{k\epsilon
C,NC}(\alpha _{k}\xi _{k}^{+}+\xi _{k}\alpha _{k}^{+}))  \notag \\
&&\times \left\{ \frac{1}{2}\sum_{k\epsilon C}\dsum\limits_{l\epsilon
NC}(C_{kl}\,\frac{\partial }{\partial \alpha _{l}^{+}}\,\frac{\partial }{%
\partial \alpha _{k}}+C_{kl}^{\ast }\,\frac{\partial }{\partial \alpha
_{k}^{+}}\,\frac{\partial }{\partial \alpha _{l}})\;P(\alpha ,\alpha
^{+},\alpha ^{\ast },\alpha ^{+\ast })\right\}
\label{Eq.TimeDerivCharFn_TimeDepModes}
\end{eqnarray}%
because using (\ref{Eq.DistnFn}) and applying integration by parts twice we
see that 
\begin{eqnarray}
&&i\xi _{l}\,i\xi _{k}^{+}\,\chi (\xi ,\xi ^{+})  \notag \\
&=&\int \int d^{2}\alpha \,d^{2}\alpha ^{+}\,\left( \frac{\partial }{%
\partial \alpha _{l}^{+}}\,\frac{\partial }{\partial \alpha _{k}}\exp
(i\sum\limits_{k\epsilon C,NC}(\alpha _{k}\xi _{k}^{+}+\xi _{k}\alpha
_{k}^{+}))\right) \,P(\alpha ,\alpha ^{+},\alpha ^{\ast },\alpha ^{+\ast }) 
\notag \\
&=&\int \int d^{2}\alpha \,d^{2}\alpha ^{+}\,\exp (i\sum\limits_{k\epsilon
C,NC}(\alpha _{k}\xi _{k}^{+}+\xi _{k}\alpha _{k}^{+}))\,\left\{ \frac{%
\partial }{\partial \alpha _{l}^{+}}\,\frac{\partial }{\partial \alpha _{k}}%
P(\alpha ,\alpha ^{+},\alpha ^{\ast },\alpha ^{+\ast })\right\}  \notag \\
&&i\xi _{k}\,i\xi _{l}^{+}\,\chi (\xi ,\xi ^{+})  \notag \\
&=&\int \int d^{2}\alpha \,d^{2}\alpha ^{+}\,\exp (i\sum\limits_{k\epsilon
C,NC}(\alpha _{k}\xi _{k}^{+}+\xi _{k}\alpha _{k}^{+}))\,\left\{ \frac{%
\partial }{\partial \alpha _{k}^{+}}\,\frac{\partial }{\partial \alpha _{l}}%
P(\alpha ,\alpha ^{+},\alpha ^{\ast },\alpha ^{+\ast })\right\}  \notag \\
&&
\end{eqnarray}

The second term on the right side of (\ref{Eq.TimeDerivCharFn}) is evaluated
using the \emph{correspondence rules} that apply when $\frac{{\LARGE %
\partial }}{{\LARGE \partial t}}\hat{\rho}$ evaluated using the
Liouville-von Neumann or master equation. The derivation of these rules is
set out in many textbooks (see for example \cite{Barnett97a}, Appendix 12)
and here we just present the results. For the \emph{non-condensat}e
annihilation, creation operators the characteristic function is of the
positive P \emph{normally ordered} type so we have%
\begin{eqnarray}
\hat{\rho} &\Rightarrow &\hat{a}_{k}\,\hat{\rho}\qquad \chi (\xi ,\xi
^{+})\Rightarrow \frac{\partial }{\partial (i\xi _{k}^{+})}\chi  \notag \\
\hat{\rho} &\Rightarrow &\hat{\rho}\,\hat{a}_{k}\qquad \chi (\xi ,\xi
^{+})\Rightarrow \left( \frac{\partial }{\partial (i\xi _{k}^{+})}+i\xi
_{k}\right) \chi  \notag \\
\hat{\rho} &\Rightarrow &\hat{a}_{k}^{\dag }\,\hat{\rho}\qquad \chi (\xi
,\xi ^{+})\Rightarrow \left( \frac{\partial }{\partial (i\xi _{k})}+i\xi
_{k}^{+}\right) \chi  \notag \\
\hat{\rho} &\Rightarrow &\hat{\rho}\,\hat{a}_{k}^{\dag }\,\qquad \chi (\xi
,\xi ^{+})\Rightarrow \frac{\partial }{\partial (i\xi _{k})}\chi
\label{Eq.BosonCharFnCorresp}
\end{eqnarray}%
For the \emph{condensate} annihilation, creation operators the
characteristic function is of the Wigner \emph{symmetrically ordered} type
so we have 
\begin{eqnarray}
\hat{\rho} &\Rightarrow &\hat{a}_{k}\hat{\rho}\qquad \chi (\xi ,\xi
^{+})\Rightarrow \left( \frac{\partial }{\partial (i\xi _{k}^{+})}-\frac{1}{2%
}i\xi _{k}\right) \chi  \notag \\
\hat{\rho} &\Rightarrow &\hat{\rho}\hat{a}_{k}\qquad \chi (\xi ,\xi
^{+})\Rightarrow \left( \frac{\partial }{\partial (i\xi _{k}^{+})}+\frac{1}{2%
}i\xi _{k}\right) \chi  \notag \\
\hat{\rho} &\Rightarrow &\hat{a}_{k}^{\dag }\hat{\rho}\qquad \chi (\xi ,\xi
^{+})\Rightarrow \left( \frac{\partial }{\partial (i\xi _{k})}+\frac{1}{2}%
i\xi _{k}^{+}\right) \chi  \notag \\
\hat{\rho} &\Rightarrow &\hat{\rho}\hat{a}_{k}^{\dag }\qquad \chi (\xi ,\xi
^{+})\Rightarrow \left( \frac{\partial }{\partial (i\xi _{k})}-\frac{1}{2}%
i\xi _{k}^{+}\right) \chi  \label{Eq.BoseSymmCharFnCorresp}
\end{eqnarray}%
From these results the corresponding changes to the distribution functions
can be deduced using (\ref{Eq.CharFn}) or (\ref{Eq.DistnFn}). This involves
integration by parts. The analyticity of the functions $\exp
i\sum_{k=1}^{n}\{\xi _{k}\alpha _{k}^{+}\}$ and $\exp
i\sum_{k=1}^{n}\{\alpha _{k}\xi _{k}^{+}\}$ enables two options for partial
differentiation ($\alpha _{kx}$ or $i\alpha _{ky}$ and $\alpha _{kx}^{+}$ or 
$i\alpha _{ky}^{+}$),to be obtained, but this need not be made explicit here
(see \cite{Drummond80a}, \cite{Gardiner91a} or \cite{Dalton10a}).

The correspondence rules for the \emph{non-condensat}e annihilation,
creation operators\emph{\ }are%
\begin{eqnarray}
\hat{\rho} &\Rightarrow &\hat{a}_{k}\,\hat{\rho}\qquad P(\alpha ,\alpha
^{+},\alpha ^{\ast },\alpha ^{+\ast })\Rightarrow \alpha _{k}\,P  \notag \\
\hat{\rho} &\Rightarrow &\hat{\rho}\,\hat{a}_{k}\qquad P(\alpha ,\alpha
^{+},\alpha ^{\ast },\alpha ^{+\ast })\Rightarrow \left( -\frac{\partial }{%
\partial \alpha _{k}^{+}}+\alpha _{k}\right) P\,  \notag \\
\hat{\rho} &\Rightarrow &\hat{a}_{k}^{\dag }\,\hat{\rho}\qquad P(\alpha
,\alpha ^{+},\alpha ^{\ast },\alpha ^{+\ast })\Rightarrow \left( -\frac{%
\partial }{\partial \alpha _{k}}+\alpha _{k}^{+}\right) P  \notag \\
\hat{\rho} &\Rightarrow &\hat{\rho}\,\hat{a}_{k}^{\dag }\,\qquad P(\alpha
,\alpha ^{+},\alpha ^{\ast },\alpha ^{+\ast })\Rightarrow \alpha _{k}^{+}P
\label{Eq.PosPCorr}
\end{eqnarray}%
The correspondence rules for the \emph{condensate} annihilation, creation
operators\emph{\ }are:%
\begin{eqnarray}
\hat{\rho} &\Rightarrow &\hat{a}_{k}\hat{\rho}\qquad P(\alpha ,\alpha
^{+},\alpha ^{\ast },\alpha ^{+\ast })\Rightarrow \left( \alpha _{k}+\frac{1%
}{2}\frac{\partial }{\partial \alpha _{k}^{+}}\right) P  \notag \\
\hat{\rho} &\Rightarrow &\hat{\rho}\hat{a}_{k}\qquad P(\alpha ,\alpha
^{+},\alpha ^{\ast },\alpha ^{+\ast })\Rightarrow \left( \alpha _{k}-\frac{1%
}{2}\frac{\partial }{\partial \alpha _{k}^{+}}\right) P  \notag \\
\hat{\rho} &\Rightarrow &\hat{a}_{k}^{\dag }\hat{\rho}\qquad P(\alpha
,\alpha ^{+},\alpha ^{\ast },\alpha ^{+\ast })\Rightarrow \left( \alpha
_{k}^{+}-\frac{1}{2}\frac{\partial }{\partial \alpha _{k}}\right) P  \notag
\\
\hat{\rho} &\Rightarrow &\hat{\rho}\hat{a}_{k}^{\dag }\qquad P(\alpha
,\alpha ^{+},\alpha ^{\ast },\alpha ^{+\ast })\Rightarrow \left( \alpha
_{k}^{+}+\frac{1}{2}\frac{\partial }{\partial \alpha _{k}}\right) P
\label{Eq.WignerCorr}
\end{eqnarray}%
where the $\alpha ,\alpha ^{+},\alpha ^{\ast },\alpha ^{+\ast }$ are
regarded as four independent complex variables.

By applying the correspondence rules (\ref{Eq.PosPCorr}) or (\ref%
{Eq.WignerCorr}) in succession to the various products of the density
operator with mode annihilation, creation operators that arise from the $%
\frac{{\LARGE \partial }}{{\LARGE \partial t}}\hat{\rho}$ term the second
term on the right hand side of Eq. (\ref{Eq.TimeDerivCharFn}) can be written
as 
\begin{eqnarray}
&&\chi (\xi ,\xi ^{+};\frac{\partial }{\partial t}\hat{\rho}\,)  \notag \\
&=&\int \int d^{2}\alpha \,d^{2}\alpha ^{+}\,\exp (i\sum\limits_{k\epsilon
C,NC}(\alpha _{k}\xi _{k}^{+}+\xi _{k}\alpha _{k}^{+}))\left\{
-\dsum\limits_{k}(\frac{\partial }{\partial \alpha _{k}}A_{k}^{-}+\frac{%
\partial }{\partial \alpha _{k}^{+}}A_{k}^{+})\right\}   \notag \\
&&\times P(\alpha ,\alpha ^{+},\alpha ^{\ast },\alpha ^{+\ast }  \notag \\
&&+\int \int d^{2}\alpha \,d^{2}\alpha ^{+}\,\exp (i\sum\limits_{k\epsilon
C,NC}(\alpha _{k}\xi _{k}^{+}+\xi _{k}\alpha _{k}^{+}))  \notag \\
&&\times \left\{ \frac{1}{2}\sum_{k}\dsum\limits_{l}(\frac{\partial }{%
\partial \alpha _{k}}\frac{\partial }{\partial \alpha _{l}}D_{kl}^{-\,-}+%
\frac{\partial }{\partial \alpha _{k}^{+}}\frac{\partial }{\partial \alpha
_{l}}D_{kl}^{+\,-}+\frac{\partial }{\partial \alpha _{k}}\frac{\partial }{%
\partial \alpha _{l}^{+}}D_{kl}^{-\,+}+\frac{\partial }{\partial \alpha
_{k}^{+}}\frac{\partial }{\partial \alpha _{l}^{+}}D_{kl}^{+\,+})\right\}  
\notag \\
&&\times P(\alpha ,\alpha ^{+},\alpha ^{\ast },\alpha ^{+\ast })
\label{Eq.TimeDerivCharFn_DRhoDt}
\end{eqnarray}

To obtain the specific forms of the drift $A_{k}^{-},A_{k}^{+}$ and
diffusion terms $D_{kl}^{-\,-},D_{kl}^{+\,-},D_{kl}^{-\,+},D_{kl}^{+\,+}$
the Liouville-von Neumann equation 
\begin{equation}
i\hbar \frac{\partial }{\partial t}\hat{\rho}=[\widehat{H},\widehat{\rho }]
\label{Eq.LVN}
\end{equation}%
is used in conjunction with the correspondence rules, with the Hamiltonian
obtained from Eq. (\ref{Eq.HamBosonFields}) by expanding the field operators
in terms of mode annihilation, creation operators using Eqs. (\ref%
{Eq.FieldOprExpn}). The Hamiltonian will then contain terms of the form $%
\hat{a}_{k}^{\dag }\hat{a}_{l}$ involving one mode creation operator and one
mode annihilation operator from the kinetic and trap potential energies, as
well as terms of the form $\hat{a}_{k}^{\dag }\hat{a}_{l}^{\dag }\hat{a}_{m}%
\hat{a}_{n}$ involving two mode creation operators and two mode annihilation
operators from the boson-boson interaction energy. These terms will be
multiplied by spatial integrals involving the mode functions - the latter
being assumed known (see Section \ref{SubSection2 - Determinination of Modes}%
). Division of the modes as condensate and non-condensate is needed to
identify whether Wigner or positive P correspondence rules apply. For the
condensate modes, third order derivatives occur in the Fokker-Planck
equation, but these are discarded as being too small to matter because of
the large occupancy of the condensate modes. It is well known that there
will only be first and second order derivatives associated with the
non-condensate modes, these being associated with the positive P
distribution.

We also need to consider the time derivative of the characteristic function
in the form Eq. (\ref{Eq.DistnFn}). Differentiation under the phase space
integral can be carried out without any issues associated with time
dependent differentials, as Eq. (\ref{Eq.PhaseSpaceIntegn}) shows. Also we
see that similar to the proof of (\ref{Eq.TimeIndepResults1})\ 
\begin{equation}
\frac{\partial }{\partial t}\sum\limits_{k}\alpha _{k}\xi _{k}^{+}=\frac{%
\partial }{\partial t}\sum\limits_{k}\xi _{k}\alpha _{k}^{+}=\frac{\partial 
}{\partial t}\sum\limits_{k}(\alpha _{k}\xi _{k}^{+}+\xi _{k}\alpha
_{k}^{+})=0  \label{Eq.TimeIndepResults2}
\end{equation}%
Hence we have%
\begin{equation}
\frac{\partial }{\partial t}\chi (\xi ,\xi ^{+})=\int \int d^{2}\alpha
\,d^{2}\alpha ^{+}\,\exp (i\sum\limits_{k\epsilon C,NC}(\alpha _{k}\xi
_{k}^{+}+\xi _{k}\alpha _{k}^{+}))\,\frac{\partial }{\partial t}P(\alpha
,\alpha ^{+},\alpha ^{\ast },\alpha ^{+\ast })\,
\label{Eq.TimeDerivCharFn_DistnFn}
\end{equation}

Substituting the results from (\ref{Eq.TimeDerivCharFn_DistnFn}), (\ref%
{Eq.TimeDerivCharFn_TimeDepModes}) and (\ref{Eq.TimeDerivCharFn_DRhoDt})
into Eq.(\ref{Eq.TimeDerivCharFn}) the \emph{Fokker-Planck equation} for the
distribution function can be obtained in the form%
\begin{eqnarray}
&&\frac{\partial }{\partial t}P(\alpha ,\alpha ^{+},\alpha ^{\ast },\alpha
^{+\ast })  \notag \\
&=&-\dsum\limits_{k}(\frac{\partial }{\partial \alpha _{k}}A_{k}^{-}+\frac{%
\partial }{\partial \alpha _{k}^{+}}A_{k}^{+})\;P(\alpha ,\alpha ^{+},\alpha
^{\ast },\alpha ^{+\ast })  \notag \\
&&+\frac{1}{2}\sum_{k}\dsum\limits_{l}(\frac{\partial }{\partial \alpha _{k}}%
\frac{\partial }{\partial \alpha _{l}}D_{kl}^{-\,-}+\frac{\partial }{%
\partial \alpha _{k}^{+}}\frac{\partial }{\partial \alpha _{l}}%
D_{kl}^{+\,-})\;P(\alpha ,\alpha ^{+},\alpha ^{\ast },\alpha ^{+\ast }) 
\notag \\
&&+\frac{1}{2}\sum_{k}\dsum\limits_{l}(\frac{\partial }{\partial \alpha _{k}}%
\frac{\partial }{\partial \alpha _{l}^{+}}D_{kl}^{-\,+}+\frac{\partial }{%
\partial \alpha _{k}^{+}}\frac{\partial }{\partial \alpha _{l}^{+}}%
D_{kl}^{+\,+})\;P(\alpha ,\alpha ^{+},\alpha ^{\ast },\alpha ^{+\ast }) 
\notag \\
&&+\frac{1}{2}\sum_{k\epsilon C}\dsum\limits_{l\epsilon NC}(C_{kl}\,\frac{%
\partial }{\partial \alpha _{l}^{+}}\,\frac{\partial }{\partial \alpha _{k}}%
+C_{kl}^{\ast }\,\frac{\partial }{\partial \alpha _{k}^{+}}\,\frac{\partial 
}{\partial \alpha _{l}})\;P(\alpha ,\alpha ^{+},\alpha ^{\ast },\alpha
^{+\ast })  \notag \\
&&  \label{Eq.Fokker-Planck0}
\end{eqnarray}%
As the expressions are now becoming rather lengthy we now change to a
simpler notation.

\subsubsection{Notation Change}

We first introduce the symbol $A$ to designate condensate and non-condensate
modes, where $A=C,NC$. The mode functions $\phi _{k}(x,t\mathbf{)}$ and
their conjugates $\phi _{k}^{\ast }(x,t\mathbf{)}$ will now be designated $%
\phi _{Ak}(x,t\mathbf{)}$ and their conjugates $\phi _{Ak}^{\ast }(x,t%
\mathbf{)}$, where $k=1,2,..,n_{A}$ lists the separate condensate or
non-condensate modes. Both the $\phi _{Ak}(x,t\mathbf{)}$ and $\phi
_{Ak}^{\ast }(x,t\mathbf{)}$ may finally be listed as $\phi _{Ak}^{\mu }(x,t%
\mathbf{)}$ with $\mu =-,+$, so that $\phi _{Ak}^{-}(x,t\mathbf{)=}\phi
_{Ak}(x,t\mathbf{),}\phi _{Ak}^{+}(x,t\mathbf{)=}\phi _{Ak}^{\ast }(x,t%
\mathbf{)}$. Note that if there are $n_{C}$ condensate modes and $n_{NC}$
non-condensate modes, the total number of $\phi _{Ak}^{\mu }(x,t\mathbf{)}$
will be $2(n_{C}+n_{NC})$, which is twice the total number of modes. This is
because there are $n=$ $n_{C}+n_{NC}$ modes $\phi _{k}(x,t\mathbf{)}$ plus
their complex conjugates $\phi _{k}^{\ast }(x,t\mathbf{)}$. The phase space
variables will now be written as $\alpha _{\mu Ak}$ and those in the
characteristic function as $\xi _{\mu Ak}$. The equations for the phase
space and the characteristic function variables now become%
\begin{eqnarray}
\frac{\partial \alpha _{\mu Ak}(t)}{\partial t} &=&\dsum\limits_{Bl}C_{Ak%
\,Bl}^{\mu }(t)\alpha _{\mu Bl}(t)  \label{Eq.TimeDerivModePhases2} \\
\frac{\partial \xi _{\mu Ak}(t)}{\partial t} &=&\dsum\limits_{Bl}C_{Ak%
\,Bl}^{\mu }(t)\xi _{\mu Bl}(t)  \label{Eq.TimeDepCharFnVar2}
\end{eqnarray}%
with 
\begin{eqnarray}
C_{Ak\,Bl}^{-}(t) &=&\dint dx\frac{\partial \phi _{Ak}^{\ast }(x,t\mathbf{)}%
}{\partial t}\phi _{Bl}(x,t\mathbf{)=}C_{Ak\,Bl}(t)  \notag \\
C_{Ak\,Bl}^{+}(t) &=&\dint dx\frac{\partial \phi _{Ak}(x,t\mathbf{)}}{%
\partial t}\phi _{Bl}^{\ast }(x,t\mathbf{)=}C_{Ak\,Bl}^{\ast }(t)
\label{Eq.Coupling Coefts2} \\
C_{Ak\,Bl}^{\mu } &=&-C_{Bl\,Ak}^{-\mu }  \label{Eq.CouplingCoeftResult2}
\end{eqnarray}%
Note that the sums in (\ref{Eq.TimeDerivModePhases2}) and (\ref%
{Eq.TimeDepCharFnVar2}) include both condensate and non-condensate modes.
The original drift, diffusion terms are now designated as $A_{Ak}^{\mu
},D_{Ak\,Bl}^{\mu \nu }$. The time derivatives of the modes are now written
as%
\begin{equation}
\frac{\partial }{\partial t}\phi _{Ak}^{\mu
}(x,t)=\dsum\limits_{Bl}C_{Ak\,Bl}^{-\mu }\phi _{Bl}^{\mu }(x,t)
\label{Eq.TimeDerivModes2}
\end{equation}

The orthonormality and completeness relationships will now be 
\begin{eqnarray}
\int dx\phi _{Ak}^{-\mu }(x,t\mathbf{)}\phi _{Bl}^{\mu }(x,t\mathbf{)}
&=&\delta _{AB}\delta _{kl}  \notag \\
\sum_{k}\phi _{Ak}^{\mu }(x,t\mathbf{)}\phi _{Ak}^{-\mu }(y,t\mathbf{)} &%
\mathbf{=}&\delta _{\mu A}(x,y)  \label{Eq.OrthonormComp3}
\end{eqnarray}%
where $\delta _{\mu A}(x,y)$ is the restricted delta function for the $\mu A$
modes and the projectors $\mathcal{P}_{x}^{\mu A}$ now are%
\begin{eqnarray}
\mathcal{P}_{x}^{\mu A}[F(x)] &=&\int dy\,\delta _{\mu A}(x,y)F(y)  \notag \\
\mathcal{P}_{x}^{\mu A}[\dsum\limits_{k}\beta _{k}\phi _{Ak}^{\mu }(x,t%
\mathbf{)+}\dsum\limits_{\substack{ Bl  \\ B\neq A}}\beta _{l}\phi
_{Bl}^{\mu }(x,t)\mathbf{]} &\mathbf{=}&\dsum\limits_{k}\beta _{k}\phi
_{Ak}^{\mu }(x,t\mathbf{)}  \label{Eq.Projectors2}
\end{eqnarray}%
so that the projector $\mathcal{P}_{x}^{\mu A}$ acting on an arbitrary
function written as a linear combination of the modes $\phi _{Ak}^{\mu }(x,t%
\mathbf{)}$ and orthogonal modes $\phi _{Bl}^{\mu }(x,t)$ $(B\neq A)$
projects the function onto just the linear combination of the modes $\phi
_{Ak}^{\mu }(x,t\mathbf{)}$.

\subsubsection{Fokker-Planck Equation Result}

In terms of the new notation the \emph{Fokker-Planck equation} will be of
the form%
\begin{eqnarray}
&&\frac{\partial }{\partial t}P(\alpha ,\alpha ^{+},\alpha ^{\ast },\alpha
^{+\ast })  \notag \\
&=&\left\{ -\dsum\limits_{\mu Ak}\frac{\partial }{\partial \alpha _{\mu Ak}}%
A_{Ak}^{\mu }+\frac{1}{2}\dsum\limits_{\mu Ak}\dsum\limits_{\nu Bl}\frac{%
\partial }{\partial \alpha _{\mu Ak}}\frac{\partial }{\partial \alpha _{\nu
Bl}}E_{Ak\,Bl}^{\mu \nu }\right\} P(\alpha ,\alpha ^{+},\alpha ^{\ast
},\alpha ^{+\ast })  \notag \\
&&  \label{Eq.NewFokker-Planck}
\end{eqnarray}%
where $A_{Ak}^{\mu }$ is the \emph{drift vector} and $E_{Ak\,Bl}^{\mu \nu }$
is the \emph{diffusion matrix.} The diffusion matrix is given by 
\begin{equation}
E_{Ak\,Bl}^{\mu \nu }=D_{Ak\,Bl}^{\mu \nu }+\frac{1}{2}(\delta
_{A\,C}\,\delta _{B\,NC}\,\delta _{\mu \,-\nu }C_{Ak\,Bl}^{\mu }+\delta
_{B\,C}\,\delta _{A\,NC}\,\delta _{\nu \,-\mu }C_{Bl\,Ak}^{\nu })
\label{Eq.NewDiffusionMatrix}
\end{equation}%
The first term $D_{Ak\,Bl}^{\mu \nu }$ is from the standard derivation of
the Fokker-Planck equation via applying the standard correspondence rules to
the terms in the characteristic function that arise from $\frac{{\LARGE %
\partial }}{{\LARGE \partial t}}\hat{\rho}$, the second arises from the time
dependence of the condensate and non-condensate mode functions. Note that
the diffusion matrix is still symmetric%
\begin{equation}
E_{Ak\,Bl}^{\mu \nu }=E_{Bl\,Ak}^{\nu \mu }  \label{Eq.SymmDiffusion}
\end{equation}

The drift $A_{Ak}^{\mu }$ vector and the diffusion matrix $E_{Ak\,Bl}^{\mu
\nu }$ are functions of the $\alpha _{\mu Ak}$. In the present approach
using time dependent phase variables the drift vector $A_{Ak}^{\mu }$ vector
is the \emph{same} as those that would be obtained if time independent modes
were used, but the diffusion matrix $E_{Ak\,Bl}^{\mu \nu }$ is \emph{changed}
from the expected $D_{Ak\,Bl}^{\mu \nu }$. The Fokker-Planck equation can
also be expressed in terms of real variables which involve the real and
imaginary components of the phase variables $\alpha _{\mu Ak}$, but we will
not do that here. In deriving the Fokker-Planck equation there are often
terms involving third and higher order derivatives arising from the
condensate modes and their Wigner representation. These are usually small,
scaling as higher powers of $1/\sqrt{N}$, and hence are discarded. Thus we
see that the Fokker-Planck equation has changed in this hybrid approach from
that for time independent modes.

\subsection{Langevin (Ito) Stochastic Equations}

\label{SubSection 2. ItoSDE}

In this section we determine the \emph{Langevin equations} that are
equivalent to the Fokker-Planck equation for the distribution function.
These will be in the form of Ito stochastic differential equations. The
derivation is based on that given by Gardiner \cite{Gardiner91a}, but
modified to allow for the phase variables being time dependent. The final
stochastic averages would determine normally ordered quantum correlation
functions associated with non-condensate operators, or symmetrically ordered
quantum correlation functions associated with condensate operators. Note
also that the Langevin equation derivation does not depend on the
distribution function having any particular properties, such as being real
or positive. We must of course assume it to be non-analytic in
general.\medskip

\subsubsection{Phase Space Average}

The phase space average of the functions, $F(\alpha ,\alpha ^{+})$ is given
by%
\begin{equation}
\left\langle F(\alpha ,\alpha ^{+})\right\rangle _{t}=\int \int d^{2}\alpha
^{+}d^{2}\alpha \,F(\alpha ,\alpha ^{+})\,P(\alpha ,\alpha ^{+},\alpha
^{\ast },\alpha ^{+\ast },t)  \label{Eq.BosePhaseSpaceAver}
\end{equation}%
For determining the quantum correlation function $%
G(l_{1},l_{2},..l_{p};m_{q},..,m_{2},m_{1}::j_{1},j_{2},..j_{r};k_{s},..,k_{2},k_{1}) 
$ in Eq.(\ref{Eq.QCFHybridCase}) the function is

$F(\alpha ,\alpha ^{+})=$ $[\alpha _{l_{1}}^{+}\,\alpha _{l_{2}}^{+}..\alpha
_{l_{p}}^{+}\,\alpha _{m_{q}}..\,\alpha _{m_{2}}\alpha _{m_{1}}][\alpha
_{j_{1}}^{+}\,\alpha _{j_{2}}^{+}..\alpha _{j_{r}}^{+}\,\alpha
_{k_{s}}..\,\alpha _{k_{2}}\alpha _{k_{1}}]$. The phase space average will
change with time not only because the distribution function $P(\alpha
,\alpha ^{+},\alpha ^{\ast },\alpha ^{+\ast },t)$ is time dependent, but
also because $F(\alpha ,\alpha ^{+})$ is now time dependent since from Eq.(%
\ref{Eq.TimeDerivModePhases}) the $\alpha ,\alpha ^{+}$ depend on time. Note
the differentials $\int \int d^{2}\alpha ^{+}d^{2}\alpha \,$\ are time
independent.

The change in the phase space average from times $t$ to $t+\delta t$ due to
all factors is given by%
\begin{eqnarray}
&&\left\langle F(\alpha ,\alpha ^{+})\right\rangle _{t+\delta
t}-\left\langle F(\alpha ,\alpha ^{+})\right\rangle _{t}  \notag \\
&=&\left( \int \int d^{2}\alpha ^{+}d^{2}\alpha \,F(\alpha ,\alpha
^{+})\,\left\{ \frac{\partial }{\partial t}P(\alpha ,\alpha ^{+},\alpha
^{\ast },\alpha ^{+\ast },t)\right\} \right) \delta t  \notag \\
&&+\left( \int \int d^{2}\alpha ^{+}d^{2}\alpha \,\left\{ \frac{\partial }{%
\partial t}F(\alpha ,\alpha ^{+})\right\} \,P(\alpha ,\alpha ^{+},\alpha
^{\ast },\alpha ^{+\ast },t)\right) \delta t
\end{eqnarray}%
The second term in the last equation is absent in the standard treatment 
\cite{Gardiner91a} based on time independent phase variables. This result
gives the change in the phase space average correct to $O(\delta t)$.

Now from (\ref{Eq.TimeDerivModePhases2}) 
\begin{eqnarray}
\frac{\partial }{\partial t}F(\alpha ,\alpha ^{+}) &=&\dsum\limits_{\mu Ak}%
\frac{\partial }{\partial \alpha _{\mu Ak}}F(\alpha ,\alpha ^{+})\frac{%
\partial \alpha _{\mu Ak}}{\partial t}  \notag \\
&=&\dsum\limits_{\mu Ak}\frac{\partial }{\partial \alpha _{\mu Ak}}F(\alpha
,\alpha ^{+})\sum_{Bl}C_{Ak\,Bl}^{\mu }\,\alpha _{\mu Bl}
\label{Eq.ResultDerivF}
\end{eqnarray}

Substituting for $\frac{{\LARGE \partial }}{{\LARGE \partial t}}P$ from the
Fokker-Planck equation (\ref{Eq.NewFokker-Planck}) and for $\frac{{\LARGE %
\partial }}{{\LARGE \partial t}}F$ from (\ref{Eq.ResultDerivF}) we get using
integration by parts and assuming that the distribution function goes to
zero fast enough on the phase space boundary%
\begin{eqnarray}
&&\left\langle F(\alpha ,\alpha ^{+})\right\rangle _{t+\delta
t}-\left\langle F(\alpha ,\alpha ^{+})\right\rangle _{t}  \notag \\
&=&\int \int d^{2}\alpha ^{+}d^{2}\alpha \,\left\{ \sum_{\mu Ak}\left[ \frac{%
\partial }{\partial \alpha _{\mu Ak}}F(\alpha ,\alpha ^{+})\right] \,\left[
A_{Ak}^{\mu }+\sum_{Bl}C_{Ak\,Bl}^{\mu }\,\alpha _{\mu Bl}\right] \right\} 
\notag \\
&&\times P(\alpha ,\alpha ^{+},\alpha ^{\ast },\alpha ^{+\ast },t)\delta t 
\notag \\
&&+\int \int d^{2}\alpha ^{+}d^{2}\alpha \,\left\{ \frac{1}{2}%
\dsum\limits_{\mu Ak}\dsum\limits_{\nu Bl}\left[ \frac{\partial }{\partial
\alpha _{\mu Ak}}\frac{\partial }{\partial \alpha _{\nu Bl}}F(\alpha ,\alpha
^{+})\right] \,E_{Ak\,Bl}^{\mu \nu }\right\}  \notag \\
&&\times P(\alpha ,\alpha ^{+},\alpha ^{\ast },\alpha ^{+\ast },t)\delta t 
\notag \\
&&  \label{Eq.ChangePhaseSpaceAverage}
\end{eqnarray}%
where the terms involving first order derivatives have been combined.

Hence 
\begin{eqnarray}
&&\frac{d}{dt}\left\langle F(\alpha ,\alpha ^{+})\right\rangle _{t}  \notag
\\
&=&\left\langle \left\{ \sum_{\mu Ak}\left[ \frac{\partial }{\partial \alpha
_{\mu Ak}}F(\alpha ,\alpha ^{+})\right] \,\left[ A_{Ak}^{\mu
}+\sum_{Bl}C_{Ak\,Bl}^{\mu }\,\alpha _{\mu Bl}\right] \right\} \right\rangle
\notag \\
&&+\left\langle \left\{ \frac{1}{2}\dsum\limits_{\mu Ak}\dsum\limits_{\nu Bl}%
\left[ \frac{\partial }{\partial \alpha _{\mu Ak}}\frac{\partial }{\partial
\alpha _{\nu Bl}}F(\alpha ,\alpha ^{+})\right] \,E_{Ak\,Bl}^{\mu \nu
}\right\} \right\rangle  \label{Eq.TimeDerivPhaseSpAverBose}
\end{eqnarray}%
giving the time derivative of the phase space average of $F(\alpha ,\alpha
^{+})$ as the phase space average of the sum of the two quantities in the $%
\{\}$ brackets. Note that no specific properties of the distribution
function were needed.\emph{\ }This result will be used later to derive
results for time derivatives of quantum correlation functions.\medskip

\subsubsection{Stochastic Variables Average}

For the Ito stochastic approach we now replace the variables $\alpha _{\mu
Ak}$ by stochastic variables $\alpha _{\mu Ak}^{s}$. In terms of these
stochastic variables the stochastic average at time $t$ of $F(\alpha ,\alpha
^{+})$ is given by%
\begin{equation}
\overline{F(\alpha ^{s}(t),\alpha ^{s+}(t))}=\frac{1}{N}\sum_{i=1}^{N}f(%
\alpha _{\mu Ak\,i}^{s}(t))  \label{Eq.BoseStochasticAver}
\end{equation}%
where $\alpha _{\mu Ak\,i}^{s}(t)$ is the $i$th member of the stochastic
ensemble of $N$ samples and $f(\alpha _{\mu Ak\,}^{s}(t))$ is the same as $%
F(\alpha ^{s}(t),\alpha ^{s+}(t))$. The key idea is that the phase space
average at any time $t$ of arbitrary functions $F(\alpha ,\alpha ^{+})$ and
the stochastic average of such functions are made to coincide when the
stochastic equation for the $\alpha _{\mu Ak}^{s}(t)$ is suitably related to
the Fokker-Planck equation for the distribution function $P(\alpha ,\alpha
^{+},\alpha ^{\ast },\alpha ^{+\ast },t)$. Thus%
\begin{eqnarray}
\left\langle F(\alpha ,\alpha ^{+})\right\rangle _{t} &=&\int \int
d^{2}\alpha ^{+}d^{2}\alpha \,F(\alpha ,\alpha ^{+})\,P(\alpha ,\alpha
^{+},\alpha ^{\ast },\alpha ^{+\ast },t)  \notag \\
&=&\overline{F(\alpha ^{s}(t),\alpha ^{s+}(t))}  \notag \\
&=&\frac{1}{N}\sum_{i=1}^{N}f(\alpha _{\mu Aki}^{s}(t))
\label{Eq.PhaseSpStochCorrespBose}
\end{eqnarray}%
In turn, the phase space averages are related to quantum averages. In
particular the normally ordered\emph{\ }quantum correlation functions are
given by the stochastic average of the product of stochastic c-number phase
space variables 
\begin{equation}
\left\langle \widehat{a}_{l_{1}}^{\dag }\,\widehat{a}_{l_{2}}^{\dag }..%
\widehat{a}_{l_{p}}^{\dag }\widehat{a}_{m_{q}}\,..\widehat{a}_{m_{2}}%
\widehat{a}_{m_{1}}\right\rangle _{t}=\overline{(\alpha
_{m_{q}}^{s}(t)\,.\,\alpha _{m_{1}}^{s}(t))\,(\alpha
_{l_{1}}^{+s}(t)\,..\alpha _{l_{p}}^{+s}(t))}
\label{Eq.BoseGeneralQCFStochastic}
\end{equation}%
For bosonic systems such stochastic averages involving c-numbers can be
carried out numerically, and this method is often more efficient than having
to determine the full distribution function. .

Now the difference in the stochastic average from time $t$ to $t+\delta t$
is given by%
\begin{eqnarray}
&&\overline{F(\alpha ^{s}(t+\delta t),\alpha ^{s+}(t+\delta t))}-\overline{%
F(\alpha ^{s}(t),\alpha ^{s+}(t))}  \notag \\
&=&\overline{\{F(\alpha ^{s}(t+\delta t),\alpha ^{s+}(t+\delta t))-F(\alpha
^{s}(t),\alpha ^{s+}(t))\}}  \notag \\
&=&\overline{\{f(\alpha _{\mu Ak\,i}^{s}(t+\delta t))-f(\alpha _{\mu
Ak\,i}^{s}(t))\}}  \notag \\
&=&\overline{\left\{ \sum_{\mu Ak}\left[ \frac{\partial }{\partial \alpha
_{\mu Ak}}F(\alpha ,\alpha ^{+})\right] \delta \alpha _{\mu
Ak}^{s}(t)\right\} }  \notag \\
&&+\overline{\left\{ \frac{1}{2}\sum_{\mu Ak}\sum_{\nu Bl}\left[ \frac{%
\partial }{\partial \alpha _{\mu Ak}}\frac{\partial }{\partial \alpha _{\nu
Bl}}F(\alpha ,\alpha ^{+})\right] \delta \alpha _{\mu Ak}^{s}(t)\delta
\alpha _{\nu Bl}^{s}(t)+..\right\} }  \notag \\
&&  \label{Eq.ChangeStochasticAver}
\end{eqnarray}%
where we have used a Taylor expansion for $f(\alpha _{\mu Ak}^{s}(t+\delta
t))$ with the notation 
\begin{equation}
\delta \alpha _{\mu Ak}^{s}(t)=\alpha _{\mu Ak}^{s}(t+\delta t)-\alpha _{\mu
Ak}^{s}(t)  \label{Eq.BoseStochFlucn}
\end{equation}%
for the \emph{fluctuation }in $\alpha _{\mu Ak}^{s}$. In obtaining the last
result the stochastic average of a sum being equal to the sum of the
stochastic averages has been used. It is necessary to consider fluctuations
of the second order because - as we will see - stochastic averages of such
terms are of $O(\delta t)$.

Now suppose $\alpha _{\mu Ak}^{s}(t)$ satisfies an \emph{Ito stochastic
equation} of the form%
\begin{eqnarray}
&&\alpha _{\mu Ak}^{s}(t+\delta t)-\alpha _{\mu Ak}^{s}(t)  \notag \\
&=&\mathcal{A}_{Ak}^{\mu }(\alpha _{\xi Dm}^{s}(t))\delta t+\sum_{a}\mathcal{%
B}_{Aka}^{\mu }(\alpha _{\xi Dm}^{s}(t))\int_{t}^{t+\delta t}dt_{1}\Gamma
_{a}(t_{1})  \label{Eq.ItoSDEBose1}
\end{eqnarray}%
which gives the change in $\alpha _{\mu Ak}^{s}(t)$ correct to $O(\delta t)$%
. The \emph{Ito stochastic equation} is also written as a differential
equation%
\begin{equation}
\frac{\partial }{\partial t}\alpha _{\mu Ak}^{s}=\mathcal{A}_{Ak}^{\mu
}(\alpha _{\xi Dm}^{s}(t))+\sum_{a}\mathcal{B}_{Ak\,a}^{\mu }(\alpha _{\xi
Dm}^{s}(t))\,\Gamma _{a}(t_{+})  \label{Eq.ItoSDEBose3}
\end{equation}%
where the stochastic behaviour is due to the \emph{Gaussian-Markoff random
noise} terms $\Gamma _{a}$. The \emph{aim} is to find expressions for the $%
\mathcal{A}_{Ak}^{\mu }$ and the $\mathcal{B}_{Ak\,a}^{\mu }$ so that the
phase space and stochastic averages coincide for an \emph{arbitrary} choice
of $F(\alpha ,\alpha ^{+})$.

The $\Gamma _{a}(t)$ are real Gaussian-Markoff random noise terms $(a=1,2,..)
$ whose stochastic averages are given by%
\begin{eqnarray}
\overline{\Gamma _{a}(t_{1})} &=&0  \notag \\
\overline{\Gamma _{a}(t_{1})\Gamma _{b}(t_{2}}) &=&\delta _{ab}\delta
(t_{1}-t_{2})  \notag \\
\overline{\Gamma _{a}(t_{1})\Gamma _{b}(t_{2})\Gamma _{c}(t_{3})} &=&0 
\notag \\
\overline{\Gamma _{a}(t_{1})\Gamma _{b}(t_{2})\Gamma _{c}(t_{3})\Gamma
_{d}(t_{4})} &=&\overline{\Gamma _{a}(t_{1})\Gamma _{b}(t_{2}})\,\overline{%
\Gamma _{c}(t_{3})\Gamma _{d}(t_{4}})+\overline{\Gamma _{a}(t_{1})\Gamma
_{c}(t_{3}})\,\overline{\Gamma _{b}(t_{2})\Gamma _{d}(t_{4}})  \notag \\
&&+\overline{\Gamma _{a}(t_{1})\Gamma _{d}(t_{4}})\,\overline{\Gamma
_{b}(t_{2})\Gamma _{c}(t_{3}})  \notag \\
&&...  \label{Eq.GaussianMarkov}
\end{eqnarray}%
so that the stochastic averages of products of odd numbers of $\Gamma $ are
zero and the stochastic averages of products of even numbers of $\Gamma $
are the sums of products of stochastic averages of pairs of $\Gamma $. For
the moment we leave the number of $\Gamma _{a}$ unspecified, the number will
turn out to be $2(n_{C}+n_{NC})$. It is also assumed that any function $h$
of the $\alpha _{\mu Ak}^{s}(t)$ at time $t$ and the $\Gamma _{a}(t)$ at
later times are \emph{uncorrelated} - this is the meaning of the $\Gamma
_{a}(t_{+})$ in (\ref{Eq.ItoSDEBose3}). Thus 
\begin{eqnarray}
&&\overline{h(\alpha _{\mu Ak}^{s}(t_{1}))\Gamma _{a}(t_{2})\Gamma
_{b}(t_{3})\Gamma _{c}(t_{4})..\Gamma _{k}(t_{l})}  \notag \\
&=&\overline{h(\alpha _{\mu Ak}^{s}(t_{1}))}\,\;\overline{\Gamma
_{a}(t_{2})\Gamma _{b}(t_{3})\Gamma _{c}(t_{4})..\Gamma _{k}(t_{l})}\qquad
t_{1}<t_{2},t_{3},..,t_{l}  \label{Eq.UnCorrelResultBose}
\end{eqnarray}

With these results we can now obtain expressions for the stochastic averages
in Eq.(\ref{Eq.ChangeStochasticAver}). Details are given in Appendix \ref%
{Appendix - Derivation of Langevin Equations}.

For the first order derivative terms we find that 
\begin{eqnarray}
&&\overline{\left\{ \sum_{\mu Ak}\left[ \frac{\partial }{\partial \alpha
_{\mu Ak}}F(\alpha ,\alpha ^{+})\right] \delta \alpha _{\mu
Ak}^{s}(t)\right\} }  \notag \\
&=&\overline{\sum_{\mu Ak}\left[ \frac{\partial }{\partial \alpha _{\mu Ak}}%
F(\alpha ,\alpha ^{+})\right] \mathcal{A}_{Ak}^{\mu }(\alpha _{\xi
Dm}^{s}(t))}\,\delta t  \label{Eq.BoseStochFirstDeriv}
\end{eqnarray}%
where the stochastic average rules for sums and products have been used, the
non-correlation between the averages of functions of $\alpha _{\mu Ak}^{s}(t)
$ at time $t$ and the $\Gamma $ at later times between $t$ to $t+\delta t$
is applied, and the term involving $\overline{\Gamma _{a}(t_{1})}$ is equal
to zero from (\ref{Eq.GaussianMarkov}). Note that the first order term is
proportional to $\delta t$.

For the second order derivative terms, we have on expanding the product $%
\delta \alpha _{\mu Ak}^{s}(t)\delta \alpha _{\nu Bl}^{s}(t)$, using the
stochastic average of a sum being the same as the sum of stochastic
averages, the result that the stochastic averages for the functions of the $%
\alpha _{\xi Dm}^{s}(t)$ and the $\Gamma _{a}(t_{+})$ are uncorrelated, and
after using the result that terms $\overline{\Gamma _{a}(t_{1})}$ involving
single $\Gamma ^{\prime }s$ give zero together with the result for terms $%
\overline{\Gamma _{a}(t_{1})\Gamma _{b}(t_{2})}$ involving two $\Gamma
^{\prime }s$ 
\begin{eqnarray}
\overline{\int_{t}^{t+\delta t}dt_{1}\Gamma _{a}(t_{1})\int_{t}^{t+\delta
t}dt_{2}\Gamma _{b}(t_{2})} &=&\int_{t}^{t+\delta t}dt_{1}\int_{t}^{t+\delta
t}dt_{2}\;\overline{\Gamma _{a}(t_{1})\Gamma _{b}(t_{2})}  \notag \\
&=&\int_{t}^{t+\delta t}dt_{1}\int_{t}^{t+\delta t}dt_{2}\;\delta
_{ab}\delta (t_{1}-t_{2})  \notag \\
&=&\delta _{ab}\,\delta t  \label{Eq.StochAverProdTwoGamma}
\end{eqnarray}%
we find that the second order derivative term is 
\begin{eqnarray}
&&\overline{\left\{ \frac{1}{2}\sum_{\mu Ak}\sum_{\nu Bl}\left[ \frac{%
\partial }{\partial \alpha _{\mu Ak}}\frac{\partial }{\partial \alpha _{\nu
Bl}}F(\alpha ,\alpha ^{+})\right] \delta \alpha _{\mu Ak}^{s}(t)\delta
\alpha _{\nu Bl}^{s}(t)\right\} }  \notag \\
&=&\overline{\frac{1}{2}\sum_{\mu Ak}\sum_{\nu Bl}\left[ \frac{\partial }{%
\partial \alpha _{\mu Ak}}\frac{\partial }{\partial \alpha _{\nu Bl}}%
F(\alpha ,\alpha ^{+})\right] \left[ \sum_{a}\mathcal{B}_{Aka}^{\mu }(\alpha
_{\xi Dm}^{s}(t))\;\mathcal{B}_{Bla}^{\nu }(\alpha _{\xi Dm}^{s}(t))\right] }%
\;\delta t  \notag \\
&=&\overline{\frac{1}{2}\sum_{\mu Ak}\sum_{\nu Bl}\left[ \frac{\partial }{%
\partial \alpha _{\mu Ak}}\frac{\partial }{\partial \alpha _{\nu Bl}}%
F(\alpha ,\alpha ^{+})\right] \left[ [\mathcal{B}(\alpha _{\xi Dm}^{s}(t))%
\mathcal{\ B}^{T}(\alpha _{\xi Dm}^{s}(t))]_{Ak,Bl}^{\mu ,\nu }\right] }%
\;\delta t  \notag \\
&&  \label{Eq.BoseStochSecondDeriv}
\end{eqnarray}%
after neglecting terms of order $\delta t^{2}$ Note that the second order
term is also proportional to $\delta t$

The remaining terms give stochastic averages correct to order $\delta t^{2}$
or higher so that we have correct to first order in $\delta t$ 
\begin{eqnarray}
&&\overline{F(\alpha ^{s}(t+\delta t),\alpha ^{s+}(t+\delta t))}-\overline{%
F(\alpha ^{s}(t),\alpha ^{s+}(t))}  \notag \\
&=&\left\{ \overline{\sum_{\mu Ak}\left[ \frac{\partial }{\partial \alpha
_{\mu Ak}}F(\alpha ,\alpha ^{+})\right] \mathcal{A}_{Ak}^{\mu }(\alpha _{\xi
Dm}^{s}(t))}\,\right\} \;\delta t  \notag \\
&&+\left\{ \overline{\frac{1}{2}\sum_{\mu Ak}\sum_{\nu Bl}\left[ \frac{%
\partial }{\partial \alpha _{\mu Ak}}\frac{\partial }{\partial \alpha _{\nu
Bl}}F(\alpha ,\alpha ^{+})\right] \left[ [\mathcal{B}(\alpha _{\xi
Dm}^{s}(t))\mathcal{\ B}^{T}(\alpha _{\xi Dm}^{s}(t))]_{Ak,Bl}^{\mu ,\nu }%
\right] }\right\} \;\delta t  \notag \\
&&  \label{Eq.ChangeStochAverBose}
\end{eqnarray}%
or 
\begin{eqnarray}
&&\frac{d}{dt}\overline{F(\alpha ^{s}(t),\alpha ^{s+}(t))}  \notag \\
&=&\overline{\sum_{\mu Ak}\left[ \frac{\partial }{\partial \alpha _{\mu Ak}}%
F(\alpha ,\alpha ^{+})\right] \mathcal{A}_{Ak}^{\mu }(\alpha _{\xi
Dm}^{s}(t))}  \notag \\
&&+\overline{\frac{1}{2}\sum_{\mu Ak}\sum_{\nu Bl}\left[ \frac{\partial }{%
\partial \alpha _{\mu Ak}}\frac{\partial }{\partial \alpha _{\nu Bl}}%
F(\alpha ,\alpha ^{+})\right] \left[ [\mathcal{B}(\alpha _{\xi Dm}^{s}(t))%
\mathcal{\ B}^{T}(\alpha _{\xi Dm}^{s}(t))]_{Ak,Bl}^{\mu ,\nu }\right] } 
\notag \\
&&  \label{Eq.TimeDerivStochAverBose}
\end{eqnarray}%
This result is exactly the same as for the standard treatment based on time
independent mode functions, since the Ito stochastic equations (\ref%
{Eq.ItoSDEBose3}) have exactly the same form in that situation.

The result (\ref{Eq.TimeDerivStochAverBose}) for the stochastic average will
be the \emph{same} as that in (\ref{Eq.TimeDerivPhaseSpAverBose}) based on
the phase space average if we have the following relationships between the
matrices $A$ and $E$ in the Fokker-Planck equation and the matrices $%
\mathcal{A}$ and $\mathcal{B}$ occurring in the Ito stochastic differential
equation. 
\begin{eqnarray}
\mathcal{A}_{Ak}^{\mu } &=&A_{Ak}^{\mu }+\sum_{Bl}C_{Ak\,Bl}^{\mu }\,\alpha
_{\mu Bl}  \notag \\
\lbrack \mathcal{B\ B}^{T}]_{Ak,Bl}^{\mu ,\nu } &=&E_{Ak\,Bl}^{\mu \nu }
\label{Eq.RelnFPEMatricesSDEMatricesBose2}
\end{eqnarray}%
If the modes were time independent, the coupling constants would all be zero
and the relationship would have been $\mathcal{A}_{Ak}^{\mu }=A_{Ak}^{\mu }$
and $[\mathcal{B\ B}^{T}]_{Ak,Bl}^{\mu ,\nu }=D_{Ak\,Bl}^{\mu \nu }$, which
relates the Ito equation quantities $\mathcal{A}$ and $\mathcal{B}$ to the
drift $A$ and diffusion $D$ terms that occur in the standard Fokker-Planck
equation obtained from just the $\frac{{\LARGE \partial }}{{\LARGE \partial t%
}}\hat{\rho}$ term in the time derivative of the characteristic function.
This is the usual condition found in the textbooks \cite{Gardiner91a}. For
the \emph{hybrid case} with \emph{time dependent modes} not only is the
relationship changed for the \emph{diffusion} terms, but it is also changed
for the \emph{drift} terms. The difference is entirely due to the presence
of the coupling terms $C_{Ak\,Bl}^{\mu }$ that depend on the time
dependences of the modes $\phi _{Ak}^{\mu }(x,t\mathbf{)}$. Clearly, if the
drift vector $A$ from the Fokker-Planck equation and the matrix $C$ from the
time dependence of the mode functions are known then the vector $\mathcal{A}$
in the Ito stochastic equation can be found. It is known \cite{Drummond80a}
that the \emph{complex} symmetric matrix $E$ may be factorised in the form $%
KK^{T}=E$, so the matrix $\mathcal{B}$ in the Ito stochastic equation can
also be determined. 
\begin{eqnarray}
\mathcal{B} &\mathcal{=}&K  \notag \\
KK^{T} &=&E  \label{Eq.Takagi}
\end{eqnarray}%
This result is known as the \emph{Takagi} factorisation \cite{Takagi25a}.
The proof is given in \cite{Horn85a} (see section 4.3). The construction
involves the eigenvectors of the matrix $EE^{\ast }$, which is hermitian
because $E$ is symmetric, and which also has non-negative real eigenvalues
This result is not well-known, and does not require $E$ to be positive
semi-definite, as is sometimes thought to be the case. Note that in general $%
\mathcal{B}$ is a \emph{complex} $2n\times 2n$ matrix. Also note that $%
\mathcal{B}$ is not unique, since with any orthogonal matrix $\mathcal{R}$
we also have $(\mathcal{BR})(\mathcal{BR})^{T}=E$. Again, the total number
of Gaussian-Markoff noise terms $\Gamma _{a}(t_{+})$ is $2n$, the total
number of $\alpha _{Ak}$ and $\alpha _{Ak}^{+}$, or equivalently twice the
total number of modes.

\subsection{Expressions for Ito Stochastic Equations}

We can also write the Ito stochastic equation (\ref{Eq.ItoSDEBose3})\ in
terms of quantities that appear in the Fokker-Planck equation. From Eq. (\ref%
{Eq.RelnFPEMatricesSDEMatricesBose2}) we have for the \emph{Ito stochastic
phase variable equation }in the hybrid case 
\begin{equation}
\frac{\partial }{\partial t}\alpha _{\mu Ak}^{s}=A_{Ak}^{\mu
}+\sum_{l}C_{Ak\,Bl}^{\mu }\,\alpha _{\mu Bl}^{s}+\sum_{a}K_{Aka}^{\mu
}\,\Gamma _{a}(t_{+})  \label{Eq.ItoSDEFinalHyb}
\end{equation}%
where $A$ is the drift vector and $K$ is related to the diffusion matrix $E$
via 
\begin{eqnarray}
\dsum\limits_{a}K_{Aka}^{\mu }K_{Bla}^{\nu } &=&E_{Ak\,Bl}^{\mu \nu }
\label{Eq.DiffusionMatrixHybridFact0} \\
&=&D_{Ak\,Bl}^{\mu \nu }+\frac{1}{2}(\delta _{A\,C}\,\delta _{B\,NC}\,\delta
_{\mu \,-\nu }C_{Ak\,Bl}^{\mu }+\delta _{B\,C}\,\delta _{A\,NC}\,\delta
_{\nu \,-\mu }C_{Bl\,Ak}^{\nu })  \notag
\end{eqnarray}%
This form of the Ito equation is the most useful as it involves the drift,
diffusion terms in the Fokker-Planck equation (\ref{Eq.NewFokker-Planck})
plus the coupling coefficients that allow for the modes being time
dependent. Note that the diffusion terms (\ref{Eq.NewDiffusionMatrix}) in
the Fokker-Planck equation also involve the coupling constants. Thus the
equivalence between the quantities $\mathcal{A}$, $\mathcal{B}$ in the Ito
stochastic equations and the drift $A$, diffusion $D$ terms in the \emph{%
standard} Fokker-Planck equation and the mode time dependence matrix $C$ can
be established. There is an extra term $\sum_{l}C_{Ak\,Bl}^{\mu }\,\alpha
_{\mu Bl}^{s}$ (which is linear in the phase variables) in \ the drift term,
and an extra term $\frac{1}{2}(\delta _{A\,C}\,\delta _{B\,NC}\,\delta _{\mu
\,-\nu }C_{Ak\,Bl}^{\mu }+\delta _{B\,C}\,\delta _{A\,NC}\,\delta _{\nu
\,-\mu }C_{Bl\,Ak}^{\nu })$ (which is independent of the phase variables) in
the diffusion term, which are not present in the standard treatment
involving time independent modes.\textbf{\ }

\subsection{Classical and Noise Terms}

We can write the Ito stochastic equation in terms of a \emph{classical} term
and a \emph{noise} term%
\begin{eqnarray}
\frac{\partial }{\partial t}\alpha _{\mu Ak}^{s} &\mathbf{=}&\left( \frac{%
\partial }{\partial t}\alpha _{\mu Ak}^{s}\right) _{class}\mathbf{+}\left( 
\frac{\partial }{\partial t}\alpha _{\mu Ak}^{s}\right) _{noise}  \notag \\
\left( \frac{\partial }{\partial t}\alpha _{\mu Ak}^{s}\right) _{class}
&=&A_{Ak}^{\mu }+\sum_{l}C_{Ak\,Bl}^{\mu }\,\alpha _{\mu Bl}^{s}  \notag \\
\left( \frac{\partial }{\partial t}\alpha _{\mu Ak}^{s}\right) _{noise}
&=&\sum_{a}K_{Aka}^{\mu }\,\Gamma _{a}(t_{+})
\label{Eq.ItoPhaseHybClassNoise}
\end{eqnarray}%
If only the classical terms were included, then the solution for $\alpha
_{\mu Ak}^{s}$ would determine classical trajectories in phase space; hence
their name.

\subsection{Properties of Noise}

The stochastic averages of the noise terms can now be evaluated. For a
single noise term and the product of two noise terms we have using Eqs. (\ref%
{Eq.GaussianMarkov}) and (\ref{Eq.DiffusionMatrixHybridFact0}) 
\begin{eqnarray}
\overline{\left( \frac{\partial }{\partial t}\alpha _{\mu Ak}^{s}(t)\right)
_{n}} &=&0  \notag \\
\overline{\left( \frac{\partial }{\partial t}\alpha _{\mu
Ak}^{s}(t_{1})\right) _{n}\left( \frac{\partial }{\partial t}\alpha _{\nu
Bl}^{s}(t_{2})\right) _{n}} &=&\delta (t_{1}-t_{2})\overline{E_{Ak\,Bl}^{\mu
\nu }(t_{1,2})}  \label{Eq.NoisePhaseHybProps1}
\end{eqnarray}%
showing that the stochastic average of a single phase noise term is zero
whilst that for the product of two phase noise terms is delta correlated in
time and equal to the appropriate diffusion matrix element.

In fact the diffusion matrix elements determine all the stochastic averages
of products of phase noise terms. With an odd number of terms the stochastic
average is zero. For an even number of terms the stochastic average involves
sums of stochastic averages of products of diffusion matrix elements, which
reflects the Gaussian-Markoff properties of the $\Gamma _{a}$. Thus for
three and four phase noise terms%
\begin{equation}
\overline{\left( \frac{\partial }{\partial t}\alpha _{\mu
Ak}^{s}(t_{1})\right) _{n}\left( \frac{\partial }{\partial t}\alpha _{\nu
Bl}^{s}(t_{2})\right) _{n}\left( \frac{\partial }{\partial t}\alpha _{\xi
Cm}^{s}(t_{3})\right) _{n}}=0  \label{Eq.NoisePhaseHybProps2}
\end{equation}%
\begin{eqnarray}
&&\overline{\left( \frac{\partial }{\partial t}\alpha _{\mu
Ak}^{s}(t_{1})\right) _{n}\left( \frac{\partial }{\partial t}\alpha _{\nu
Bl}^{s}(t_{2})\right) _{n}\left( \frac{\partial }{\partial t}\alpha _{\xi
Cm}^{s}(t_{3})\right) _{n}\left( \frac{\partial }{\partial t}\alpha
_{\lambda Dp}^{s}(t_{4})\right) _{n}}  \notag \\
&=&\delta (t_{1}-t_{2})\delta (t_{3}-t_{4})\;\overline{E_{AkBl}^{\mu \nu
}(t_{1,2})E_{CmDp}^{\xi \lambda }(t_{3,4})}  \notag \\
&&+\delta (t_{1}-t_{3})\delta (t_{2}-t_{4})\;\overline{E_{AkCm}^{\mu \xi
}(t_{1,3})E_{BlDp}^{\nu \lambda }(t_{2,4})}  \notag \\
&&+\delta (t_{1}-t_{4})\delta (t_{2}-t_{3})\;\overline{E_{AkDp}^{\mu \lambda
}(t_{1,4})E_{BlCm}^{\nu \xi }(t_{2,3})}  \label{Eq.NoisePhaseHybProps3}
\end{eqnarray}%
Note that the phase noise terms are not themselves Gaussian-Markoff
processes.\pagebreak

\section{Theory - Quantum Field Treatment}

\label{Section 3 - Quantum Field Case}

In this section the field operator and functional phase space approach is
developed for the situation treated in Section \ref{Section 2 Separate Modes
Case}, where the modes are divided into condensate and non-condensate modes.
In general, the phase space functional approach is based on representing
quantum density operator by a \emph{quasi-distribution functional} in a
phase space involving \emph{field functions} which replace the bosonic field
annihilation, creation operators. Functionals and key properties are
summarised in Appendix \ref{Appendix - Functional Calculus}. In this hybrid
approach time dependent \emph{condensate} and \emph{non-condensate field
operators} are involved and these are represented by time dependent
condensate and non-condensate field functions. The distribution functional
will be of the Wigner type for the condensate field and of the positive P
type for the non-condensate field. As shown in Appendix \ref{Appendix
Equivalence of Separate Mode and Field Theory}, this treatment is equivalent
to the treatment in Section \ref{Section 2 Separate Modes Case} based on 
\emph{quasi-distribution functions} of \emph{phase space variables} that
replace the annihilation, creation operators for the separate modes. Details
of derivations are contained in Appendices \ref{Appendix Derivation of
Functional Fokker-Planck Equation} and \ref{Appendix Derivation of Ito
Stochastic Field Equations}.

\subsection{Condensate and Non-Condensate Field Operators}

Field operators $\hat{\Psi}_{C}(x\mathbf{)}$ and $\hat{\Psi}_{NC}(x\mathbf{)}
$ for condensate and non-condensate modes respectively may also be defined
in which the sums in (\ref{Eq.FieldOprExpn}) are over \emph{restricted} sets
of modes \cite{Dalton10a}. The \emph{condensate} and \emph{non-condensate
field operators} are 
\begin{eqnarray}
\hat{\Psi}_{C}(x,t\mathbf{)} &\mathbf{=}&\sum_{k\epsilon C}\widehat{a}%
_{k}(t)\phi _{k}(x,t\mathbf{)},\qquad \hat{\Psi}_{C}^{\dag }(x,t\mathbf{)=}%
\sum_{k\epsilon C}\widehat{a}_{_{k}}^{\dag }(t)\phi _{k}^{\ast }(x,t\mathbf{)%
}  \notag \\
\hat{\Psi}_{NC}(x,t\mathbf{)} &\mathbf{=}&\sum_{k\epsilon NC}\widehat{a}%
_{k}(t)\phi _{k}(x,t\mathbf{)},\qquad \hat{\Psi}_{NC}^{\dag }(x,t\mathbf{)=}%
\sum_{k\epsilon NC}\widehat{a}_{_{k}}^{\dag }(t)\phi _{k}^{\ast }(x,t\mathbf{%
)}  \label{Eq.CondNonCondFldOprs}
\end{eqnarray}%
where the mode sums are restricted to be \emph{only} over condensate or
non-condensate modes. Their sum gives the total field operators 
\begin{equation}
\hat{\Psi}(x)=\hat{\Psi}_{C}(x,t)+\hat{\Psi}_{NC}(x,t)\qquad \hat{\Psi}%
^{\dag }(x)=\hat{\Psi}_{C}^{\dag }(x,t)+\hat{\Psi}_{NC}^{\dag }(x,t)
\label{Eq.TotalFldOprs}
\end{equation}%
which satisfy (\ref{Eq.FieldOprExpn}).

These operators satisfy commutation rules involving \emph{restricted delta
functions} (see (\ref{Eq.CompletenessRestrict})). 
\begin{eqnarray}
\lbrack \hat{\Psi}_{C}(x,t),\hat{\Psi}_{C}^{\dagger }(y,t)] &=&\delta
_{C}(x,y)\qquad \lbrack \hat{\Psi}_{NC}(x,t),\hat{\Psi}_{NC}^{\dagger
}(y,t)]=\delta _{NC}(x,y)  \notag \\
\lbrack \hat{\Psi}_{C}(x,t),\hat{\Psi}_{NC}^{\dagger }(y,t)] &=&[\hat{\Psi}%
_{NC}(x,t),\hat{\Psi}_{C}^{\dagger }(y,t)]=0  \label{Eq.CondNonCondCommRules}
\\
\delta _{C}(x,y)+\delta _{C}(x,y) &=&\delta (x-y)  \label{Eq.DeltaSum}
\end{eqnarray}%
and condensate and non-condensate operators commute. In terms of projectors 
\begin{equation}
\hat{\Psi}_{C}(x\mathbf{)=}\mathcal{P}_{x}^{C}[\hat{\Psi}(x)]\qquad \hat{\Psi%
}_{NC}(x\mathbf{)=}\mathcal{P}_{x}^{NC}[\hat{\Psi}(x)]
\label{Eq.CondNonCondProj}
\end{equation}

We note that the condensate and non-condensate field operators are \emph{%
time dependent} - it is only their sum that is time independent, as shown in
(\ref{Eq.TimeDerivTotalFldOprs}). The feature that the separate field
operators are time dependent requires significant modifications to the
standard derivations of the FFPE and Ito stochastic field equations. We find
that 
\begin{eqnarray}
\frac{\partial }{\partial t}\hat{\Psi}_{C}(x,t) &\mathbf{=}%
&\sum\limits_{k\epsilon C}\dsum\limits_{l\epsilon NC}C_{kl}\widehat{a}%
_{l}(t)\phi _{k}(x,t)+\sum\limits_{l\epsilon C}\dsum\limits_{k\epsilon NC}%
\widehat{a}_{l}(t)C_{lk}^{\ast }\phi _{k}(x,t)  \notag \\
&\neq &0  \label{Eq.TimeDepCondFldOpr}
\end{eqnarray}%
because now there are pairs of $k,l$ such as $k\epsilon C,l\epsilon NC$ for
which the cancellation of $C_{kl}$ via the required $C_{lk}^{\ast }$ using (%
\ref{Eq.RelnCouplingCoefts}) cannot occur, since for the latter $l\epsilon
C,k\epsilon NC$. The other three condensate and non-condensate field
operators are also time dependent.

\subsection{Condensate, Non-Condensate and Total Fields}

The \emph{total} field operators $\hat{\Psi}(x),\hat{\Psi}^{\dagger }(x)$ in
Eqs. (\ref{Eq.FieldOprExpn}) are represented by \emph{total} field
functions, which are defined by equations analogous to (\ref{Eq.FieldOprExpn}%
) for the field operators 
\begin{equation}
\psi (x,t\mathbf{)=}\sum_{k}\alpha _{k}(t)\phi _{k}(x,t\mathbf{)=}\psi (x%
\mathbf{)}\qquad \psi ^{+}(x,t\mathbf{)=}\sum_{k}\alpha _{_{k}}^{+}(t)\phi
_{k}^{\ast }(x,t\mathbf{)=}\psi ^{+}(x\mathbf{)}  \label{Eq.FieldFns}
\end{equation}%
where $\alpha _{k}(t),\alpha _{k}^{+}(t)$ are time dependent mode phase
variables that satisfy Eq. (\ref{Eq.TimeDerivModePhases}). Using Eq.(\ref%
{Eq.TimeDerivModes1}) and $C_{kl}+C_{lk}^{\ast }=0$ we see that the total
field functions are actually \emph{time independent}.%
\begin{equation}
\frac{\partial }{\partial t}\psi (x,t\mathbf{)}=0\qquad \frac{\partial }{%
\partial t}\psi ^{+}(x,t\mathbf{)}=0  \label{Eq.TimeIndepFieldFns}
\end{equation}%
This result depended on \emph{all} of the $\phi _{k}(x,t\mathbf{)}$ or $\phi
_{k}^{\ast }(x,t\mathbf{)}$ being involved in the field functions, since
expanding $\frac{{\LARGE \partial }}{{\LARGE \partial t}}\phi _{k}(x,t)$ or $%
\frac{{\LARGE \partial }}{{\LARGE \partial t}}\phi _{k}^{\ast }(x,t)$
involves all modes. Modifications are needed when \emph{restricted sets} of
modes are involved for condensate and non-condensate field functions.

However, in the \emph{hybrid approach} where condensate and non-condensate
modes are treated differently, the phase space functional approach is based
on representing quantum density operator by a \emph{quasi-distribution
functional} in a phase space involving four \emph{field functions} which
replace the bosonic field annihilation, creation operators $\hat{\Psi}%
_{C}(x,t),\hat{\Psi}_{C}^{\dagger }(x,t),\hat{\Psi}_{NC}(x,t),\hat{\Psi}%
_{NC}^{\dagger }(x,t)$. The \emph{condensate} and \emph{non-condensate field
functions} are defined by equations analogous to Eq. (\ref%
{Eq.CondNonCondFldOprs}) for the field operators 
\begin{eqnarray}
\psi _{C}(x,t\mathbf{)} &\mathbf{=}&\sum_{k\epsilon C}\alpha _{k}(t)\phi
_{k}(x,t\mathbf{)}\qquad \psi _{C}^{+}(x,t\mathbf{)=}\sum_{k\epsilon
C}\alpha _{_{k}}^{+}(t)\phi _{k}^{\ast }(x,t\mathbf{)}  \notag \\
\psi _{NC}(x,t\mathbf{)} &\mathbf{=}&\sum_{k\epsilon NC}\alpha _{k}(t)\phi
_{k}(x,t\mathbf{)}\qquad \psi _{NC}^{+}(x,t\mathbf{)=}\sum_{k\epsilon
NC}\alpha _{_{k}}^{+}(t)\phi _{k}^{\ast }(x,t\mathbf{)}  \label{Eq.FieldFns2}
\end{eqnarray}%
where $\alpha _{k}(t),\alpha _{k}^{+}(t)$ are time dependent mode phase
variables that satisfy Eq. (\ref{Eq.TimeDerivModePhases}). In this hybrid
case the field functions are \emph{time dependent}. Similar to Eq. (\ref%
{Eq.TimeDepCondFldOpr}) the time dependences of the field functions are 
\begin{eqnarray}
\frac{\partial }{\partial t}\psi _{C}(x,t) &\mathbf{=}&\sum\limits_{k%
\epsilon C}\dsum\limits_{l\epsilon NC}C_{kl}\alpha _{l}(t)\phi
_{k}(x,t)+\sum\limits_{l\epsilon C}\dsum\limits_{k\epsilon NC}\alpha
_{l}(t)C_{lk}^{\ast }\phi _{k}(x,t)  \notag \\
\frac{\partial }{\partial t}\psi _{C}^{+}(x,t) &\mathbf{=}%
&\sum\limits_{k\epsilon C}\dsum\limits_{l\epsilon NC}C_{kl}^{\ast }\alpha
_{l}^{+}(t)\phi _{k}^{\ast }(x,t)+\sum\limits_{l\epsilon
C}\dsum\limits_{k\epsilon NC}\alpha _{l}^{+}(t)C_{lk}\phi _{k}^{\ast }(x,t) 
\notag \\
\frac{\partial }{\partial t}\psi _{NC}(x,t) &\mathbf{=}&\sum\limits_{k%
\epsilon NC}\dsum\limits_{l\epsilon C}C_{kl}\alpha _{l}(t)\phi
_{k}(x,t)+\sum\limits_{l\epsilon NC}\dsum\limits_{k\epsilon C}\alpha
_{l}(t)C_{lk}^{\ast }\phi _{k}(x,t)  \notag \\
\frac{\partial }{\partial t}\psi _{NC}^{+}(x,t) &\mathbf{=}%
&\sum\limits_{k\epsilon NC}\dsum\limits_{l\epsilon C}C_{kl}^{\ast }\alpha
_{l}^{+}(t)\phi _{k}^{\ast }(x,t)+\sum\limits_{l\epsilon
NC}\dsum\limits_{k\epsilon C}\alpha _{l}^{+}(t)C_{lk}\phi _{k}^{\ast }(x,t) 
\notag \\
&&  \label{Eq.TimeDepCondNonCondFldFns}
\end{eqnarray}%
However, in this case there is incomplete cancellation because $C_{kl}$ and $%
C_{lk}^{\ast }$ involve different pairs of $k,l$ so these derivatives are
non-zero.

\subsection{Characteristic and Distribution Functionals}

The field theory phase space approach involves introducing \emph{%
characteristic functionals} that can be used to specify all the quantum
field correlation functions for a given density operator.We first define
four time dependent \emph{characteristic field functions} via 
\begin{eqnarray}
\Xi _{C}(x,t\mathbf{)} &\mathbf{=}&\sum_{k\epsilon C}\xi _{k}(t)\phi _{k}(x,t%
\mathbf{)}\qquad \Xi _{C}^{+}(x,t\mathbf{)=}\sum_{k\epsilon C}\xi
_{_{k}}^{+}(t)\phi _{k}^{\ast }(x,t\mathbf{)}  \notag \\
\Xi _{NC}(x,t\mathbf{)} &\mathbf{=}&\sum_{k\epsilon NC}\xi _{k}(t)\phi
_{k}(x,t\mathbf{)}\qquad \Xi _{NC}^{+}(x,t\mathbf{)=}\sum_{k\epsilon NC}\xi
_{_{k}}^{+}(t)\phi _{k}^{\ast }(x,t\mathbf{)}
\label{Eq.ConNonCondCharFldFns}
\end{eqnarray}%
with the same time dependent c-numbers $\xi _{k}(t),\xi _{k}^{+}(t)$ as
before. These satisfy Eq.(\ref{Eq.TimeDepCharFnVar}). These sum to total
characteristic field functions%
\begin{equation}
\Xi ^{+}=\Xi _{C}^{+}+\Xi _{NC}^{+}\qquad \Xi ^{-}=\Xi _{C}^{-}+\Xi _{NC}^{-}
\label{Eq.TotalCharFldFns}
\end{equation}

The \emph{hybrid characteristic functional} $\chi \lbrack \Xi _{C},\Xi
_{C}^{+},\Xi _{NC},\Xi _{NC}^{+}]$ is defined via 
\begin{eqnarray}
\chi \lbrack \Xi _{C},\Xi _{C}^{+},\Xi _{NC},\Xi _{NC}^{+}] &=&Tr(\hat{\Omega%
}^{W}[\Xi _{C},\Xi _{C}^{+}]\,\hat{\Omega}^{+}[\Xi _{NC}^{+}]\,\hat{\rho}\,%
\hat{\Omega}^{-}[\Xi _{NC}])  \notag \\
\hat{\Omega}^{+}[\Xi _{NC}^{+}] &=&\exp i\dint dx\,\hat{\Psi}_{NC}(x,t%
\mathbf{)\,}\Xi _{NC}^{+}(x,t\mathbf{)}  \notag \\
\hat{\Omega}^{-}[\Xi _{NC}] &=&\exp i\dint dx\,\Xi _{NC}(x,t\mathbf{)}\,\hat{%
\Psi}_{NC}^{\dag }(x,t\mathbf{)}  \notag \\
\hat{\Omega}^{W}[\Xi _{C},\Xi _{C}^{+}] &=&\exp i\dint dx\,(\hat{\Psi}%
_{C}(x,t\mathbf{)\,}\Xi _{C}^{+}(x,t\mathbf{)+}\,\Xi _{C}(x,t\mathbf{)}\,%
\hat{\Psi}_{C}^{\dag }(x,t\mathbf{))}  \notag \\
&&  \label{Eq.HybridCharFnal}
\end{eqnarray}%
By applying the Baker-Hausdorff theorem we find that%
\begin{eqnarray}
\hat{\Omega}^{W}[\Xi _{C},\Xi _{C}^{+}] &=&\hat{\Omega}^{-}[\Xi _{C}]\,\hat{%
\Omega}^{+}[\Xi _{C}^{+}]\,\exp (-\frac{1}{2}\dint dx\,\Xi _{C}(x,t\mathbf{%
)\,}\Xi _{C}^{+}(x,t\mathbf{))}  \notag \\
\hat{\Omega}^{+}[\Xi _{C}^{+}] &=&\exp i\dint dx\,\hat{\Psi}_{C}(x,t\mathbf{%
)\,}\Xi _{C}^{+}(x,t\mathbf{)}  \notag \\
\hat{\Omega}^{-}[\Xi _{C}] &=&\exp i\dint dx\,\Xi _{C}(x,t\mathbf{)}\,\hat{%
\Psi}_{C}^{\dag }(x,t\mathbf{)}  \label{Eq.BakerHauss3}
\end{eqnarray}%
we find that the overall characteristic functional is given by 
\begin{equation}
\chi \lbrack \Xi _{C},\Xi _{C}^{+},\Xi _{NC},\Xi _{NC}^{+}]=\exp \left\{ -%
\frac{1}{2}\dint dx\,\Xi _{C}(x,t\mathbf{)\,}\Xi _{C}^{+}(x,t\mathbf{)}%
\right\} \,\chi _{P+}[\Xi _{C},\Xi _{C}^{+},\Xi _{NC},\Xi _{NC}^{+}]
\label{Eq.CharFnReln2}
\end{equation}%
where%
\begin{eqnarray}
&&\chi _{P+}[\Xi _{C},\Xi _{C}^{+},\Xi _{NC},\Xi _{NC}^{+}]  \notag \\
&=&Tr(\left\{ \exp i\dint dx\,(\hat{\Psi}_{C}(x,t\mathbf{)\,}\Xi _{C}^{+}(x,t%
\mathbf{)+}\hat{\Psi}_{NC}(x,t\mathbf{)\,}\Xi _{NC}^{+}(x,t\mathbf{))}%
\right\} \,\hat{\rho}\,  \notag \\
&&\times \left\{ \exp i\dint dx(\Xi _{C}(x,t\mathbf{)}\,\hat{\Psi}_{C}^{\dag
}(x,t\mathbf{)+}\Xi _{NC}(x,t)\,\hat{\Psi}_{NC}^{\dag }(x,t))\right\}  
\notag \\
&=&Tr(\left\{ \exp i\dint dx\,(\hat{\Psi}(x\mathbf{)\,}\Xi ^{+}(x\mathbf{))}%
\right\} \,\hat{\rho}\,\left\{ \exp i\dint dx(\Xi (x\mathbf{)}\,\hat{\Psi}%
^{\dag }(x\mathbf{)})\right\}   \label{Eq.AuxCharFnal}
\end{eqnarray}%
is an auxiliary characteristic functional. Similar to the separate mode
situation, this characteristic functional is that applying if both the
condensate and non-condensate fields were treated via a \emph{normally
ordered} characteristic functional. Here $\Xi (x\mathbf{),}\Xi ^{+}(x\mathbf{%
)}$ are as in Eq. (\ref{Eq.TotalCharFldFns}) for the total field and $\hat{%
\Psi}(x\mathbf{),}\hat{\Psi}^{\dag }(x\mathbf{)}$ are the total field
operators as in Eq.(\ref{Eq.FieldOprExpn}), both involving \emph{all} the
modes and both of which are \emph{time independent}. Details in deriving (%
\ref{Eq.AuxCharFnal}) are set out in Appendix \ref{Appendix Derivation of
Functional Fokker-Planck Equation}. Apart from the density operator the only
quantity in the expression (\ref{Eq.CharFnReln2}) for the characteristic
functional that is time dependent is the factor $\exp \left\{ -\frac{1}{2}%
\dint dx\,\Xi _{C}(x,t\mathbf{)\,}\Xi ^{+}(x,t\mathbf{)}\right\} $, and this
simplifies the derivation of the functional Fokker-Planck equation. Eq. (\ref%
{Eq.CharFnReln2}) relates the \emph{actual} characteristic functional to
that for the case where all modes are treated via a \emph{normally ordered}
characteristic functional.

By comparing the expression (\ref{Eq.CharFnReln2}) for the characteristic
functional with (\ref{Eq.CharFnReln}) for the hybrid characteristic function
it follows that the characteristic functionals and characteristic functions
are \emph{equivalent}. They both contain the same information - in one case
as a functional of the fields $\Xi _{C},\Xi _{C}^{+},\Xi _{NC},\Xi _{NC}^{+}$
in the other as a function of all the expansion coefficients $\{\xi
_{k}(t),\xi _{k}^{+}(t)\}$ for both condensate and non-condensate modes. 
\begin{equation}
\chi \lbrack \Xi _{C},\Xi _{C}^{+},\Xi _{NC},\Xi _{NC}^{+}]\equiv \chi (\xi
,\xi ^{+})  \label{Eq.CharFnalCharFnReln2}
\end{equation}

The phase space integral that relates the distribution function to the
characteristic function can be expressed as a phase space functional
integral in which the distribution function $P(\alpha ,\alpha ^{+},\alpha
^{\ast },\alpha ^{+\ast })$ is replaced by equivalent distribution
functionals 
\begin{equation}
P[\underrightarrow{\psi },\underrightarrow{\psi }^{\ast }]=P(\alpha ,\alpha
^{+},\alpha ^{\ast },\alpha ^{+\ast })  \label{Eq.DistFnalDistFnReln2}
\end{equation}%
where for short we introduce the notation 
\begin{equation}
\underrightarrow{\psi }\equiv \{\psi _{C},\psi _{C}^{+},\psi _{NC},\psi
_{NC}^{+}\}\qquad \underrightarrow{\psi }^{\ast }\equiv \{\psi _{C}^{\ast
},\psi _{C}^{+\ast },\psi _{NC}^{\ast },\psi _{NC}^{+\ast }\}
\label{Eq.NotationCondNonCondFlds}
\end{equation}%
The equivalence of functional and phase space integration based on separate
modes can be established using the methods of functional calculus (see
Appendix \ref{Appendix - Functional Calculus}, see Eq.(\ref%
{Eq.PhaseSpaceIntegral3})). From Eq.(\ref{Eq.PhaseSpaceIntegn}) the phase
space integration is time independent even though the phase variables depend
on time. Thus 
\begin{eqnarray}
&&\int \int \int \int D^{2}\psi _{C}\,\,D^{2}\psi _{C}^{+}D^{2}\psi
_{NC}\,D^{2}\psi _{NC}^{+}  \notag \\
&=&\int \int \int \int d^{2}\alpha _{C}\,d^{2}\alpha _{C}^{+}d^{2}\alpha
_{NC}\,d^{2}\alpha _{NC}^{+}  \notag \\
&=&\int \int \int \int d^{2}\alpha _{C}(0)\,d^{2}\alpha
_{C}^{+}(0)\,d^{2}\alpha _{NC}(0)\,d^{2}\alpha _{NC}^{+}(0)
\label{Eq.PhaseSpaceFunctionalIntegn2}
\end{eqnarray}%
indicating that functional integration is also essentially non time
dependent. Note that all modes are involved as are the four fields $\psi
_{C},\psi _{C}^{+},\psi _{NC},\psi _{NC}^{+}$ which are equivalent to the
set of all $\alpha _{k},\alpha _{k}^{+}$. Note that the functional
integration $\int D^{2}\psi _{C}$ and $\int D^{2}\psi _{C}^{+}$ involve a
space grid with the same number of intervals as condensate modes. ).

The characteristic function is still related to the quasi-distribution
function $P(\alpha ,\alpha ^{+},\alpha ^{\ast },\alpha ^{+\ast })$ via Eq.(%
\ref{Eq.DistnFn}) so writing this relationship in terms of functional
integration we have an equivalent relation between the hybrid characteristic
and \emph{distribution functionals}%
\begin{eqnarray}
&&\chi \lbrack \Xi _{C},\Xi _{C}^{+},\Xi _{NC},\Xi _{NC}^{+}]  \notag \\
&=&\int \int \int \int D^{2}\psi _{C}\,\,D^{2}\psi _{C}^{+}\,D^{2}\psi
_{NC}\,D^{2}\psi _{NC}^{+}  \notag \\
&&\times \exp (i\dint dx\,(\psi _{C}(x,t\mathbf{)\,}\Xi _{C}^{+}(x,t\mathbf{%
)+}\,\Xi _{C}(x,t\mathbf{)}\,\psi _{C}^{+}(x,t\mathbf{)))}\,  \notag \\
&&\times \exp (i\dint dx\,\psi _{NC}(x,t\mathbf{)\,}\Xi _{NC}^{+}(x,t\mathbf{%
)})\,P[\underrightarrow{\psi },\underrightarrow{\psi }^{\ast }]\,\exp
(i\dint dx\,\Xi _{NC}(x,t\mathbf{)}\,\psi _{NC}^{+}(x,t\mathbf{)})  \notag \\
&=&\int \int \int \int D^{2}\psi _{C}\,\,D^{2}\psi _{C}^{+}D^{2}\psi
_{NC}\,D^{2}\psi _{NC}^{+}  \notag \\
&&\times \exp (i\dint dx\,(\psi _{C}(x,t\mathbf{)\,}\Xi _{C}^{+}(x,t\mathbf{%
)+}\psi _{NC}(x,t)\,\Xi _{NC}^{+}(x,t)\mathbf{+}\,\Xi _{C}(x,t\mathbf{)}%
\,\psi _{C}^{+}(x,t\mathbf{)+\,}\Xi _{NC}(x,t)\,\psi _{NC}^{+}(x,t)\mathbf{)}%
)  \notag \\
&&\times P[\psi _{C},\psi _{C}^{+},\psi _{NC},\psi _{NC}^{+},\psi _{C}^{\ast
},\psi _{C}^{+\ast },\psi _{NC}^{\ast },\psi _{NC}^{+\ast }]
\label{Eq.DistnFnal2}
\end{eqnarray}%
where in the second line all the exponentials have been combined. The
quasi-distribution functional is of the Wigner type for the condensate
fields and the positive P type for the non-condensate fields, as in \cite%
{Dalton10a}.

These results are the same as those for the case of time independent phase
space variables. We note that 
\begin{eqnarray}
\dint dx\,(\psi _{C}(x,t\mathbf{)\,}\Xi _{C}^{+}(x,t\mathbf{)+}\psi
_{NC}(x,t)\,\Xi _{NC}^{+}(x,t)) &\mathbf{=}&\sum\limits_{k\epsilon
C,NC}\alpha _{k}\xi _{k}^{+}  \notag \\
\dint dx\,(\Xi _{C}(x,t\mathbf{)}\,\psi _{C}^{+}(x,t\mathbf{)+}\Xi
_{NC}(x,t)\,\psi _{NC}^{+}(x,t)) &\mathbf{=}&\sum\limits_{k\epsilon C,NC}\xi
_{k}\alpha _{k}^{+}\,  \label{Eq.TimeIndepResults4}
\end{eqnarray}%
which are time independent as before. Other details are given in Appendix %
\ref{Appendix Derivation of Functional Fokker-Planck Equation}.

\subsubsection{Notation Change}

As in Section \ref{Section 2 Separate Modes Case} we introduce the notation
where we list the modes as $\phi _{Ak}^{\mu }(x,t\mathbf{)}$ with $\mu =-,+$
and $A=C,NC$. Thus $\phi _{Ak}^{-}(x,t\mathbf{)=}\phi _{Ak}(x,t\mathbf{),}%
\phi _{Ak}^{+}(x,t\mathbf{)=}\phi _{Ak}^{\ast }(x,t\mathbf{)}$. The
condensate and non-condensate fields are listed as $\psi _{A}^{\mu }(x,t)$
and $\Xi _{A}^{\mu }(x,t)$, with $\psi _{C}^{-}(x,t)=\psi _{C}(x,t),\psi
_{C}^{+}(x,t)=\psi _{C}^{+}(x,t),\psi _{NC}^{-}(x,t)=\psi _{NC}(x,t)$ and $%
\psi _{NC}^{+}(x,t)=\psi _{NC}^{+}(x,t)$. The total fields are $\psi ^{\mu
}(x,t)=\psi _{C}^{\mu }(x,t)+\psi _{NC}^{\mu }(x,t)$ and $\Xi ^{\mu
}(x,t)=\Xi _{C}^{\mu }(x,t)+\Xi _{NC}^{\mu }(x,t)$. The coupling constants
are listed as $C_{Ak,Bl}^{\mu }$ as in Eq.(\ref{Eq.Coupling Coefts2}).
Writing $\underrightarrow{\psi }$.$\equiv \{\psi _{C},\psi _{C}^{+},\psi
_{NC},\psi _{NC}^{+}\}$ and $\underrightarrow{\psi }^{\ast }$.$\equiv \{\psi
_{C}^{\ast },\psi _{C}^{+\ast },\psi _{NC}^{\ast },\psi _{NC}^{+\ast }\}$
then we have

$\int \int \int \int D^{2}\psi _{C}\,\,D^{2}\psi _{C}^{+}D^{2}\psi
_{NC}\,D^{2}\psi _{NC}^{+}\;$will be written as $\int D^{2}\underrightarrow{%
\psi }$. In the new notation with phase space variables now $\alpha _{\mu
Ak}(t)$ and the field functions are 
\begin{equation}
\psi _{A}^{\mu }(x,t)=\dsum\limits_{k}\alpha _{\mu Ak}(t)\phi _{Ak}^{\mu
}(x,t\mathbf{)}  \label{Eq.FieldFns3}
\end{equation}%
Inverting Eq.(\ref{Eq.FieldFns3}) gives%
\begin{equation}
\alpha _{\mu Ak}=\int dx\,\phi _{Ak}^{-\mu }(x,t\mathbf{)}\psi _{A}^{\mu
}(x,t)  \label{Eq.PhaseVariablesHybrid}
\end{equation}%
Also from Eqs. (\ref{Eq.TimeDerivModePhases2}), (\ref{Eq.TimeDerivModes2})
and (\ref{Eq.PhaseVariablesHybrid}) we see that with 
\begin{eqnarray}
&&\frac{\partial }{\partial t}\psi _{A}^{\mu }(x,t)  \notag \\
&=&\int dy\sum_{kBl}\phi _{Ak}^{-\mu }(y,t)C_{Ak\,Bl}^{-\mu }\phi _{Bl}^{\mu
}(x,t)\;\psi _{A}^{\mu }(y,t)+\int dy\sum_{kBl}\phi _{Ak}^{\mu
}(x,t)C_{Ak\,Bl}^{\mu }\phi _{Bl}^{-\mu }(y,t)\;\psi _{B}^{\mu }(y,t)  \notag
\\
&=&-\int dy\sum_{\substack{ kBl  \\ B\neq A}}\phi _{Bl}^{\mu
}(x,t)C_{Bl\,Ak}^{\mu }\phi _{Ak}^{-\mu }(y,t)\;\psi _{A}^{\mu }(y,t)+\int
dy\sum_{\substack{ kBl  \\ B\neq A}}\phi _{Ak}^{\mu }(x,t)C_{Ak\,Bl}^{\mu
}\phi _{Bl}^{-\mu }(y,t)\;\psi _{B}^{\mu }(y,t)  \notag \\
&&  \label{Eq.TimeDerivHybridFields}
\end{eqnarray}%
showing that the time derivative of the field function $(C,NC)$ is a
functional of both the original field function $(C,NC)$ and the other field
function $(NC,C)$. Note that the $B=A$ terms cancel because of $%
C_{Ak\,Al}^{\mu }+C_{Al\,Ak}^{-\mu }=0$.

\subsection{Quantum Correlation Functions}

The \emph{quantum correlation functions} for the field operators can be
obtained from the characteristic functionals via taking \emph{functional
derivatives} with respect to $\Xi _{C},\Xi _{C}^{+},\Xi _{NC},\Xi _{NC}^{+}$
and then setting these quantities to be zero. The result gives the quantum
correlation functions as phase space functional integrals involving the
distribution functionals. Thus the quantum correlation function is%
\begin{eqnarray}
&&G(r_{1}\cdots r_{p};s_{q}\cdots \ s_{1};;u_{1}\cdots u_{r};v_{s}\cdots \
v_{1})  \notag \\
&=&\langle \{\hat{\Psi}_{C}(r_{1})^{\dag }\cdots \hat{\Psi}_{C}(r_{p})^{\dag
}\hat{\Psi}_{C}(s_{q})\cdots \hat{\Psi}_{C}(s_{1})\}\hat{\Psi}%
_{NC}(u_{1})^{\dag }\cdots \hat{\Psi}_{NC}(u_{r})^{\dag }\hat{\Psi}%
_{NC}(v_{s})\cdots \hat{\Psi}_{NC}(v_{1})\rangle   \notag \\
&=&\int \int \int \int D^{2}\psi _{C}\,D^{2}\psi _{C}^{+}\,D^{2}\psi
_{NC}\,D^{2}\psi _{NC}^{+}\,\;  \notag \\
&&\times \psi _{C}^{+}(r_{1})\cdots \psi _{C}^{+}(r_{p})\psi
_{C}(s_{q})\cdots \psi _{C}(s_{1})\;\psi _{NC}^{+}(u_{1})\cdots \psi
_{NC}^{+}(u_{r})\psi _{NC}(v_{s})\cdots \psi _{NC}(v_{1})\;\,P[%
\underrightarrow{\psi },\underrightarrow{\psi }^{\ast }]  \notag \\
&&  \label{Eq.QCorrFnHybrid}
\end{eqnarray}%
For simplicity the $t$ dependences of the condensate and non-condensate
field operators and field functions have been left out.

\subsection{Functional Fokker-Planck Equations}

To derive the functional Fokker-Planck equation we first differentiate the
expression in Eq. (\ref{Eq.CharFnReln2}) for the characteristic functional
with respect to $t$ and second do the same to the expression (\ref%
{Eq.DistnFnal2}) involving the distribution functional.

Differentiating the characteristic functional in its first form (\ref%
{Eq.CharFnReln2}) we have

\begin{eqnarray}
&&\frac{\partial }{\partial t}\chi \lbrack \Xi _{C},\Xi _{C}^{+},\Xi
_{NC},\Xi _{NC}^{+}]  \notag \\
&=&\exp \left\{ -\frac{1}{2}\dint dx\,\Xi _{C}(x,t\mathbf{)\,}\Xi
_{C}^{+}(x,t\mathbf{)}\right\}   \notag \\
&&\times Tr(\left\{ \exp i\dint dx\,(\hat{\Psi}(x\mathbf{)\,}\Xi ^{+}(x%
\mathbf{))}\right\} \,\left( \frac{\partial }{\partial t}\hat{\rho}\right)
\,\left\{ \exp i\dint dx(\Xi (x\mathbf{)}\,\hat{\Psi}^{\dag }(x\mathbf{)}%
)\right\} )  \notag \\
&&+\left( \frac{\partial }{\partial t}\exp \left\{ -\frac{1}{2}\dint dx\,\Xi
_{C}(x,t\mathbf{)\,}\Xi _{C}^{+}(x,t\mathbf{)}\right\} \right)   \notag \\
&&\times Tr(\left\{ \exp i\dint dx\,(\hat{\Psi}(x\mathbf{)\,}\Xi ^{+}(x%
\mathbf{))}\right\} \,\hat{\rho}\,\left\{ \exp i\dint dx(\Xi (x\mathbf{)}\,%
\hat{\Psi}^{\dag }(x\mathbf{)})\right\} )
\label{Eq.TimeDerivCharFnalConNonCond}
\end{eqnarray}%
This result gives the time derivative of the characteristic functional as
the sum of two terms. The first term in (\ref{Eq.TimeDerivCharFnalConNonCond}%
) only involves the time derivative of the density operator since from Eq. (%
\ref{Eq.ModeSumConNonConIntegRelns}) and (\ref{Eq.TimeIndepResults1}) the
integrals involving $\Xi (x\mathbf{)}$ and $\Xi ^{+}(x\mathbf{)}$ are time
independent. This term equals the characteristic functional that would apply
if $\hat{\rho}$ is replaced by $\frac{{\LARGE \partial }}{{\LARGE \partial t}%
}\hat{\rho}$. The first term produces the standard terms in the functional
Fokker-Planck equation. The second term depends on the time dependence of
the $\dint dx\,\Xi _{C}(x,t\mathbf{)\,}\Xi _{C}^{+}(x,t\mathbf{)}$ factor
due to the time dependent modes and is equal to the characteristic
functional multiplied by functionals of the condensate and non-condensate
characteristic fields. The second term leads to additional diffusion terms
in the functional Fokker-Planck equation.

For the second term in $\frac{{\LARGE \partial }}{{\LARGE \partial t}}\chi
\lbrack \Xi _{C},\Xi _{C}^{+},\Xi _{NC},\Xi _{NC}^{+}]$ the quantity
involving the time derivative is 
\begin{eqnarray}
&&\frac{\partial }{\partial t}\exp \left\{ -\frac{1}{2}\dint dx\,\Xi _{C}(x,t%
\mathbf{)\,}\Xi _{C}^{+}(x,t\mathbf{)}\right\}   \notag \\
&=&\left\{ -\frac{1}{2}\frac{\partial }{\partial t}\dint dx\,\Xi _{C}(x,t%
\mathbf{)\,}\Xi _{C}^{+}(x,t\mathbf{)}\right\} \times \exp \left\{ -\frac{1}{%
2}\dint dx\,\Xi _{C}(x,t\mathbf{)\,}\Xi _{C}^{+}(x,t\mathbf{)}\right\}  
\notag \\
&&  \label{Eq.TimeDerivIntegCondCharFlds00}
\end{eqnarray}%
and using 
\begin{eqnarray}
&&-\frac{1}{2}\frac{\partial }{\partial t}\dint dx\,\Xi _{C}(x,t\mathbf{)\,}%
\Xi _{C}^{+}(x,t\mathbf{)}  \notag \\
&=&\frac{1}{2}\left\{ \dint \int dx\,dy\,\left\{ \sum_{k\epsilon
C}\dsum\limits_{l\epsilon NC}\phi _{l}^{\ast }(x,t\mathbf{)}C_{kl}(t)\phi
_{k}(y,t\mathbf{)}\right\} \mathbf{(}i\Xi _{NC}(x,t\mathbf{))}(i\Xi
_{C}^{+}(y,t\mathbf{))}\right\}   \notag \\
&&+\frac{1}{2}\left\{ \dint \int dx\,dy\,\left\{ \sum_{l\epsilon
C}\dsum\limits_{k\epsilon NC}\phi _{l}^{\ast }(x,t\mathbf{)}C_{lk}^{\ast
}(t)\phi _{k}(y,t\mathbf{)}\right\} \,(i\Xi _{C}(x,t\mathbf{))}\,(i\Xi
_{NC}^{+}(y,t\mathbf{))}\right\}   \notag \\
&&  \label{Eq.TimeDerivIntegCondCharFlds}
\end{eqnarray}%
we find that the second term is given as the sum of two contributions as 
\begin{eqnarray}
&&\left( -\frac{1}{2}\frac{\partial }{\partial t}\dint dx\,\Xi _{C}(x,t%
\mathbf{)\,}\Xi _{C}^{+}(x,t\mathbf{)}\right) \chi \lbrack \Xi _{C},\Xi
_{C}^{+},\Xi _{NC},\Xi _{NC}^{+}]  \notag \\
&=&\int D^{2}\underrightarrow{\psi }\;\exp (i\dint dx\,\sum_{\mu }\psi ^{\mu
}(x,t\mathbf{)\,}\Xi ^{-\mu }(x,t\mathbf{))}  \notag \\
&&\times \dint \int dx\,dy\,\left\{ \frac{1}{2}\sum_{k\epsilon
C}\dsum\limits_{l\epsilon NC}\phi _{NCl}^{+}(x,t\mathbf{)}%
C_{Ck\,NCl}^{-}(t)\phi _{Ck}^{-}(y,t\mathbf{)}\frac{\delta }{\delta \psi
_{NC}^{+}(x,t)}\frac{\delta }{\delta \psi _{C}^{-}(y,t\mathbf{)}}P[%
\underrightarrow{\psi },\underrightarrow{\psi }^{\ast }]\right\}   \notag \\
&&+\int D^{2}\underrightarrow{\psi }\;\exp (i\dint dx\,\sum_{\mu }\psi ^{\mu
}(x,t\mathbf{)\,}\Xi ^{-\mu }(x,t\mathbf{))}  \notag \\
&&\times \dint \int dx\,dy\,\left\{ \frac{1}{2}\sum_{k\epsilon
NC}\dsum\limits_{l\epsilon C}\phi _{Cl}^{+}(x,t\mathbf{)}C_{Cl\,NCk}^{+}(t)%
\phi _{NCk}^{-}(y,t\mathbf{)}\frac{\delta }{\delta \psi _{C}^{+}(x,t)}\frac{%
\delta }{\delta \psi _{NC}^{-}(y,t\mathbf{)}}P[\underrightarrow{\psi },%
\underrightarrow{\psi }^{\ast }]\right\}   \notag \\
&&  \label{Eq.TimeDerivCharFnalSecondTerm}
\end{eqnarray}%
in terms of the new notation. In obtaining this result the exponential
factor in Eq. (\ref{Eq.TimeDerivIntegCondCharFlds00}) is recombined via (\ref%
{Eq.CharFnReln2}) with the normally ordered characteristic functional $%
Tr(\left\{ \exp i\dint dx\,(\hat{\Psi}(x\mathbf{)\,}\Xi ^{+}(x\mathbf{))}%
\right\} \,\hat{\rho}\,\left\{ \exp i\dint dx(\Xi (x\mathbf{)}\,\hat{\Psi}%
^{\dag }(x\mathbf{)})\right\} )$ to produce the original characteristic
functional $\chi \lbrack \Xi _{C},\Xi _{C}^{+},\Xi _{NC},\Xi _{NC}^{+}]$.
Writing the characteristic functional in the form in Eq. (\ref{Eq.DistnFnal2}%
) and noting that multiplication by $\mathbf{(}i\Xi _{NC}(x,t\mathbf{))}%
(i\Xi _{C}^{+}(y,t\mathbf{))}$ etc. can be replaced by functional
differentiations such as $\frac{{\LARGE \delta }}{{\LARGE \delta \psi }%
_{NC}^{+}{\LARGE (x,t)}}\frac{{\LARGE \delta }}{{\LARGE \delta \psi }_{C}%
{\LARGE (y,t)}}$ result (\ref{Eq.TimeDerivCharFnalSecondTerm}) follows after
applying functional integration by parts twice. Details for the derivation
of (\ref{Eq.TimeDerivIntegCondCharFlds}) are set out in Appendix \ref%
{Appendix Derivation of Functional Fokker-Planck Equation}.

For the first term in $\frac{{\LARGE \partial }}{{\LARGE \partial t}}\chi
\lbrack \Xi _{C},\Xi _{C}^{+},\Xi _{NC},\Xi _{NC}^{+}]$ we use the standard
results for the \emph{correspondence rules} associated with replacing the
density operator by its product with the condensate and non-condensate field
annihilation, creation operators $\hat{\Psi}_{C}(x,t),\hat{\Psi}%
_{C}^{\dagger }(x,t),\hat{\Psi}_{NC}(x,t),\hat{\Psi}_{NC}^{\dagger }(x,t)$
that occur when $\frac{{\LARGE \partial }}{{\LARGE \partial t}}\hat{\rho}$
is evaluated using the Liouville-von Neumann or master equation. For the 
\emph{non-condensate field }operators we have%
\begin{eqnarray}
\hat{\rho} &\Rightarrow &\hat{\Psi}_{NC}(x,t)\,\hat{\rho}\qquad \chi \lbrack
\Xi _{C},\Xi _{C}^{+},\Xi _{NC},\Xi _{NC}^{+}]\Rightarrow \frac{\delta }{%
\delta (i\Xi _{NC}^{+}(x,t\mathbf{)})}\chi   \notag \\
\hat{\rho} &\Rightarrow &\hat{\rho}\,\hat{\Psi}_{NC}(x,t)\qquad \chi \lbrack
\Xi _{C},\Xi _{C}^{+},\Xi _{NC},\Xi _{NC}^{+}]\Rightarrow \left( \frac{%
\delta }{\delta (i\Xi _{NC}^{+}(x,t\mathbf{)})}+i\Xi _{NC}(x,t\mathbf{)}%
\right) \chi   \notag \\
\hat{\rho} &\Rightarrow &\hat{\Psi}_{NC}^{\dagger }(x,t)\,\hat{\rho}\qquad
\chi \lbrack \Xi _{C},\Xi _{C}^{+},\Xi _{NC},\Xi _{NC}^{+}]\Rightarrow
\left( \frac{\delta }{\delta (i\Xi _{NC}(x,t\mathbf{)})}+i\Xi _{NC}^{+}(x,t%
\mathbf{)}\right) \chi   \notag \\
\hat{\rho} &\Rightarrow &\hat{\rho}\,\hat{\Psi}_{NC}^{\dagger }(x,t)\,\qquad
\chi \lbrack \Xi _{C},\Xi _{C}^{+},\Xi _{NC},\Xi _{NC}^{+}]\Rightarrow \frac{%
\delta }{\delta (i\Xi _{NC}(x,t\mathbf{)})}\chi 
\end{eqnarray}%
For the \emph{condensate field }operators we have 
\begin{eqnarray}
\hat{\rho} &\Rightarrow &\hat{\Psi}_{C}(x,t)\hat{\rho}\qquad \chi \lbrack
\Xi _{C},\Xi _{C}^{+},\Xi _{NC},\Xi _{NC}^{+}]\Rightarrow \left( \frac{%
\delta }{\delta (i\Xi _{C}^{+}(x,t\mathbf{)})}-\frac{1}{2}i\Xi _{C}(x,t%
\mathbf{)}\right) \chi   \notag \\
\hat{\rho} &\Rightarrow &\hat{\rho}\hat{\Psi}_{C}(x,t)\qquad \chi \lbrack
\Xi _{C},\Xi _{C}^{+},\Xi _{NC},\Xi _{NC}^{+}]\Rightarrow \left( \frac{%
\delta }{\delta (i\Xi _{C}^{+}(x,t\mathbf{)})}+\frac{1}{2}i\Xi _{C}(x,t%
\mathbf{)}\right) \chi   \notag \\
\hat{\rho} &\Rightarrow &\hat{\Psi}_{C}^{\dagger }(x,t)\hat{\rho}\qquad \chi
\lbrack \Xi _{C},\Xi _{C}^{+},\Xi _{NC},\Xi _{NC}^{+}]\Rightarrow \left( 
\frac{\delta }{\delta (i\Xi _{C}(x,t\mathbf{)})}+\frac{1}{2}i\Xi _{C}^{+}(x,t%
\mathbf{)}\right) \chi   \notag \\
\hat{\rho} &\Rightarrow &\hat{\rho}\hat{\Psi}_{C}^{\dagger }(x,t)\qquad \chi
\lbrack \Xi _{C},\Xi _{C}^{+},\Xi _{NC},\Xi _{NC}^{+}]\Rightarrow \left( 
\frac{\delta }{\delta (i\Xi _{C}(x,t\mathbf{)})}-\frac{1}{2}i\Xi _{C}^{+}(x,t%
\mathbf{)}\right) \chi   \notag \\
&&
\end{eqnarray}%
From these results the corresponding changes to the distribution functionals
can be deduced using (\ref{Eq.HybridCharFnal}) and (\ref{Eq.DistnFnal2}).
This involves functional integration by parts.

The correspondence rules for the \emph{non-condensate field }operators are%
\begin{eqnarray}
\hat{\rho} &\Rightarrow &\hat{\Psi}_{NC}(x,t)\,\hat{\rho}\qquad P[%
\underrightarrow{\psi },\underrightarrow{\psi }^{\ast }]\Rightarrow \psi
_{NC}(x,t\mathbf{)}\,P  \notag \\
\hat{\rho} &\Rightarrow &\hat{\rho}\,\hat{\Psi}_{NC}(x,t)\qquad P[%
\underrightarrow{\psi },\underrightarrow{\psi }^{\ast }]\Rightarrow \left( -%
\frac{\delta }{\delta \psi _{NC}^{+}(x,t\mathbf{)}}+\psi _{NC}(x,t\mathbf{)}%
\right) P\,  \notag \\
\hat{\rho} &\Rightarrow &\hat{\Psi}_{NC}^{\dagger }(x,t)\,\hat{\rho}\qquad P[%
\underrightarrow{\psi },\underrightarrow{\psi }^{\ast }]\Rightarrow \left( =%
\frac{\delta }{\delta \psi _{NC}(x,t\mathbf{)}}+\psi _{NC}^{+}(x,t\mathbf{)}%
\right) P  \notag \\
\hat{\rho} &\Rightarrow &\hat{\rho}\,\,\hat{\Psi}_{NC}^{\dagger }(x,t)\qquad
P[\underrightarrow{\psi },\underrightarrow{\psi }^{\ast }]\Rightarrow \psi
_{NC}^{+}(x,t\mathbf{)}P  \label{Eq.NonCondOprsFnalCorr}
\end{eqnarray}%
The correspondence rules for the \emph{condensate field }operators\emph{\ }%
are:%
\begin{eqnarray}
\hat{\rho} &\Rightarrow &\hat{\Psi}_{C}(x,t)\hat{\rho}\qquad P[%
\underrightarrow{\psi },\underrightarrow{\psi }^{\ast }]\Rightarrow \left(
\psi _{C}(x,t\mathbf{)}+\frac{1}{2}\frac{\delta }{\delta \psi _{C}^{+}(x,t%
\mathbf{)}}\right) P  \notag \\
\hat{\rho} &\Rightarrow &\hat{\rho}\hat{\Psi}_{C}(x,t)\qquad P[%
\underrightarrow{\psi },\underrightarrow{\psi }^{\ast }]\Rightarrow \left(
\psi _{C}(x,t\mathbf{)}-\frac{1}{2}\frac{\delta }{\delta \psi _{C}^{+}(x,t%
\mathbf{)}}\right) P  \notag \\
\hat{\rho} &\Rightarrow &\hat{\Psi}_{C}^{\dagger }(x,t)\hat{\rho}\qquad P[%
\underrightarrow{\psi },\underrightarrow{\psi }^{\ast }]\Rightarrow \left(
\psi _{C}^{+}(x,t\mathbf{)}-\frac{1}{2}\frac{\delta }{\delta \psi _{C}(x,t%
\mathbf{)}}\right) P  \notag \\
\hat{\rho} &\Rightarrow &\hat{\rho}\hat{\Psi}_{C}^{\dagger }(x,t)\qquad P[%
\underrightarrow{\psi },\underrightarrow{\psi }^{\ast }]\Rightarrow \left(
\psi _{C}^{+}(x,t\mathbf{)}+\frac{1}{2}\frac{\delta }{\delta \psi _{C}(x,t%
\mathbf{)}}\right) P  \label{Eq.CondOprsFnalCorr}
\end{eqnarray}%
where the $\psi _{C},\psi _{C}^{+},\psi _{C}^{\ast },\psi _{C}^{+\ast },\psi
_{NC},\psi _{NC}^{+},\psi _{NC}^{\ast },\psi _{NC}^{+\ast }$ are regarded as
eight independent complex fields. The equivalence of these correspondence
rules to those applying for separate modes can be established using the
methods of functional calculus (see Appendix \ref{Appendix - Functional
Calculus}, see Eq.(\ref{Eq.FnalDiffnOrdDiffn}) which relates ordinary and
functional differentiation).

By applying the correspondence rules (\ref{Eq.NonCondOprsFnalCorr}) or (\ref%
{Eq.CondOprsFnalCorr}) in succession to the various products of the density
operator with field annihilation, creation operators that arise from $\frac{%
{\LARGE \partial }}{{\LARGE \partial t}}\hat{\rho}$ we get for the first
term in $\frac{{\LARGE \partial }}{{\LARGE \partial t}}\chi \lbrack \Xi
_{C},\Xi _{C}^{+},\Xi _{NC},\Xi _{NC}^{+}]$

\begin{eqnarray}
&&\chi \lbrack \Xi _{C},\Xi _{C}^{+},\Xi _{NC},\Xi _{NC}^{+};\frac{\partial 
}{\partial t}\hat{\rho}]  \notag \\
&=&\int D^{2}\underrightarrow{\psi }\;\exp (i\dint dx\,\sum_{\mu }\psi ^{\mu
}(x,t\mathbf{)\,}\Xi ^{-\mu }(x,t\mathbf{))}\,  \notag \\
&&\times \left\{ -\dsum\limits_{\mu A}\dint dx\frac{\delta }{\delta \psi
_{A}^{\mu }(x)}A_{A}^{\mu }(x)+\frac{1}{2}\dsum\limits_{\mu
A}\dsum\limits_{\nu B}\dint \dint dxdy\frac{\delta }{\delta \psi _{A}^{\mu
}(x)}\frac{\delta }{\delta \psi _{B}^{\nu }(y)}D_{A\,B}^{\mu \,\nu
}(x,y)\right\} P[\underrightarrow{\psi },\underrightarrow{\psi }^{\ast }] 
\notag \\
&&  \label{Eq.TimeDerivDensOprTerm}
\end{eqnarray}%
in terms of the new notation, where $A_{A}^{\mu }(x),D_{A\,B}^{\mu \,\nu
}(x,y)$ are the drift, diffusion terms associated with the standard
functional Fokker-Planck equation. The integrals over $x,y$ arise because
the Hamiltonian involves spatial integrals over the field operators.
Specific forms for the drift, diffusion terms are given for the cases of a
single or two mode single component BEC in \cite{Dalton10a}.

Differentiating the characteristic functional in its second form (\ref%
{Eq.DistnFnal2}) we have 
\begin{eqnarray}
&&\frac{\partial }{\partial t}\chi \lbrack \Xi _{C},\Xi _{C}^{+},\Xi
_{NC},\Xi _{NC}^{+};\hat{\rho}]  \notag \\
&=&\int \int \int \int D^{2}\psi _{C}\,\,D^{2}\psi _{C}^{+}D^{2}\psi
_{NC}\,D^{2}\psi _{NC}^{+}  \notag \\
&&\times \exp (i\dint dx\,(\psi _{C}(x,t\mathbf{)\,}\Xi _{C}^{+}(x,t\mathbf{%
)+}\psi _{NC}(x,t)\,\Xi _{NC}^{+}(x,t)\mathbf{+}\,\Xi _{C}(x,t\mathbf{)}%
\,\psi _{C}^{+}(x,t\mathbf{)+\,}\Xi _{NC}(x,t)\,\psi _{NC}^{+}(x,t)\mathbf{)}
\notag \\
&&\times \frac{\partial }{\partial t}P[\psi _{C},\psi _{C}^{+},\psi
_{NC},\psi _{NC}^{+},\psi _{C}^{\ast },\psi _{C}^{+\ast },\psi _{NC}^{\ast
},\psi _{NC}^{+\ast }]  \label{Eq.TimeDerivCharFnalCondNonCond2} \\
&=&\int D^{2}\underrightarrow{\psi }\;\exp (i\dint dx\,\sum_{\mu }\psi ^{\mu
}(x,t\mathbf{)\,}\Xi ^{-\mu }(x,t\mathbf{))}\,\frac{\partial }{\partial t}P[%
\underrightarrow{\psi },\underrightarrow{\psi }^{\ast }]\,  \notag
\end{eqnarray}%
in terms of the new notation. From Eq. (\ref{Eq.ModeSumConNonConIntegRelns})
and (\ref{Eq.TimeIndepResults1}) the integrals involving $\Xi _{C}(x,t%
\mathbf{)}$, $\Xi _{NC}(x,t\mathbf{)}$, $\Xi _{C}^{+}(x,t\mathbf{)}$, and $%
\Xi _{NC}^{+}(x,t\mathbf{)}$ are time independent, whilst the functional
integration $\int \int \int \int D^{2}\psi _{C}\,\,D^{2}\psi
_{C}^{+}D^{2}\psi _{NC}\,D^{2}\psi _{NC}^{+}\,$\ being the same as the phase
space integration $\int \int \int \int d^{2}\alpha _{C}\,d^{2}\alpha
_{C}^{+}d^{2}\alpha _{NC}\,d^{2}\alpha _{NC}^{+}=\int \int \int \int
d^{2}\alpha _{C}(0)\,d^{2}\alpha _{C}^{+}(0)\,d^{2}\alpha
_{NC}(0)\,d^{2}\alpha _{NC}^{+}(0)$ - is also time independent. It follows
that the time derivative of the characteristic functional is now determined
from the time derivative of the distribution functional.

Equating both sides of $\frac{{\LARGE \partial }}{{\LARGE \partial t}}\chi
\lbrack \Xi _{C},\Xi _{C}^{+},\Xi _{NC},\Xi _{NC}^{+}]$ via Eqs. (\ref%
{Eq.TimeDerivCharFnalSecondTerm}), (\ref{Eq.TimeDerivDensOprTerm}) and (\ref%
{Eq.TimeDerivCharFnalCondNonCond2}) gives the \emph{functional Fokker-Planck
equation }for the hybrid distribution function. All the diffusion terms are
combined, and there are new diffusion terms involving condensate fields
paired with non-condensate fields arising due to the coupling coefficients.
We have

\begin{eqnarray}
&&\frac{\partial }{\partial t}P[\underrightarrow{\psi },\underrightarrow{%
\psi }^{\ast }]\,  \notag \\
&=&\left\{ -\dsum\limits_{\mu A}\dint dx\frac{\delta }{\delta \psi _{A}^{\mu
}(x,t)}A_{A}^{\mu }(x)+\frac{1}{2}\dsum\limits_{\mu A}\dsum\limits_{\nu
B}\dint \dint dxdy\frac{\delta }{\delta \psi _{A}^{\mu }(x,t)}\frac{\delta }{%
\delta \psi _{B}^{\nu }(y,t)}E_{A\,B}^{\mu \,\nu }(x,y)\right\} P[%
\underrightarrow{\psi },\underrightarrow{\psi }^{\ast }]  \notag \\
&&  \label{Eq.FnalFokkerPlanckHybridDistnFnal2}
\end{eqnarray}%
where the new diffusion term is given by%
\begin{eqnarray}
E_{A\,B}^{\mu \,\nu }(x,y) &=&D_{A\,B}^{\mu \,\nu }(x,y)  \notag \\
&&+\frac{1}{2}\left\{ \delta _{A\,C}\delta _{B\,NC}\,\delta _{\mu \,-\nu
}\left( \sum_{k}\dsum\limits_{l}\phi _{Ak}^{\mu }(x,t\mathbf{)}%
C_{Ak\,Bl}^{\mu }(t)\phi _{Bl}^{\nu }(y,t\mathbf{)}\right) \right\}  \notag
\\
&&+\frac{1}{2}\left\{ \delta _{B\,C}\delta _{A\,NC}\,\delta _{\nu \,-\mu
}\left( \sum_{k}\dsum\limits_{l}\phi _{Bl}^{\nu }(y,t\mathbf{)}%
C_{Bl\,Ak}^{\nu }(t)\phi _{Ak}^{\mu }(x,t\mathbf{)}\right) \right\}  \notag
\\
&&  \label{Eq.NewDiffusionMatrixHybridDistnFnal}
\end{eqnarray}%
Details are given in Appendix \ref{Appendix Derivation of Functional
Fokker-Planck Equation}. Here the additional diffusion contribution has been
rewritten in a manifestly symmetric form. The drift $A_{A}^{\mu }(x)$ vector
and the diffusion matrix $E_{A\,B}^{\mu \,\nu }(x,y)$ are functions of the $%
\psi _{A}^{\mu }$ and ultimately depend on $x$ or $x,y$. The diffusion
matrix is symmetric $E_{A\,B}^{\mu \,\nu }(x,y)=E_{B\,A}^{\nu \,\mu }(y,x)$.
Thus for the hybrid case the functional Fokker-Planck equation has the same
drift vector $A_{A}^{\mu }(x)$ but a different diffusion matrix $%
E_{A\,B}^{\mu \,\nu }(x,y)$ to the diffusion matrix $D_{A\,B}^{\mu \,\nu
}(x,y)$ that would be obtained from the characteristic functional by
applying the standard correspondence rules to the $\frac{{\LARGE \partial }}{%
{\LARGE \partial t}}\hat{\rho}$ contribution. The additional terms involved
the time dependent mode functions and the coupling coefficients. The
functional Fokker-Planck equation can also be expressed in terms of real
variables which involve the real and imaginary components of the phase
variables $\psi _{\mu }$, but we will not do that here. In deriving the
functional Fokker-Planck equation there are often terms involving third and
higher order derivatives arising from the $\frac{{\LARGE \partial }}{{\LARGE %
\partial t}}\hat{\rho}$ contribution. These are usually discarded on the
basis of being small due to scaling as higher powers of $1/\sqrt{N}$.

The functional Fokker-Planck equation (\ref%
{Eq.FnalFokkerPlanckHybridDistnFnal2}) for the distribution functional $P[%
\underrightarrow{\psi },\underrightarrow{\psi }^{\ast }]\,$must be
equivalent to the ordinary Fokker-Planck equation Eq. (\ref%
{Eq.NewFokker-Planck}) for the distribution function $P(\alpha ,\alpha
^{+},\alpha ^{\ast },\alpha ^{+\ast })$. This is shown in Appendix \ref%
{Appendix Equivalence of Separate Mode and Field Theory}.

\subsection{Langevin (Ito) Stochastic Field Equations}

To derive the Ito stochastic field equations we follow a similar approach to
Section \ref{SubSection 2. ItoSDE}. We consider a functional $F[%
\underrightarrow{\psi }]$ of the condensate and non-condensate fields and
find conditions where the phase space functional average of the functional $%
F[\underrightarrow{\psi }]$ is the same as the stochastic average when the
fields $\psi _{A}^{\mu }$ are replaced by stochastic fields $\psi _{A}^{\mu
\,s}$.

\subsubsection{Phase Space Functional Average}

The phase space functional average is given by%
\begin{equation}
\left\langle F[\underrightarrow{\psi }]\right\rangle _{t}=\int D^{2}%
\underrightarrow{\psi }\,F[\underrightarrow{\psi }]\,P[\underrightarrow{\psi 
},\underrightarrow{\psi }^{\ast }]  \label{Eq.FnalPhaseAverHybrid}
\end{equation}%
For determining the quantum correlation function $G(r_{1}\cdots
r_{p};s_{q}\cdots \ s_{1};;u_{1}\cdots u_{r};v_{s}\cdots \ v_{1})$ the
functional is $F[\underrightarrow{\psi }]=$ $\psi _{C}^{+}(r_{1})\cdots \psi
_{C}^{+}(r_{p})\psi _{C}(s_{q})\cdots \psi _{C}(s_{1})\;\psi
_{NC}^{+}(u_{1})\cdots \psi _{NC}^{+}(u_{r})\psi _{NC}(v_{s})\cdots \psi
_{NC}(v_{1})$ - a product of functionals. In the phase space functional
average both the distribution functional and $F[\underrightarrow{\psi }]$
are time dependent, because from Eq.(\ref{Eq.TimeDepCondNonCondFldFns}) the $%
\psi _{C},\psi _{C}^{+},\psi _{NC},\psi _{NC}^{+}$ depend on time. However
from (\ref{Eq.PhaseSpaceFunctionalIntegn2}) the functional integration is
not time dependent.

The time rate of change in the phase space functional average then consists
of two contributions%
\begin{eqnarray}
&&\frac{\partial }{\partial t}\left\langle F[\underrightarrow{\psi }%
]\right\rangle _{t}  \notag \\
&=&\int D^{2}\underrightarrow{\psi }\,\frac{\partial }{\partial t}F[%
\underrightarrow{\psi }]\,P[\underrightarrow{\psi },\underrightarrow{\psi }%
^{\ast }]+\int D^{2}\underrightarrow{\psi }\,F[\underrightarrow{\psi }]\,%
\frac{\partial }{\partial t}P[\underrightarrow{\psi },\underrightarrow{\psi }%
^{\ast }]  \label{Eq.TimeDerivFnalPhaseAverHybrid}
\end{eqnarray}

To evaluate the first term we first consider changes $\delta \psi _{A}^{\mu
}(x,t)$ in $\psi _{A}^{\mu }(x,t)$ during an interval $\delta t$ that
results in a change $\delta F[\underrightarrow{\psi }]$ to the functional $F[%
\underrightarrow{\psi }]$. From Eq.(\ref{Eq.FnalDeriv}) we find that 
\begin{equation}
\frac{\partial }{\partial t}F[\underrightarrow{\psi }]=\dsum\limits_{\mu
A}\int dx\,\left( \frac{\delta }{\delta \psi _{A}^{\mu }(x,t)}F[%
\underrightarrow{\psi }]\right) \frac{\partial \psi _{A}^{\mu }(x,t)}{%
\partial t}  \label{Eq.TimeDerivFnalF}
\end{equation}%
so substituting from Eq. (\ref{Eq.TimeDerivHybridFields}) we get

\begin{eqnarray}
&&\frac{\partial }{\partial t}F[\underrightarrow{\psi }]  \notag \\
&=&\dsum\limits_{\mu A}\int dx\,\left( \frac{\delta }{\delta \psi _{A}^{\mu
}(x,t)}F[\underrightarrow{\psi }]\right) \left( \dint
dy\,\dsum\limits_{B}\left( \dsum\limits_{kl}\phi _{Ak}^{\mu
}(x,t)C_{Ak\,Bl}^{\mu }\phi _{Bl}^{-\mu }(y,t)\right) \;\psi _{B}^{\mu
}(y,t)\right)  \notag \\
&&+\dsum\limits_{\mu A}\int dx\,\left( \frac{\delta }{\delta \psi _{A}^{\mu
}(x,t)}F[\underrightarrow{\psi }]\right) \left( \dint
dy\,\dsum\limits_{B}\left( \dsum\limits_{kl}\phi _{Bl}^{\mu
}(x,t)C_{Ak\,Bl}^{-\mu }\phi _{Ak}^{-\mu }(y,t)\right) \;\psi _{A}^{\mu
}(y,t)\right)  \notag \\
&&  \label{Eq.TimeDerivFnalF2}
\end{eqnarray}%
Hence the first term in (\ref{Eq.TimeDerivFnalPhaseAverHybrid}) is given by%
\begin{eqnarray}
&&\int D^{2}\underrightarrow{\psi }\,\frac{\partial }{\partial t}F[%
\underrightarrow{\psi }]\,P[\underrightarrow{\psi },\underrightarrow{\psi }%
^{\ast }]  \notag \\
&=&\left\langle \sum_{\mu A}\dint dx\left( \frac{\delta }{\delta \psi
_{A}^{\mu }(x,t)}F[\underrightarrow{\psi }]\right) \left\{ \int
dy\,\dsum\limits_{B}\left( \dsum\limits_{kl}\phi _{Ak}^{\mu
}(x,t)C_{Ak\,Bl}^{\mu }\phi _{Bl}^{-\mu }(y,t)\right) \;\psi _{B}^{\mu
}(y,t)\right\} \right\rangle  \notag \\
&&+\left\langle \sum_{\mu A}\dint dx\left( \frac{\delta }{\delta \psi
_{A}^{\mu }(x,t)}F[\underrightarrow{\psi }]\right) \left\{ \dint
dy\,\dsum\limits_{B}\left( \dsum\limits_{kl}\phi _{Bl}^{\mu
}(x,t)C_{Ak\,Bl}^{-\mu }\phi _{Ak}^{-\mu }(y,t)\right) \;\psi _{A}^{\mu
}(y,t)\right\} \right\rangle  \notag \\
&&  \label{Eq.Term1TimeDerivFnalPhaseAver}
\end{eqnarray}%
Details are given in Appendix \ref{Appendix Derivation of Ito Stochastic
Field Equations}.

Using the functional Fokker-Planck equation (\ref%
{Eq.FnalFokkerPlanckHybridDistnFnal2})\ we find after using integration by
parts that the second term in (\ref{Eq.TimeDerivFnalPhaseAverHybrid}) is 
\begin{eqnarray}
&&\int D^{2}\underrightarrow{\psi }\,F[\underrightarrow{\psi }]\,\frac{%
\partial }{\partial t}P[\underrightarrow{\psi },\underrightarrow{\psi }%
^{\ast }]  \notag \\
&=&\left\langle \dsum\limits_{\mu A}\dint dx\left( \frac{\delta }{\delta
\psi _{A}^{\mu }(x,t)}F[\underrightarrow{\psi }]\right) A_{A}^{\mu
}(x)\right\rangle  \notag \\
&&+\left\langle \frac{1}{2}\dsum\limits_{\mu A}\dsum\limits_{\nu B}\dint
\dint dxdy\left( \frac{\delta }{\delta \psi _{A}^{\mu }(x,t)}\frac{\delta }{%
\delta \psi _{B}^{\nu }(y,t)}F[\underrightarrow{\psi }]\right) E_{A\,B}^{\mu
\,\nu }(x,y)\right\rangle  \notag \\
&&  \label{Eq.Term2TimeDerivFnalPhaseAver}
\end{eqnarray}%
where functional integration by parts has been used. Details are given in
Appendix \ref{Appendix Derivation of Ito Stochastic Field Equations}.

Overall we now have%
\begin{eqnarray}
&&\frac{\partial }{\partial t}\left\langle F[\underrightarrow{\psi }%
]\right\rangle _{t}  \notag \\
&=&\left\langle \sum_{\mu A}\dint dx\left( \frac{\delta }{\delta \psi
_{A}^{\mu }(x,t)}F[\underrightarrow{\psi }]\right) \left\{ A_{A}^{\mu
}(x)\right\} \right\rangle   \notag \\
&&+\left\langle \sum_{\mu A}\dint dx\left( \frac{\delta }{\delta \psi
_{A}^{\mu }(x,t)}F[\underrightarrow{\psi }]\right) \left\{ \int
dy\,\dsum\limits_{B}\left( \dsum\limits_{kl}\phi _{Ak}^{\mu
}(x,t)C_{Ak\,Bl}^{\mu }\phi _{Bl}^{-\mu }(y,t)\right) \;\psi _{B}^{\mu
}(y,t)\right\} \right\rangle   \notag \\
&&+\left\langle \sum_{\mu A}\dint dx\left( \frac{\delta }{\delta \psi
_{A}^{\mu }(x,t)}F[\underrightarrow{\psi }]\right) \left\{ \dint
dy\,\dsum\limits_{B}\left( \dsum\limits_{kl}\phi _{Bl}^{\mu
}(x,t)C_{Ak\,Bl}^{-\mu }\phi _{Ak}^{-\mu }(y,t)\right) \;\psi _{A}^{\mu
}(y,t)\right\} \right\rangle   \notag \\
&&+\left\langle \frac{1}{2}\dsum\limits_{\mu A}\dsum\limits_{\nu B}\dint
\dint dxdy\left( \frac{\delta }{\delta \psi _{A}^{\mu }(x,t)}\frac{\delta }{%
\delta \psi _{B}^{\nu }(y,t)}F[\underrightarrow{\psi }]\right) E_{A\,B}^{\mu
\,\nu }(x,y)\right\rangle   \label{Eq.HybridTimeDerivFnalPhaseAver}
\end{eqnarray}

This result can be compared to that for a function $F(\alpha ,\alpha ^{+})$
as set out in Section \ref{Section 2 Separate Modes Case} where from Eq. (%
\ref{Eq.TimeDerivPhaseSpAverBose}) 
\begin{eqnarray}
&&\frac{\partial }{\partial t}\left\langle F(\alpha ,\alpha
^{+})\right\rangle _{t}  \notag \\
&=&\left\langle \left\{ \sum_{\mu Ak}\left[ \frac{\partial }{\partial \alpha
_{\mu Ak}}F(\alpha ,\alpha ^{+})\right] \;\left[ A_{Ak}^{\mu
}+\sum_{Bl}C_{Ak\,Bl}^{\mu }\,\alpha _{\mu Bl}\right] \right\} \right\rangle 
\notag \\
&&+\left\langle \left\{ \frac{1}{2}\dsum\limits_{\mu Ak}\dsum\limits_{\nu Bl}%
\left[ \frac{\partial }{\partial \alpha _{\mu Ak}}\frac{\partial }{\partial
\alpha _{\nu Bl}}F(\alpha ,\alpha ^{+})\right] \;E_{Ak\,Bl}^{\mu \nu
}\right\} \right\rangle 
\end{eqnarray}%
We see that the derivative of the functional phase space average is not the
same form as that for the original phase space average involving separate
modes, since a term corresponding to one of the first order functional
derivative terms is missing. As a specific case consider the functional $F[%
\underrightarrow{\psi }]=\psi _{C}^{+}(r_{1},t)\psi _{C}(s_{1},t)$ that is
involved in working out the symmetrically ordered correlation function $%
\langle \{\hat{\Psi}_{C}(r_{1},t)^{\dag }\hat{\Psi}_{C}(s_{1},t)\}\rangle $.
By substituting we get $F[\underrightarrow{\psi }]=\dsum\limits_{k}\alpha
_{Ck}^{+}(t)\phi _{Ck}^{\ast }(r_{1},t)\;\dsum\limits_{l}\alpha _{Cl}(t)\phi
_{Cl}(s_{1},t)=\dsum\limits_{kl}\phi _{Ck}^{\ast }(r_{1},t)\;\alpha
_{Ck}^{+}(t)\alpha _{Cl}(t)\;\phi _{Cl}(s_{1},t)$. The last equation is
useful for working out the time derivatives of the various $F(\alpha ,\alpha
^{+})=\alpha _{Ck}^{+}\alpha _{Cl}$, but it does not include the
contribution to the time derivative of $F[\underrightarrow{\psi }]$ from the
various $\phi _{Ck}^{\ast }(r_{1},t)\phi _{Cl}(s_{1},t)$. These
contributions can of course be worked out and included later, but clearly
the advantage of using field functions rather than separate phase variables
is in terms of focusing \emph{directly} on the quantum correlation functions
that involve condensate and non-condensate field operators rather than just
mode operators. As we will see there is no paradox associated with the
missing term - there is a compensating term arising from the stochastic
field average.

\subsubsection{Stochastic Fields Average}

For the Ito stochastic field approach we now replace the fields $\psi
_{A}^{\mu }$ by stochastic fields $\psi _{A}^{\mu \,s}$. We suppose $\psi
_{A}^{\mu \,s}(x,t)$ satisfies an \emph{Ito stochastic field equation} of
the form 
\begin{equation}
\delta \psi _{A}^{\mu \,s}(x,t\mathbf{)=}\mathcal{G}_{A}^{\mu }(x)\delta
t+\sum_{a}\mathcal{N}_{Aa}^{\mu }(x)\int_{t}^{t+\delta t}dt_{1}\Gamma
_{a}(t_{1})  \label{Eq.ItoSDEFieldsHyb}
\end{equation}%
or equivalently 
\begin{equation}
\frac{\partial }{\partial t}\psi _{A}^{\mu \,s}(x,t\mathbf{)=}\mathcal{G}%
_{A}^{\mu }(x)+\sum_{a}\mathcal{N}_{Aa}^{\mu }(x)\Gamma _{a}(t_{+})
\label{Eq.ItoSDEFieldsHyb2}
\end{equation}%
where expansions of $\psi _{A}^{\mu \,s}(x,t),\mathcal{G}_{A}^{\mu }(x)$ and 
$\mathcal{N}_{Aa}^{\mu }(x)$ in terms of the mode functions $\phi _{Ak}^{\mu
}(x,t)$ and stochastic phases $\alpha _{\mu Ak}^{s}(t\mathbf{)}$ will be
determined later.

Using an obvious generalisation of (\ref{Eq.Taylor}) we can write the change
in the stochastic functional $F[\psi ^{s},\psi ^{s+}]$ due to changes $%
\delta \psi _{A}^{\mu \,s}(x,t\mathbf{)}$ in the stochastic fields as%
\begin{eqnarray}
&&F[\psi ^{s}(x,t)+\delta \psi ^{s}(x,t),\psi ^{s+}(x,t)+\delta \psi
^{s+}(x,t)]-F[\psi ^{s}(x,t),\psi ^{s+}(x,t)]  \notag \\
&=&\dint dx\,\dsum\limits_{\mu A}\delta \psi _{A}^{\mu s}(x,t)\,\left( \frac{%
\delta F[\psi ^{s},\psi ^{s+}]}{\delta \psi _{A}^{\mu s}(x,t)}\right) _{x}
\label{Eq.ChangeStochasticFnHybrid} \\
&&+\frac{1}{2}\dint \dint dxdy\,\dsum\limits_{\mu A,\nu B}\delta \psi
_{A}^{\mu s}(x,t)\,\delta \psi _{B}^{\nu s}(y,t)\,\left( \frac{\delta
^{2}F[\psi ^{s},\psi ^{s+}]}{\delta \psi _{A}^{\mu s}(x,t)\delta \psi
_{B}^{\nu s}(y,t)}\right) _{x,y}  \notag
\end{eqnarray}%
correct to second order. Using the result that for any function $H(\psi
_{A}^{\mu s}(x,t))$ of the stochastic fields we have 
\begin{eqnarray}
&&\overline{H(\psi _{A}^{\mu s}(x,t_{1}))\,\Gamma _{a}(t_{2})\Gamma
_{b}(t_{3})\Gamma _{c}(t_{4})..\Gamma _{k}(t_{l})}  \notag \\
&=&\overline{H(\psi _{A}^{\mu s}(x,t_{1}))}\;\,\overline{\Gamma
_{a}(t_{2})\Gamma _{b}(t_{3})\Gamma _{c}(t_{4})..\Gamma _{k}(t_{l})}\qquad
t_{1}<t_{2},t_{3},..,t_{l}  \label{Eq.UncorrelStochFlds3}
\end{eqnarray}%
we can then evaluate $\overline{F[\psi ^{s}(x,t)+\delta \psi ^{s}(x,t),\psi
^{s+}(x,t)+\delta \psi ^{s+}(x,t)]-F[\psi ^{s}(x,t),\psi ^{s+}(x,t)]}$.

For the first order derivative terms we have after substituting from (\ref%
{Eq.ItoSDEFieldsHyb}), expanding and carrying out the stochastic averaging 
\begin{eqnarray}
&&\overline{\dint dx\,\left\{ \dsum\limits_{\mu A}\delta \psi _{A}^{\mu
s}(x,t)\,\left( \frac{\delta F[\psi ^{s},\psi ^{s+}]}{\delta \psi _{A}^{\mu
s}(x,t)}\right) _{x}\right\} }  \notag \\
&=&\overline{\dint dx\,\sum_{\mu A}\left( \frac{\delta F[\psi ^{s},\psi
^{s+}]}{\delta \psi _{A}^{\mu s}(x,t)}\right) _{x}\mathcal{G}_{A}^{\mu }(x)}%
\,\delta t  \label{Eq.BoseStochFirstDeriv2Hyb}
\end{eqnarray}%
where the stochastic average rules for sums and products have been used, the
non-correlation between the averages of functions of $\psi _{\mu }^{s}(x,t)$
at time $t$ and the $\Gamma $ at later times between $t$ to $t+\delta t$ is
applied and the term involving $\overline{\Gamma _{a}(t_{1})}$ is equal to
zero from (\ref{Eq.GaussianMarkov}). Note that this term is proportional to $%
\delta t$. Details are set out in Appendix \ref{Appendix Derivation of Ito
Stochastic Field Equations}.

For the second order derivative terms we have after substituting from (\ref%
{Eq.ItoSDEFieldsHyb}), expanding and carrying out the stochastic averaging 
\begin{eqnarray}
&&\overline{\left\{ \frac{1}{2}\dint \dint dxdy\,\dsum\limits_{\mu A,\nu
B}\delta \psi _{A}^{\mu s}(x,t)\,\delta \psi _{B}^{\nu s}(y,t)\,\left( \frac{%
\delta ^{2}F[\psi ^{s},\psi ^{s+}]}{\delta \psi _{A}^{\mu s}(x,t)\delta \psi
_{B}^{\nu s}(y,t)}\right) _{x,y}\right\} }  \notag \\
&=&\overline{\frac{1}{2}\dint \dint dxdy\,\dsum\limits_{\mu A,\nu B}\left( 
\frac{\delta ^{2}F[\psi ^{s},\psi ^{s+}]}{\delta \psi _{A}^{\mu
s}(x,t)\delta \psi _{B}^{\nu s}(y,t)}\right) _{x,y}\left[ \sum_{a}\mathcal{N}%
_{Aa}^{\mu }(x)\;\mathcal{N}_{Ba}^{\nu }(y)\right] }\;\delta t  \notag \\
&=&\overline{\frac{1}{2}\dint \dint dxdy\,\dsum\limits_{\mu A,\nu B}\left( 
\frac{\delta ^{2}F[\psi ^{s},\psi ^{s+}]}{\delta \psi _{A}^{\mu
s}(x,t)\delta \psi _{B}^{\nu s}(y,t)}\right) _{x,y}\left[ [\mathcal{N}(x)%
\mathcal{\ N}^{T}(y)]_{A,B}^{\mu ,\nu }\right] }\;\delta t  \notag \\
&&  \label{Eq.BoseStochSecondDeriv2Hyb}
\end{eqnarray}%
where the stochastic average rules for sums and products have been used, the
non-correlation between the averages of functions of $\psi _{\mu }^{s}(x,t)$
at time $t$ and the $\Gamma $ at later times between $t$ to $t+\delta t$ is
applied. The terms involving a single $\Gamma $ have a zero stochastic
average, whilst from Eq.(\ref{Eq.GaussianMarkov}) the terms with two $\Gamma 
$ give a stochastic average proportional to $\delta t$. In evaluating the
latter term Eq. (\ref{Eq.StochAverProdTwoGamma}) is used. Details are set
out in Appendix \ref{Appendix Derivation of Ito Stochastic Field Equations}.

The remaining terms give stochastic averages correct to order $\delta t^{2}$
or higher so that we have correct to first order in $\delta t$ 
\begin{eqnarray}
&&\overline{F[\psi ^{s}(x,t)+\delta \psi ^{s}(x,t),\psi ^{s+}(x,t)+\delta
\psi ^{s+}(x,t)]}-\overline{F[\psi ^{s}(x,t),\psi ^{s+}(x,t)]}  \notag \\
&=&\left\{ \overline{\dint dx\,\sum_{\mu A}\left( \frac{\delta F[\psi
^{s},\psi ^{s+}]}{\delta \psi _{A}^{\mu s}(x,t)}\right) _{x}\mathcal{G}%
_{A}^{\mu }(x)}\,\right\} \;\delta t  \label{Eq.AverageChangeStochasticFnHyb}
\\
&&+\left\{ \overline{\frac{1}{2}\dint \dint dxdy\,\dsum\limits_{\mu A,\nu
B}\left( \frac{\delta ^{2}F[\psi ^{s},\psi ^{s+}]}{\delta \psi _{A}^{\mu
s}(x,t)\delta \psi _{v}(y,t)}\right) _{x,y}\left[ [\mathcal{N}(x)\mathcal{\ N%
}^{T}(y)]_{A,B}^{\mu ,\nu }\right] }\right\} \;\delta t  \notag
\end{eqnarray}%
or 
\begin{eqnarray}
&&\frac{\partial }{\partial t}\overline{F[\psi ^{s}(x,t),\psi ^{s+}(x,t)]} 
\notag \\
&=&\overline{\dint dx\,\sum_{\mu A}\left( \frac{\delta F[\psi ^{s},\psi
^{s+}]}{\delta \psi _{A}^{\mu s}(x,t)}\right) _{x}\mathcal{G}_{A}^{\mu }(x)}
\label{Eq.HybridTimeDerivStochAver} \\
&&+\overline{\frac{1}{2}\dint \dint dxdy\,\dsum\limits_{\mu A,\nu B}\left( 
\frac{\delta ^{2}F[\psi ^{s},\psi ^{s+}]}{\delta \psi _{A}^{\mu
s}(x,t)\delta \psi _{v}(y,t)}\right) _{x,y}\left[ [\mathcal{N}(x)\mathcal{\ N%
}^{T}(y)]_{A,B}^{\mu ,\nu }\right] }  \notag
\end{eqnarray}

For the phase space functional average (\ref{Eq.HybridTimeDerivFnalPhaseAver}%
) and the stochastic average (\ref{Eq.HybridTimeDerivStochAver}) to agree we
require two conditions to be satisfied, as follows.%
\begin{eqnarray}
&&\mathcal{G}_{A}^{\mu }(x)=A_{A}^{\mu }(x)  \notag \\
&&+\int dy\,\dsum\limits_{B}\left( \dsum\limits_{kl}\phi _{Ak}^{\mu
}(x,t)C_{Ak\,Bl}^{\mu }\phi _{Bl}^{-\mu }(y,t)\right) \;\psi _{B}^{\mu }(y,t)
\notag \\
&&+\dint dy\,\dsum\limits_{B}\left( \dsum\limits_{kl}\phi _{Bl}^{\mu
}(x,t)C_{Ak\,Bl}^{-\mu }\phi _{Ak}^{-\mu }(y,t)\right) \;\psi _{A}^{\mu
}(y,t)  \notag \\
&&  \notag \\
&&[\mathcal{N}(x)\mathcal{\ N}^{T}(y)]_{A,B}^{\mu ,\nu }=E_{A\,B}^{\mu \,\nu
}(x,y)  \notag \\
&=&D_{A\,B}^{\mu \,\nu }(x,y)  \label{Eq.HybRelnFFPEStochFldDE} \\
&&+\frac{1}{2}\left\{ \delta _{A\,C}\delta _{B\,NC}\,\delta _{\mu \,-\nu
}\left( \sum_{k}\dsum\limits_{l}\phi _{Ak}^{\mu }(x,t\mathbf{)}%
C_{Ak\,Bl}^{\mu }(t)\phi _{Bl}^{\nu }(y,t\mathbf{)}\right) \right\}  \notag
\\
&&+\frac{1}{2}\left\{ \delta _{B\,C}\delta _{A\,NC}\,\delta _{\nu \,-\mu
}\left( \sum_{k}\dsum\limits_{l}\phi _{Bl}^{\nu }(y,t\mathbf{)}%
C_{Bl\,Ak}^{\nu }(t)\phi _{Ak}^{\mu }(x,t\mathbf{)}\right) \right\}  \notag
\end{eqnarray}%
These conditions determine the quantities $\mathcal{G}_{A}^{\mu }(x)$ and $%
\mathcal{N}_{Aa}^{\mu }(x)$ that occur in the Ito stochastic field equations
(\ref{Eq.ItoSDEFieldsHyb}) in terms of the quantities that occur in the
functional Fokker-Planck equation (\ref{Eq.FnalFokkerPlanckHybridDistnFnal2}%
). Note that the $B=A$ terms in the result for $\mathcal{G}_{A}^{\mu }(x)$
actually cancel out because $C_{Ak\,Al}^{\mu }+C_{Al\,Ak}^{-\mu }=0$. These
relationships are to be interpreted as replacing the stochastic quantities $%
\mathcal{G}_{A}^{\mu }(x)$, $\mathcal{N}_{Aa}^{\mu }(x)$ in the Ito
stochastic field equation (\ref{Eq.ItoSDEFieldsHyb2}) with the quantities $%
A_{A}^{\mu }(x)$, $N_{Aa}^{\mu }(x)$ and $\psi _{A}^{\mu }(x,t)$ originating
from the functional \ Fokker-Planck equation (\ref%
{Eq.FnalFokkerPlanckHybridDistnFnal2}), which are then regarded as depending
on the stochastic fields $\psi _{\mu }^{s}(x,t)$.

We next show that these conditions can always be satisfied. In Appendix \ref%
{Appendix Equivalence of Separate Mode and Field Theory} we have shown that
the Fokker-Planck equation for the distribution function and the functional
Fokker-Planck equation for the distribution functional are equivalent, and
we can make use of the results obtained there.

First, we consider the condition for $\mathcal{G}_{A}^{\mu }(x)$. From (\ref%
{Eq.DriftDiffnRelnsHybrid}) the drift term $A_{A}^{\mu }(x)$ in the
functional Fokker-Planck equation is related to the drift vector $%
A_{Ak}^{\mu }$ in the Fokker-Planck equation via 
\begin{eqnarray}
A_{A}^{\mu }(x) &=&\dsum\limits_{k}\phi _{Ak}^{\mu }(x,t)A_{Ak}^{\mu } 
\notag \\
A_{Ak}^{\mu } &=&\int dx\,\phi _{Ak}^{-\mu }(x,t\mathbf{)}A_{A}^{\mu }(x)
\end{eqnarray}%
We can also use Eqs. (\ref{Eq.FieldFns3}), (\ref{Eq.PhaseVariablesHybrid})
and (\ref{Eq.TimeDerivModes2}) to show that 
\begin{eqnarray}
&&\int dy\,\dsum\limits_{B}\left( \dsum\limits_{kl}\phi _{Ak}^{\mu
}(x,t)C_{Ak\,Bl}^{\mu }\phi _{Bl}^{-\mu }(y,t)\right) \;\psi _{B}^{\mu }(y,t)
\notag \\
&&+\dint dy\,\dsum\limits_{B}\left( \dsum\limits_{kl}\phi _{Bl}^{\mu
}(x,t)C_{Ak\,Bl}^{-\mu }\phi _{Ak}^{-\mu }(y,t)\right) \;\psi _{A}^{\mu
}(y,t)  \notag \\
&=&\left( \dsum\limits_{k}\phi _{Ak}^{\mu
}(x,t)\,\dsum\limits_{Bl}C_{Ak\,Bl}^{\mu }\alpha _{\mu Bl}\right) +\left(
\dsum\limits_{k}\frac{\partial }{\partial t}\phi _{Ak}^{\mu }(x,t)\,\alpha
_{\mu Ak}\right) 
\end{eqnarray}%
Clearly the first condition is satisfied via the choice 
\begin{eqnarray}
\mathcal{G}_{A}^{\mu }(x) &=&\dsum\limits_{k}\phi _{Ak}^{\mu
}(x,t)A_{Ak}^{\mu }+\dsum\limits_{k}\phi _{Ak}^{\mu
}(x,t)\,\dsum\limits_{Bl}C_{Ak\,Bl}^{\mu }\,\alpha _{\mu
Bl}^{s}+\dsum\limits_{k}\frac{\partial }{\partial t}\phi _{Ak}^{\mu
}(x,t)\,\alpha _{\mu Ak}^{s}  \notag \\
&&  \label{Eq.GeeHyb}
\end{eqnarray}%
written in terms of mode functions and quantities derived from the
Fokker-Planck equation (\ref{Eq.NewFokker-Planck}).

Second, for the condition for $\mathcal{N}_{Aa}^{\mu }(x)$ we need to be
able to find $\mathcal{N}_{Aa}^{\mu }(x)$ such that 
\begin{equation}
\dsum\limits_{a}\mathcal{N}_{Aa}^{\mu }(x)\mathcal{N}_{Ba}^{\nu
}(y)=E_{A\,B}^{\mu \,\nu }(x,y)
\end{equation}%
Using the relationship in Eq.(\ref{Eq.DriftDiffnRelnsHybrid}) between the
diffusion matrix $E_{A\,B}^{\mu \,\nu }(x,y)$ in the functional
Fokker-Planck equation (\ref{Eq.FnalFokkerPlanckHybridDistnFnal2}) and the
diffusion matrix $E_{Ak\,Bl}^{\mu \nu }$ in the Fokker-Planck equation (\ref%
{Eq.NewFokker-Planck}) we have%
\begin{eqnarray}
E_{A\,B}^{\mu \,\nu }(x,y) &=&\dsum\limits_{k\,l}\phi _{Ak}^{\mu }(x,t%
\mathbf{)}E_{Ak\,Bl}^{\mu \nu }\phi _{Bl}^{\nu }(y,t\mathbf{)}  \notag \\
E_{Ak\,Bl}^{\mu \,\nu } &=&\dint \dint dxdy\,\phi _{Ak}^{-\mu
}(x,t)E_{A\,B}^{\mu \,\nu }(x,y)\phi _{Bl}^{-\nu }(y,t)
\end{eqnarray}%
The matrix $E$ is symmetric $(E_{Ak\,Bl}^{\mu \nu }=E_{Bl\,Ak}^{\nu \mu })$,
so from the Takagi factorisation \cite{Takagi25a}, \cite{Horn85a} it can be
written as $KK^{T}$, where the matrix $K$ is as in Section \ref{Section 2
Separate Modes Case}, Eq.(\ref{Eq.Takagi}). Thus the diffusion matrix $%
E_{A\,B}^{\mu \,\nu }(x,y)$ may be factorised as%
\begin{equation}
E_{A\,B}^{\mu \,\nu }(x,y)=\dsum\limits_{a}N_{Aa}^{\mu }(x)N_{Ba}^{\nu }(y)
\end{equation}%
where 
\begin{eqnarray}
N_{Aa}^{\mu }(x) &=&\dsum\limits_{k}K_{Aka}^{\mu }\phi _{Ak}^{\mu }(x,t%
\mathbf{)}  \label{Eq.NetaHyb0} \\
E_{Ak\,Bl}^{\mu \,\nu } &=&\dsum\limits_{a}K_{Aka}^{\mu }K_{Bla}^{\nu }
\label{Eq.DiffusionMatrixHybFac}
\end{eqnarray}%
Hence the second condition is satisfied with the choice%
\begin{eqnarray}
\mathcal{N}_{Aa}^{\mu }(x) &=&N_{Aa}^{\mu }(x)  \notag \\
&=&\dsum\limits_{k}K_{Aka}^{\mu }\phi _{Ak}^{\mu }(x,t\mathbf{)}
\label{Eq.NetaHyb}
\end{eqnarray}%
again written in terms of mode functions and quantities derived from the
Fokker-Planck equation (\ref{Eq.NewFokker-Planck}).

Finally, we can show that the Ito stochastic field equations (\ref%
{Eq.ItoSDEFieldsHyb2}) for the stochastic fields $\psi _{A}^{\mu s}(x,t%
\mathbf{)}$ in the field theory treatment are entirely equivalent to the Ito
stochastic equations (\ref{Eq.ItoSDEFinalHyb}) for the stochastic phase
variables $\alpha _{\mu Ak}^{s}$ in the separate mode treatment. The proof
is given in Appendix \ref{Appendix Equivalence of Separate Mode and Field
Theory}.

This proof shows that the \emph{stochastic fields} are given by 
\begin{equation}
\psi _{A}^{\mu s}(x,t\mathbf{)=}\sum_{k}\alpha _{\mu Ak}^{s}(t)\phi
_{Ak}^{\mu }(x,t\mathbf{)}  \label{Eq.StochasticFldHyb2}
\end{equation}%
where the stochastic feature results \emph{only} from the phase variables
being replaced by stochastic variables. Thus the stochastic field equations
for $\psi _{A}^{\mu s}(x,t\mathbf{)}$ are equivalent to the stochastic phase
variable equations for the $\alpha _{\mu Ak}^{s}(t)$.

\subsection{Expressions for Ito Stochastic Field Equations}

The original Ito stochastic field equation (\ref{Eq.ItoSDEFieldsHyb2}) can
also be written in terms of the drift, diffusion terms that are derived
directly from the functional Fokker-Planck equation (\ref%
{Eq.FnalFokkerPlanckHybridDistnFnal2}) itself. Substituting from (\ref%
{Eq.NetaHyb}) and (\ref{Eq.HybRelnFFPEStochFldDE}) we have for the \emph{Ito
stochastic field equation} 
\begin{eqnarray}
&&\frac{\partial }{\partial t}\psi _{A}^{\mu s}(x,t\mathbf{)=}A_{A}^{\mu }(x)
\notag \\
&&+\int dy\,\dsum\limits_{B\neq A}\left( \dsum\limits_{kl}\phi _{Ak}^{\mu
}(x,t)C_{Ak\,Bl}^{\mu }\,\phi _{Bl}^{-\mu }(y,t\mathbf{)}\right) \mathbf{\;}%
\psi _{B}^{\mu s}(y,t)  \notag \\
&&+\dint dy\,\dsum\limits_{B\neq A}\left( \dsum\limits_{kl}\phi _{Ak}^{-\mu
}(y,t)C_{Ak\,Bl}^{-\mu }\phi _{Bl}^{\mu }(x,t)\right) \;\psi _{A}^{\mu
s}(y,t)  \notag \\
&&+\sum_{a}N_{Aa}^{\mu }(x)\Gamma _{a}(t_{+})
\label{Eq.ItoSDEFieldsHybFinal}
\end{eqnarray}%
where $A_{A}^{\mu }(x)$ is the drift term and $N_{Aa}^{\mu }(x)$ is related
to the diffusion term $E_{A\,B}^{\mu \,\nu }(x,y)$ that occur in the
functional Fokker-Planck equation (\ref{Eq.FnalFokkerPlanckHybridDistnFnal2}%
) via 
\begin{eqnarray}
\dsum\limits_{a}N_{Aa}^{\mu }(x)N_{Ba}^{\mu }(y) &=&E_{A\,B}^{\mu \,\nu
}(x,y)  \notag \\
&=&D_{A\,B}^{\mu \,\nu }(x,y)  \label{Eq.ItoSDEFinalHybNetas} \\
&&+\frac{1}{2}\left\{ \delta _{A\,C}\delta _{B\,NC}\,\delta _{\mu \,-\nu
}\left( \sum_{k}\dsum\limits_{l}\phi _{Ak}^{\mu }(x,t\mathbf{)}%
C_{Ak\,Bl}^{\mu }(t)\phi _{Bl}^{\nu }(y,t\mathbf{)}\right) \right\}  \notag
\\
&&+\frac{1}{2}\left\{ \delta _{B\,C}\delta _{A\,NC}\,\delta _{\nu \,-\mu
}\left( \sum_{k}\dsum\limits_{l}\phi _{Bl}^{\nu }(y,t\mathbf{)}%
C_{Bl\,Ak}^{\nu }(t)\phi _{Ak}^{\mu }(x,t\mathbf{)}\right) \right\}  \notag
\end{eqnarray}%
The $B=A$ terms have been cancelled out using $C_{Ak\,Al}^{\mu
}\,+C_{Ak\,Al}^{-\mu }=0$. The result for $\frac{{\LARGE \partial }}{{\LARGE %
\partial t}}\psi _{A}^{\mu s}(x,t\mathbf{)}$ may be compared to the Ito
stochastic equation (\ref{Eq.ItoSDEFinalHyb}) for the phase variables. The
term that is added to $A_{A}^{\mu }(x)$ involves a non-local contribution
with a kernel that depends on the mode functions and their time dependence
via the coupling coefficients. It has the effect of coupling the condensate
stochastic fields $\psi _{C}^{\mu s}$ to the non-condensate stochastic
fields $\psi _{NC}^{\mu s}$ $(\mu =-,+)$. This term is not present when time
independent modes are used. Note that it is of the same form as in (\ref%
{Eq.TimeDerivHybridFields}) for the time derivative of the (non-stochastic)
field functions $\psi _{A}^{\mu }(x,t\mathbf{)}$. This form for the Ito
stochastic field equation is the most useful, since it involves the drift $%
A_{A}^{\mu }(x)$ and diffusion terms $D_{A\,B}^{\mu \,\nu }(x,y)$ in the
standard functional Fokker-Planck equation obtained by applying the
correspondence rules for the $\frac{{\LARGE \partial }}{{\LARGE \partial t}}%
\hat{\rho}$ contribution in the characteristic functional, plus the time
dependent modes $\phi _{Ak}^{\mu }(x,t\mathbf{)}$ and the coupling
coefficients $C_{Ak\,Bl}^{\mu }(t)$.\textbf{\ }

\subsection{Classical and Noise Field Terms}

We can write the Ito stochastic field equation in terms of a \emph{classical}
term and a \emph{noise} term%
\begin{eqnarray}
\frac{\partial }{\partial t}\psi _{A}^{\mu s}(x,t\mathbf{)} &\mathbf{=}%
&\left( \frac{\partial }{\partial t}\psi _{A}^{\mu s}(x,t\mathbf{)}\right)
_{class}\mathbf{+}\left( \frac{\partial }{\partial t}\psi _{A}^{\mu s}(x,t%
\mathbf{)}\right) _{noise}  \notag \\
\left( \frac{\partial }{\partial t}\psi _{A}^{\mu s}(x,t\mathbf{)}\right)
_{class} &=&A_{A}^{\mu }(x)  \notag \\
&&+\int dy\,\dsum\limits_{B\neq A}\left( \dsum\limits_{kl}\phi _{Ak}^{\mu
}(x,t)C_{Ak\,Bl}^{\mu }\,\phi _{Bl}^{-\mu }(y,t\mathbf{)}\right) \mathbf{\;}%
\psi _{B}^{\mu s}(y,t)  \notag \\
&&+\dint dy\,\dsum\limits_{B\neq A}\left( \dsum\limits_{kl}\phi _{Ak}^{-\mu
}(y,t)C_{Ak\,Bl}^{-\mu }\phi _{Bl}^{\mu }(x,t)\right) \;\psi _{A}^{\mu
s}(y,t)  \notag \\
\left( \frac{\partial }{\partial t}\psi _{A}^{\mu s}(x,t\mathbf{)}\right)
_{noise} &=&\sum_{a}N_{Aa}^{\mu }(x)\Gamma _{a}(t_{+})
\label{Eq.ItoFieldsHybClassNoise}
\end{eqnarray}%
If only the classical terms were included, then the solution for $\psi
_{A}^{\mu s}(x,t\mathbf{)}$ would determine classical field functions; hence
their name. Note that the number of noise terms will be equal to $%
2(n_{C}+n_{NC})$, or twice the total number of condensate and non-condensate
modes.

\subsection{Properties of Noise Fields}

The stochastic averages of the noise field terms can now be evaluated. For a
single noise term and the product of two noise terms we have using Eqs. (\ref%
{Eq.GaussianMarkov}) and (\ref{Eq.UncorrelStochFlds3}) 
\begin{eqnarray}
\overline{\left( \frac{\partial }{\partial t}\psi _{\mu A}^{s}(x,t)\right)
_{n}} &=&0  \notag \\
\overline{\left( \frac{\partial }{\partial t}\psi _{\mu
A}^{s}(x_{1},t_{1})\right) _{n}\left( \frac{\partial }{\partial t}\psi _{\nu
B}^{s}(x_{2},t_{2})\right) _{n}} &=&\delta (t_{1}-t_{2})\overline{%
E_{A\,B}^{\mu \nu }(x_{1},x_{2};t_{1,2})}  \label{Eq.NoiseFieldHybProps1}
\end{eqnarray}%
showing that the stochastic average of a single noise field term is zero
whilst that for the product of two noise field terms is delta correlated in
time and equal to the appropriate diffusion matrix element. Note that it is
not necessarily delta correlate in space.

In fact the diffusion matrix elements determine all the stochastic averages
of products of noise field terms. With an odd number of terms the stochastic
average is zero. For an even number of terms the stochastic average involves
sums of stochastic averages of products of diffusion matrix elements, which
reflects the Gaussian-Markoff properties of the $\Gamma _{a}$. Thus for
three and four noise field terms%
\begin{equation}
\overline{\left( \frac{\partial }{\partial t}\psi _{\mu
A}^{s}(x_{1},t_{1})\right) _{n}\left( \frac{\partial }{\partial t}\psi _{\nu
B}^{s}(x_{2},t_{2})\right) _{n}\left( \frac{\partial }{\partial t}\psi _{\xi
C}^{s}(x_{3},t_{3})\right) _{n}}=0  \label{Eq.NoiseFieldHybProps2}
\end{equation}%
\begin{eqnarray}
&&\overline{\left( \frac{\partial }{\partial t}\psi _{\mu
A}^{s}(x_{1},t_{1})\right) _{n}\left( \frac{\partial }{\partial t}\psi _{\nu
B}^{s}(x_{2},t_{2})\right) _{n}\left( \frac{\partial }{\partial t}\psi _{\xi
C}^{s}(x_{3},t_{3})\right) _{n}\left( \frac{\partial }{\partial t}\psi
_{\lambda D}^{s}(x_{4},t_{4})\right) _{n}}  \notag \\
&=&\delta (t_{1}-t_{2})\delta (t_{3}-t_{4})\;\overline{E_{AB}^{\mu \nu
}(x_{1},x_{2};t_{1,2})E_{CD}^{\xi \lambda }(x_{3},x_{4};t_{3,4})}  \notag \\
&&+\delta (t_{1}-t_{3})\delta (t_{2}-t_{4})\;\overline{E_{AC}^{\mu \xi
}(x_{1},x_{3};t_{1,3})E_{BD}^{\nu \lambda }(x_{2},x_{4};t_{2,4})}  \notag \\
&&+\delta (t_{1}-t_{4})\delta (t_{2}-t_{3})\;\overline{E_{AD}^{\mu \lambda
}(x_{1},x_{4};t_{1,4})E_{BC}^{\nu \xi }(x_{2},x_{3};t_{2,3})}
\label{Eq.NoiseFieldHybProps3}
\end{eqnarray}%
Note that the noise field terms are not themselves Gaussian-Markoff
processes.

\textbf{\pagebreak }

\section{Conclusions}

\label{Section 5 - Conclusions}

This paper has set out a phase space theory approach to treating the
dynamical behaviour of Bose-Einstein condensates which could be applied to a
variety of experimental situations, including interferometry with BEC
involving time-dependent double well potentials. The key feature is the use
of time-dependent mode functions, chosen so that a few (one, two,..) highly
occupied modes provide a good approximation to the physics of interacting
bosons in the time dependent potential at temperatures well below the BEC
transition tememperature. The motivation is that when the effect the lightly
occupied non-condensate modes are taken into account in treatments of
decoherence effects allowing for Bogoliubov excitations, collisions with
thermal bosons etc., the previously determined condensate modes are still
important in treating these extensions. To allow for time dependent mode
functions, the major innovation here is the introduction of time dependent
phase variables, the time dependence being chosen to match that of the time
dependent mode annihilation, creation operators. This means that in this
approach we are representing time dependent \emph{mode} annihilation,
creation operators by time dependent phase variables, whilst time
independent \emph{total} \emph{field} annihilation, creation operators are
represented by time independent field functions. In the more usual approach
using time independent phase variables the opposite situation applies, which
is rather unsatisfactory. The theory presented here treats the two
situations, one (\emph{mode theory}) being where mode annihilation, creation
operators and their related phase variables and distribution functions are
dealt with specifically, the other (\emph{field theory}) being where field
creation, annihilation operators and their related field functions and
distribution functionals involve a description where individual modes are
not distinguished. Though each situation is treated separately, they are
shown to be equivalent. Within each of these two situations the modes are
divided up between condensate (highly occupied) modes and non-condensate
(sparsely occupied) modes. This is referred to as the \emph{hybrid approach}%
. For the hybrid case a Wigner type distribution function or functional is
used for the condensate bosons and a positive P type for the non-condensate
bosons. Fokker-Planck or functional Fokker-Planck equations are derived,
along with the Ito stochastic equations for stochastic phases or stochastic
fields, and the relationship between the Ito equation quantities and those
in the Fokker-Planck equation is obtained. The stochastic equations contain
the sum of a classical term related to the drift vector and a noise term,
and the stochastic properties of the noise terms are related to the
diffusion matrix. Also, the expressions for the stochastic fields involve
the same expansion in terms of time dependent mode functions and stochastic
phase variables, as had applied to the field function expansions involving
non-stochastic phase variables.

The key results are set out below. For the hybrid approach the condensate
and non-condensate field operators are time dependent, as are the
corresponding field functions. Also, both the Fokker-Planck and functional
Fokker-Planck equations differ from those derived using the usual
correspondence rules, the drift vectors are unchanged but the diffusion
matrices contain additional terms relating to the time dependence of the
mode functions and involving the \emph{coupling coefficients}, which are
defined via overlap integrals between mode functions and their time
derivatives. In addition, there are extra terms in the Ito stochastic
equations both for the stochastic phases and stochastic fields, again
related to the time dependence of the mode functions and involving the
coupling coefficients. For the hybrid case the final form for the Ito
stochastic field equation in Eqs. (\ref{Eq.ItoSDEFieldsHybFinal2}), (\ref%
{Eq.NetaHyb2}) is the most useful, since it involves the drift $A_{A}^{\mu
}(x)$ and diffusion terms $D_{A\,B}^{\mu \,\nu }(x,y)$ in the standard
functional Fokker-Planck equation obtained by applying the correspondence
rules for the $\frac{{\LARGE \partial }}{{\LARGE \partial t}}\hat{\rho}$
contribution in the characteristic functional, plus the time dependent modes 
$\phi _{Ak}^{\mu }(x,t\mathbf{)}$ and the coupling coefficients $%
C_{Ak\,Bl}^{\mu }(t)$.\textbf{\ }

From the point of view of best describing the physics of BECs even near zero
temperature the treatment preferred here is the hybrid one involving
condensate and non-condensate fields rather than considering a large number
of separate modes, but based on time dependent mode functions. For this
reason, the field theory treatment has been given an important emphasis in
this paper. That treatment of the hybrid approach capitalises on the simple
but accurate picture based on highly occupied condensate modes and lightly
occupied non-condensate modes, but avoids a mode-by-mode approach in
determining the quantum correlation functions - which after all only involve
field operators. There are of course some costs, for numerical work the set
of time dependent mode functions and the coupling constants would need to be
worked out first, but hopefully only a few condensate modes would be needed
and the non-condensate modes might be introduced via physically based
variational methods involving Fock states where only one boson is in a
non-condensate mode, or by just applying Schmidt orthogonalisation
procedures.

In Ref. \cite{Dalton10a} the hybrid approach was applied based on including
time dependent condensate modes and the functional Fokker-Planck equations
derived using the correspondence rules as set out here in Eqs. (\ref%
{Eq.NonCondOprsFnalCorr}) and (\ref{Eq.CondOprsFnalCorr}). In terms of this
paper (see Eq. (\ref{Eq.SummHyb1})) this gives the drift vector $A_{A}^{\mu
}(x)$ and the diffusion matrix contribution $D_{A\,B}^{\mu \,\nu }(x,y)$.
However, no terms were included in either the functional Fokker-Planck
equation or the Ito stochastic field equations involving the coupling
coefficients $C_{Ak\,Bl}^{\mu }(t)$. The theory in Ref. \cite{Dalton10a} was
based on using time independent phase variables and characteristic function
variables, but the time dependences of both the condensate and
non-condensate field operators and field functions consequential on using
time dependent condensate modes were ignored. Eqs. (192), (193) and (207) in
Ref. \cite{Dalton10a} do not include contributions involving the coupling
coefficients. The results for the Ito stochastic field equations in Ref. 
\cite{Dalton10a} are correct if time independent (or slowly varying)
condensate and non-condensate modes were used, but then the condensate would
require many more than one or two modes for an adequate description. Note
however that the non-local term involving $F(\mathbf{r},\mathbf{s})$ in the
diffusion matrices in Ref. \cite{Dalton10a} would still be present, as this
arose from the $\frac{{\LARGE \partial }}{{\LARGE \partial t}}\hat{\rho}$
contribution in the characteristic functional and does not involve mode time
dependence. However, results from the present paper are really needed to
supplement those in Ref. \cite{Dalton10a} if the physically based idea of
just having one or two genuinely time-dependent condensate modes is to be
implemented.

The results in this paper can be easily converted back to the corresponding
results based on using a convenient set of time independent mode functions.
In this case all the coupling constants become zero and the phase variables,
mode and field operators become time independent again. The functional
Fokker-Planck and Fokker-Planck equations would revert to those obtainable
by standard methods using the correspondence rules, the Ito stochastic
equations for fields and phase variables are also the same as in standard
methods, as also are the relationships between these equations and the
Fokker-Planck equations. However, such mode functions are not closely
aligned to those that provide a good approximation to the near zero
temperature physics, so any advantages in terms of simplicity in applying
the theory and convenience for numerical work is likely to be negated.

As well as the hybrid approach treated in this paper, situations where time
dependent mode functions are involved may also be treated using the
so-called \emph{combined approach}, where the modes are not divided up into
condensate or non-condensate types and only the total field operators and
field functions are considered. For completeness, the results for the
combined approach are also set out below. The same method involving time
dependent phase space variables can be used. Again, two situations may be
considered, one (mode theory) being where mode annihilation, creation
operators and their related phase variables and distribution functions are
dealt with specifically, the other (field theory) being where total field
creation, annihilation operators and their related field functions and
distribution functionals involve a description where individual modes are
not distinguished. The distribution function or functional may be either of
the Wigner or positive P type. For the combined approach both the
Fokker-Planck and functional Fokker-Planck equations are exactly the same as
those derived using the usual correspondence rules, even though time
dependent modes are present. The field functions are time independent.
However, there are extra terms in the Ito stochastic equations for the
stochastic phases though not for the stochastic fields. These terms relate
to the time dependence of the mode functions and involve the coupling
coefficients. For the combined case, the final form for the Ito stochastic
field equation in Eqs. (\ref{Eq.ItoSDEFinalTotalFlds2}) is the most useful,
since it involves the drift $A^{\mu }(x)$ and diffusion terms $D^{\mu \,\nu
}(x,y)$ in the standard functional Fokker-Planck equation obtained by
applying the correspondence rules for the $\frac{{\LARGE \partial }}{{\LARGE %
\partial t}}\hat{\rho}$ contribution in the characteristic functional, plus
the time dependent modes $\phi _{k}^{\mu }(x,t\mathbf{)}$ and the coupling
coefficients $C_{k\,l}^{\mu }(t)$.\textbf{\ } \pagebreak 

\section{Summary of Key Results}

\label{Section 4 - Summary of Key Results}

\subsection{Hybrid Approach - Condensate and Non-Condensate Modes and Fields}

Here we summarise the key results obtained in this paper. Results for
functional Fokker-Planck, Ito stochastic field equations and the
Fokker-Planck, Ito stochastic phase variable equations for the \emph{hybrid }%
approach where the time dependent modes are divided into $n_{C}$ condensate
modes and $n_{NC}$ non-condensate modes, plus conversions and key results
are as follows.

For the field theory results:%
\begin{eqnarray}
\frac{\partial }{\partial t}P[\underrightarrow{\psi },\underrightarrow{\psi }%
^{\ast }] &=&\left\{ -\dsum\limits_{\mu A}\dint dx\frac{\delta }{\delta \psi
_{A}^{\mu }(x,t)}A_{A}^{\mu }(x)\right\} P[\underrightarrow{\psi },%
\underrightarrow{\psi }^{\ast }]  \notag \\
&&+\left\{ \frac{1}{2}\dsum\limits_{\mu A}\dsum\limits_{\nu B}\dint \dint
dxdy\frac{\delta }{\delta \psi _{A}^{\mu }(x,t)}\frac{\delta }{\delta \psi
_{B}^{\nu }(y,t)}E_{A\,B}^{\mu \,\nu }(x,y)\right\} P[\underrightarrow{\psi }%
,\underrightarrow{\psi }^{\ast }]  \notag \\
\psi _{A}^{\mu }(x,t) &=&\dsum\limits_{k}\alpha _{\mu Ak}(t)\phi _{Ak}^{\mu
}(x,t\mathbf{)}\qquad \qquad \alpha _{\mu Ak}=\int dy\,\phi _{Ak}^{-\mu }(y,t%
\mathbf{)}\psi _{A}^{\mu }(y,t)  \notag \\
\frac{\partial }{\partial t}\psi _{A}^{\mu s}(x,t) &\mathbf{=}&A_{A}^{\mu
}(x)+\int dy\,\dsum\limits_{B\neq A}\left( \dsum\limits_{kl}\phi _{Ak}^{\mu
}(x,t)C_{Ak\,Bl}^{\mu }\,\phi _{Bl}^{-\mu }(y,t\mathbf{)}\right) \mathbf{\;}%
\psi _{B}^{\mu s}(y,t)  \notag \\
&&+\dint dy\,\dsum\limits_{B\neq A}\left( \dsum\limits_{kl}\phi _{Ak}^{-\mu
}(y,t)C_{Ak\,Bl}^{-\mu }\phi _{Bl}^{\mu }(x,t)\right) \;\psi _{A}^{\mu
s}(y,t)  \notag \\
&&+\sum_{a}N_{Aa}^{\mu }(x)\Gamma _{a}(t_{+})
\label{Eq.ItoSDEFieldsHybFinal2} \\
E_{A\,B}^{\mu \,\nu }(x,y) &=&D_{A\,B}^{\mu \,\nu }(x,y)  \notag \\
&&+\frac{1}{2}\left\{ \delta _{A\,C}\delta _{B\,NC}\,\delta _{\mu \,-\nu
}\left( \sum_{k}\dsum\limits_{l}\phi _{Ak}^{\mu }(x,t\mathbf{)}%
C_{Ak\,Bl}^{\mu }(t)\phi _{Bl}^{\nu }(y,t\mathbf{)}\right) \right\}  \notag
\\
&&+\frac{1}{2}\left\{ \delta _{B\,C}\delta _{A\,NC}\,\delta _{\nu \,-\mu
}\left( \sum_{k}\dsum\limits_{l}\phi _{Bl}^{\nu }(y,t\mathbf{)}%
C_{Bl\,Ak}^{\nu }(t)\phi _{Ak}^{\mu }(x,t\mathbf{)}\right) \right\}  \notag
\\
&=&\dsum\limits_{a}N_{Aa}^{\mu }(x)N_{Ba}^{\mu }(y)  \label{Eq.NetaHyb2} \\
\psi _{A}^{\mu s}(x,t\mathbf{)} &\mathbf{=}&\sum_{k}\alpha _{\mu
Ak}^{s}(t)\phi _{Ak}^{\mu }(x,t\mathbf{)}\qquad \qquad \alpha _{\mu
Ak}^{s}=\int dy\,\phi _{Ak}^{-\mu }(y,t\mathbf{)}\psi _{A}^{\mu s}(y,t)
\label{Eq.SummHyb1}
\end{eqnarray}%
For the mode theory results:%
\begin{eqnarray}
\frac{\partial }{\partial t}P(\alpha ,\alpha ^{+},\alpha ^{\ast },\alpha
^{+\ast }) &=&\left\{ -\dsum\limits_{\mu Ak}\frac{\partial }{\partial \alpha
_{\mu Ak}}A_{Ak}^{\mu }+\frac{1}{2}\dsum\limits_{\mu Ak}\dsum\limits_{\nu Bl}%
\frac{\partial }{\partial \alpha _{\mu Ak}}\frac{\partial }{\partial \alpha
_{\nu Bl}}E_{Ak\,Bl}^{\mu \nu }\right\} P(\alpha ,\alpha ^{+},\alpha ^{\ast
},\alpha ^{+\ast })  \notag \\
\frac{\partial }{\partial t}\alpha _{\mu Ak}^{s} &=&A_{Ak}^{\mu
}+\sum_{Bl}C_{Ak\,Bl}^{\mu }\,\alpha _{\mu Bl}^{s}+\sum_{a}K_{Aka}^{\mu
}\,\Gamma _{a}(t_{+})  \notag \\
E_{Ak\,Bl}^{\mu \nu } &=&D_{Ak\,Bl}^{\mu \nu }+\frac{1}{2}(\delta
_{A\,C}\,\delta _{B\,NC}\,\delta _{\mu \,-\nu }C_{Ak\,Bl}^{\mu }+\delta
_{B\,C}\,\delta _{A\,NC}\,\delta _{\nu \,-\mu }C_{Bl\,Ak}^{\nu })  \notag \\
&=&\dsum\limits_{a}K_{Aka}^{\mu }K_{Bla}^{\nu }  \label{Eq.SummHybModes1}
\end{eqnarray}%
Conversions between the quantities are as follows%
\begin{eqnarray}
A_{Ak}^{\mu } &=&\dint dx\,\phi _{Ak}^{-\mu }(x,t)A_{A}^{\mu }(x)\qquad
A_{A}^{\mu }(x)=\dsum\limits_{k}\phi _{Ak}^{\mu }(x,t)A_{Ak}^{\mu }  \notag
\\
E_{Ak\,Bl}^{\mu \,\nu } &=&\dint \dint dxdy\,\phi _{Ak}^{-\mu
}(x,t)E_{A\,B}^{\mu \,\nu }(x,y)\phi _{Bl}^{-\nu }(y,t)\qquad E_{A\,B}^{\mu
\,\nu }(x,y)=\dsum\limits_{kl}\phi _{Ak}^{\mu }(x,t)E_{Ak\,Bl}^{\mu \nu
}\phi _{Bl}^{\nu }(y,t)  \notag \\
N_{Aa}^{\mu }(x) &=&\dsum\limits_{k}K_{Aka}^{\mu }\phi _{Ak}^{\mu }(x,t%
\mathbf{)}  \label{Eq.SummHyb2}
\end{eqnarray}%
Key results for the time dependences are%
\begin{eqnarray}
\frac{\partial }{\partial t}\phi _{Ak}^{\mu }(x,t)
&=&\dsum\limits_{Bl}C_{Ak\,Bl}^{-\mu }\phi _{Bl}^{\mu }(x,t)  \notag \\
\frac{\partial }{\partial t}\alpha _{\mu Ak} &=&\sum_{Bl}C_{Ak\,Bl}^{\mu
}\,\alpha _{\mu Bl}  \label{Eq.SummHyb3}
\end{eqnarray}%
with coupling coefficients%
\begin{eqnarray}
C_{Ak\,Bl}^{-}(t) &=&\dint dx\frac{\partial \phi _{Ak}^{\ast }(x,t\mathbf{)}%
}{\partial t}\phi _{Bl}(x,t\mathbf{)=}C_{Ak\,Bl}(t)  \notag \\
C_{Ak\,Bl}^{+}(t) &=&\dint dx\frac{\partial \phi _{Ak}(x,t\mathbf{)}}{%
\partial t}\phi _{Bl}^{\ast }(x,t\mathbf{)=}C_{Ak\,Bl}^{\ast }(t)
\label{Eq,.SummHyb4}
\end{eqnarray}

\subsection{Combined Approach - Modes and Fields}

A much simpler treatment applies if all $n$ condensate and non-condensate
modes are considered together in the so-called $\emph{combined}$ approach,
even though the mode functions are time dependent. Here we will just
summarise the results for completeness, but without proof since the proofs
are easily obtained from the previous sections. Many of the extra terms for
the hybrid case just become zero. The phase variables are again considered
to be time dependent, as in Eq. (\ref{Eq.TimeDerivModePhases}) as are the
characteristic function variables, as in Eq. (\ref{Eq.TimeDepCharFnVar}).
The phase space treatment is based on the positive P distribution for all
modes, but similar results would apply if the double space Wigner
representation was used. It then turns out that the Fokker-Planck equations
involving the phase variables $\alpha _{\mu k}$ $(\mu =-,+,k=1,..,n)$ have
the same drift $A_{k}^{\mu }$ and diffusion $D_{kl}^{\mu \nu }$ terms as in
the standard Fokker-Planck equations resulting from just the $\frac{{\LARGE %
\partial }}{{\LARGE \partial t}}\hat{\rho}$ contribution in the
characteristic function. However, the Ito stochastic equations have an
additional contribution $\sum_{l}C_{kl}^{\mu }\,\alpha _{\mu l}^{s}$ to the
classical field term, though the noise term just involves the matrix $B$
related to the diffusion matrix via $D=BB^{T}$. The functional Fokker-Planck
for the time independent total fields $\psi _{\mu }(x\mathbf{)}$ turns out
to be exactly the same as in the standard functional Fokker-Planck equations
resulting from just the $\frac{{\LARGE \partial }}{{\LARGE \partial t}}\hat{%
\rho}$ contribution in the characteristic functional, involving drift $%
A^{\mu }(x)$ and diffusion $D^{\mu \nu }(x,y)$ terms. Also, the Ito
stochastic field equations are of the same form as if the mode functions
were time independent. The classical field term is given by the drift term
and the noise field term is related in the standard way to the diffusion
term in the functional Fokker-Planck equation. Furthermore, the stochastic
field $\psi _{\mu }^{s}(x,t\mathbf{)}$ can be expanded in terms of the time
dependent mode functions with the stochastic phase variables as
coefficients, with the same form as for the non-stochastic fields $\psi
_{\mu }(x\mathbf{)}$.

Results for functional Fokker-Planck, Ito stochastic field equations and the
Fokker-Planck, Ito stochastic phase variable equations for the \emph{combined%
} approach where all time dependent modes are considered together, plus
conversions and key results are as follows.

For the field theory results:%
\begin{eqnarray}
\frac{\partial }{\partial t}P[\psi ,\psi ^{+},\psi ^{\ast },\psi ^{+\ast }]
&=&\left\{ -\dsum\limits_{\mu }\dint dx\frac{\delta }{\delta \psi _{\mu }(x)}%
A^{\mu }(x)\right\} P[\psi ,\psi ^{+},\psi ^{\ast },\psi ^{+\ast }]  \notag
\\
&&+\left\{ \frac{1}{2}\dsum\limits_{\mu }\dsum\limits_{\nu }\dint \dint dxdy%
\frac{\delta }{\delta \psi _{\mu }(x)}\frac{\delta }{\delta \psi _{\nu }(y)}%
D^{\mu \nu }(x,y)\right\} P[\psi ,\psi ^{+},\psi ^{\ast },\psi ^{+\ast }] 
\notag \\
\psi _{\mu }(x\mathbf{)} &\mathbf{=}&\sum_{k}\alpha _{\mu k}(t)\phi
_{k}^{\mu }(x,t\mathbf{)}\qquad \qquad \alpha _{\mu k}(t)=\dint dy\,\phi
_{k}^{-\mu }(y,t\mathbf{)}\psi _{\mu }(y\mathbf{)}  \notag \\
\frac{\partial }{\partial t}\psi _{\mu }^{s}(x,t\mathbf{)} &\mathbf{=}%
&A^{\mu }(x)+\sum_{a}N_{a}^{\mu }(x)\Gamma _{a}(t_{+})
\label{Eq.ItoSDEFinalTotalFlds2} \\
D^{\mu \nu }(x,y) &=&\dsum\limits_{a}N_{a}^{\mu }(x)N_{a}^{\nu }(y)  \notag
\\
\psi _{\mu }^{s}(x,t\mathbf{)} &\mathbf{=}&\sum_{k}\alpha _{\mu
k}^{s}(t)\phi _{k}^{\mu }(x,t\mathbf{)\qquad \qquad }\alpha _{\mu
k}^{s}(t)=\dint dy\,\phi _{k}^{-\mu }(y,t\mathbf{)}\psi _{\mu }^{s}(y,t%
\mathbf{)}  \label{Eq.SummComb1}
\end{eqnarray}%
For the mode theory results:%
\begin{eqnarray}
\frac{\partial }{\partial t}P(\alpha ,\alpha ^{+},\alpha ^{\ast },\alpha
^{+\ast }) &=&\left\{ -\dsum\limits_{\mu k}\frac{\partial }{\partial \alpha
_{\mu k}}A_{k}^{\mu }+\frac{1}{2}\dsum\limits_{\mu k}\dsum\limits_{\nu l}%
\frac{\partial }{\partial \alpha _{\mu k}}\frac{\partial }{\partial \alpha
_{\nu l}}D_{kl}^{\mu \nu }\right\} P(\alpha ,\alpha ^{+},\alpha ^{\ast
},\alpha ^{+\ast })  \notag \\
\frac{\partial }{\partial t}\alpha _{\mu k}^{s} &=&A_{k}^{\mu
}+\sum_{l}C_{kl}^{\mu }\,\alpha _{\mu l}^{s}+\sum_{a}B_{ka}^{\mu }\,\Gamma
_{a}(t_{+})  \notag \\
D_{kl}^{\mu \nu } &=&\dsum\limits_{a}B_{ka}^{\mu }B_{la}^{\nu }
\label{Eq.SummModes}
\end{eqnarray}%
Conversions between the quantities are as follows%
\begin{eqnarray}
A_{k}^{\mu } &=&\dint dx\,\phi _{k}^{-\mu }(x,t)A^{\mu }(x)\qquad A^{\mu
}(x)=\dsum\limits_{k}\phi _{k}^{\mu }(x,t)A_{k}^{\mu }  \notag \\
D_{kl}^{\mu \nu } &=&\dint \dint dxdy\,\phi _{k}^{-\mu }(x,t)D^{\mu \nu
}(x,y)\phi _{l}^{-\nu }(y,t)\qquad D^{\mu \nu }(x,y)=\dsum\limits_{kl}\phi
_{k}^{\mu }(x,t)D_{kl}^{\mu \nu }\phi _{l}^{\nu }(y,t)  \notag \\
N_{a}^{\mu }(x) &=&\dsum\limits_{k}B_{ka}^{\mu }\phi _{k}^{\mu }(x,t\mathbf{)%
}  \label{Eq.SummComb2}
\end{eqnarray}%
Key results for the time dependences are%
\begin{eqnarray}
\frac{\partial }{\partial t}\phi _{k}^{\mu }(x,t)
&=&\dsum\limits_{l}C_{kl}^{-\mu }\phi _{l}^{\mu }(x,t)  \notag \\
\frac{\partial }{\partial t}\alpha _{\mu k} &=&\sum_{l}C_{kl}^{\mu }\,\alpha
_{\mu l}  \label{Eq.SummComb3}
\end{eqnarray}%
with coupling coefficients%
\begin{eqnarray}
C_{kl}^{-}(t) &=&\dint dx\frac{\partial \phi _{k}^{\ast }(x,t\mathbf{)}}{%
\partial t}\phi _{l}(x,t\mathbf{)=}C_{kl}(t)  \notag \\
C_{kl}^{+}(t) &=&\dint dx\frac{\partial \phi _{k}(x,t\mathbf{)}}{\partial t}%
\phi _{l}^{\ast }(x,t\mathbf{)=}C_{kl}^{\ast }(t)  \label{Eq.SummComb4}
\end{eqnarray}

\pagebreak

\subsection{Acknowledgements}

This work was supported by the Australian Research Council Centre of
Excellence for Quantum Atom Optics (ACQAO). The author thanks J. Corney, P.
Deuar for helpful discussions and a referee for valuable advice on improving
the presentation. .

\smallskip \pagebreak

\section{Appendix - Functional Calculus}

\label{Appendix - Functional Calculus}

\subsection{Functionals}

Essentially a \emph{functional} $F[\psi (x)]$ of a \emph{field function} $%
\psi (x)$ just defines a \emph{process} that results in a c-number which
depends on \emph{all} the values of the field function, that is over the
entire range of positions $x$. Unless otherwise stated the field function is
complex. Generalisations to cases involving several functions $\psi (x),\psi
^{+}(x)$ or $\psi (x),\psi ^{+}(x),\psi ^{\ast }(x),\psi ^{+\ast }(x)$ etc
are straightforward, as also when $x$ refers to position $\mathbf{r}$ in a
3D space. The field functions may also sometimes be time-dependent.

Examples of functionals are numerous. If $\chi (x)$ is fixed then the
integral $\dint dx\chi ^{\ast }(x)\psi (x)$ is a functional of $\psi (x)$,
often written $\chi \lbrack \psi (x)]$. This functional defines the scalar
product, often written $(\chi ,\psi )$. Here$\chi ^{\ast }(x)$ acts as a 
\emph{kernel}. A function $\psi (y)$ can also be considered as a functional
of $\psi (x)$, with the delta function acting as the kernel. This specific
functional is written as $F_{y}[\psi (x)]=\dint dx\delta (y-x)\psi (x)=\psi
(y)$.

If the function $\psi (x)$ is expanded in terms of orthonormal \emph{mode
functions} $\phi _{k}(x)$ with \emph{expansion coefficients} $\alpha _{k}$ 
\begin{eqnarray}
\psi (x) &=&\dsum\limits_{k}\alpha _{k}\phi _{k}(x)  \label{Eq.ModeExpn} \\
\alpha _{k} &=&\dint dx\,\phi _{k}^{\ast }(x)\psi (x)  \label{Eq.ExpnCoefts}
\end{eqnarray}%
then the functional $F[\psi (x)]$ can always be determined from the $\alpha
_{k}$, so is the same as the function $f(\alpha _{1},\alpha _{2},..,\alpha
_{k},..)$ obtained by replacing $\psi (x)$ by its mode expansion in the
process that determines the functional.%
\begin{equation}
F[\psi (x)]=f(\alpha _{1},\alpha _{2},..,\alpha _{k},..)
\label{Eq.FnalFnReln}
\end{equation}

The \emph{orthonormality} and \emph{completeness} relationships are 
\begin{eqnarray}
\int dx\phi _{k}^{\ast }(x,t\mathbf{)}\phi _{l}(x,t\mathbf{)} &=&\delta _{kl}
\label{Eq.Orthonorm2} \\
\sum_{k}\phi _{k}(x,t\mathbf{)}\phi _{k}^{\ast }(y,t\mathbf{)} &\mathbf{=}%
&\delta (x-y)  \label{Eq.Completeness2}
\end{eqnarray}

\subsection{Functional Differentiation}

A summary of \emph{functional differentiation} is as follows. For a
functional $F[\psi (x)]$ of a field $\psi (x)$ the functional derivative $%
\frac{{\LARGE \delta F[\psi (x)]}}{{\LARGE \delta \psi (x)}}$ is defined by 
\begin{equation}
F[\psi (x)+\delta \psi (x)]\doteqdot F[\psi (x)]+\dint dx\,\delta \psi
(x)\,\left( \frac{\delta F[\psi (x)]}{\delta \psi (x)}\right) _{x}
\label{Eq.FnalDeriv}
\end{equation}%
where $\delta \psi (x)$ is \emph{any} small but arbitrary change in $\psi (x)
$. In this equation the left side is a functional of $\psi (x)+\delta \psi
(x)$ and the first term on the right side is a functional of $\psi (x)$. The
second term on the right side is a functional of $\delta \psi (x)$ and thus
the functional derivative must be a function of $x$; hence the subscript $x$%
. In most situations this subscript will be left understood. If we write $%
\delta \psi (x)=\epsilon \delta (x-y)$ for small $\epsilon $ then an
equivalent result for the functional derivative at $x=y$ is%
\begin{equation}
\left( \frac{\delta F[\psi (x)]}{\delta \psi (x)}\right)
_{x=y}=\lim_{\epsilon \rightarrow 0}\left( \frac{F[\psi (x)+\epsilon \delta
(x-y)]-F[\psi (x)]}{\epsilon }\right) .  \label{Eq.FnalDeriv2}
\end{equation}%
Note that for a fixed function $\xi (x)$ 
\begin{equation}
\frac{\delta \xi (x)}{\delta \psi (x)}=0  \label{Eq.FnalDerivFixedFn}
\end{equation}%
since the fixed function does not change when $\psi (x)$ does. On the other
hand 
\begin{equation}
\left( \frac{\delta \psi (y)}{\delta \psi (x)}\right) _{x}=\left( \frac{%
\delta F_{y}[\psi (x)]}{\delta \psi (x)}\right) _{x}=\delta (y-x)
\label{Eq.FnalDerivFldFn}
\end{equation}

Functional differentiation satisfies many of the rules of ordinary
differentiation including a \emph{product rule}

\begin{equation}
\left( \frac{\delta \{F[\psi (x)]G[\psi (x)]\}}{\delta \psi (x)}\right)
=\left( \frac{\delta F[\psi (x)]}{\delta \psi (x)}\right) G[\psi (x)]+F[\psi
(x)]\left( \frac{\delta G[\psi (x)]}{\delta \psi (x)}\right)
\label{Eq.ProdRule}
\end{equation}%
Note that for a fixed function $\xi (x)$ we have from Eqs.(\ref{Eq.ProdRule}%
) and (\ref{Eq.FnalDerivFixedFn}) 
\begin{equation}
\frac{\delta \{\xi (x)F[\psi (x)]\}}{\delta \psi (x)}=\xi (x)\left( \frac{%
\delta F[\psi (x)]}{\delta \psi (x)}\right)  \label{Eq.ProdRule2}
\end{equation}%
so a fixed function may be moved through the functional differentiation.

Mode expansions can be used to \emph{relate} functional differentiation and
ordinary differentiation.%
\begin{eqnarray}
\left( \frac{\delta F[\psi (x)]}{\delta \psi (x)}\right) _{x}
&=&\sum\limits_{k}\phi _{k}^{\ast }(x)\,\frac{\partial f(\alpha _{1},\alpha
_{2},..,\alpha _{k},..)}{\partial \alpha _{k}}  \notag \\
\frac{\partial f(\alpha _{1},\alpha _{2},..,\alpha _{k},..)}{\partial \alpha
_{k}} &=&\int dx\,\phi _{k}(x)\left( \frac{\delta F[\psi (x)]}{\delta \psi
(x)}\right) _{x}  \label{Eq.FnalDiffnOrdDiffn}
\end{eqnarray}%
Note that the first of these simple results are obtained by considering in (%
\ref{Eq.FnalDeriv}) the change in $\delta \psi (x)=\dsum\limits_{k}\delta
\alpha _{k}\,\phi _{k}(x)$ given by changes in the $\alpha _{k}$, and then
equating $F[\psi (x)+\delta \psi (x)]-F[\psi (x)]$ to the corresponding
result $f(\alpha _{1}+\delta \alpha _{1},\alpha _{2}+\delta \alpha
_{2},..,\alpha _{k}+\delta \alpha _{k},..)-f(\alpha _{1},\alpha
_{2},..,\alpha _{k},..)$. The second is then obtained via the orthogonality
result (\ref{Eq.Orthonorm2}).

A \emph{Taylor series expansion} to higher orders can also be obtained. Thus
to second order%
\begin{equation}
F[\psi (x)+\delta \psi (x)]\doteqdot F[\psi (x)]+\dint dx\,\delta \psi
(x)\,\left( \frac{\delta F[\psi (x)]}{\delta \psi (x)}\right) _{x}+\frac{1}{2%
}\dint \dint dxdy\,\delta \psi (x)\delta \psi (y)\,\left( \frac{\delta
^{2}F[\psi (x)]}{\delta \psi (x)\delta \psi (y)}\right) _{x,y}
\label{Eq.Taylor}
\end{equation}%
and can be obtained via a Taylor expansion of $f(\alpha _{1},\alpha
_{2},..,\alpha _{k},..)$ plus the relationships in (\ref%
{Eq.FnalDiffnOrdDiffn}).

\subsection{Functional Integration}

A brief summary of functional integration is as follows. If there are $n$
modes then the range for each function $\psi (x)$ is divided up into $n$
small intervals $\Delta x_{i}=x_{i+1}-x_{i}$ (the $i$th interval, where $%
\epsilon >|\Delta x_{i}|$), then we may specify the \emph{value} $\psi _{i}$
of the function $\psi (x)$ in the $i$th interval via the \emph{average}%
\begin{equation}
\psi _{i}=\frac{1}{\Delta x_{i}}\dint\limits_{\Delta x_{i}}dx\,\psi (x)
\label{Eq.AverageField}
\end{equation}%
and then any functional $F[\psi (x)]$ may be regarded as a \emph{function} $%
F(\psi _{1},\psi _{2},..,\psi _{i},..,\psi _{n})$ of all the $\psi _{i}$,
and ordinary integration over the $\psi _{i}$ is used to define the
functional integral. If each function $\psi (x)=\psi _{x}(x)+i\psi _{y}(x)$%
.is written in terms of its real and imaginary parts, then the functional
integration becomes an ordinary integration over the values $\psi _{ix}$, $%
\psi _{iy}$ of these components in each interval $i$ of the function $F(\psi
_{1},\psi _{2},..,\psi _{i},..,\psi _{n})$ multiplied by a suitably chosen
weight function $w(\psi _{1},\psi _{2},..,\psi _{i},..,\psi _{n})$. Thus the
functional integral is defined by 
\begin{eqnarray}
\dint D^{2}\psi \,F[\psi (x)] &=&\lim_{n\rightarrow \infty }\lim_{\epsilon
\rightarrow 0}\didotsint d^{2}\psi _{1}d^{2}\psi _{2}..d^{2}\psi
_{i}..d^{2}\psi _{n}\,w(\psi _{1},\psi _{2},..,\psi _{i},..,\psi _{n})\, 
\notag \\
&&\times F(\psi _{1},\psi _{2},..,\psi _{i},..,\psi _{n})
\label{Eq.FnalIntegral}
\end{eqnarray}%
where the number of modes is increased to infinity along with the space
interval decreasing to zero. The symbol $D^{2}\psi $ stands for $d^{2}\psi
_{1}d^{2}\psi _{2}..d^{2}\psi _{i}..d^{2}\psi _{n}\,w(\psi _{1},\psi
_{2},..,\psi _{i},..,\psi _{n})$, where the quantity $d^{2}\psi _{i}$ means $%
d\psi _{ix}d\psi _{iy}$.

A useful \emph{integration by parts }rule can often be established from Eq.(%
\ref{Eq.ProdRule}). Consider the functional $H[\psi (x)]=F[\psi (x)]G[\psi
(x)]$. Then%
\begin{equation}
\dint D^{2}\psi \,F[\psi (x)]\left( \frac{\delta G[\psi (x)]}{\delta \psi (x)%
}\right) =\dint D^{2}\psi \,\left( \frac{\delta H[\psi (x)]}{\delta \psi (x)}%
\right) -\dint D^{2}\psi \,\left( \frac{\delta F[\psi (x)]}{\delta \psi (x)}%
\right) G[\psi (x)]  \label{Eq.IntegParts}
\end{equation}

Functional integration and \emph{phase space integration} are inter-related
via the mode expansion, so we can relate the value $\psi _{i}$ of the
function in the $i$th interval to that of the mode function $\phi _{ki}$ via%
\begin{equation}
\psi _{i}=\dsum\limits_{k}\alpha _{k}\phi _{ki}  \label{Eq.RelnModeField}
\end{equation}%
Choosing the number of intervals to coincide with the number of modes the
integration $\didotsint d^{2}\psi _{1}d^{2}\psi _{2}..d^{2}\psi
_{i}..d^{2}\psi _{n}$ can be changed to an integration $\didotsint
d^{2}\alpha _{1}d^{2}\alpha _{2}..d^{2}\alpha _{k}..d^{2}\alpha _{n}$ over
the expansion coefficients $\alpha _{1},\alpha _{2},..,\alpha _{k},..\alpha
_{n}$. We have 
\begin{eqnarray}
\dint D^{2}\psi \,F[\psi (x)] &=&\lim_{n\rightarrow \infty }\lim_{\epsilon
\rightarrow 0}\didotsint d^{2}\alpha _{1}d^{2}\alpha _{2}..d^{2}\alpha
_{k}..d^{2}\alpha _{n}\,||J(\alpha _{1},\alpha _{2},..,\alpha _{k},..\alpha
_{n})||  \notag \\
&&\times v(\alpha _{1},\alpha _{2},..,\alpha _{k},..\alpha _{n})\,f(\alpha
_{1},\alpha _{2},..,\alpha _{k},..\alpha _{n})
\label{Eq.PhaseSpaceIntegral1}
\end{eqnarray}%
where $v(\alpha _{1},\alpha _{2},..,\alpha _{k},..\alpha _{n})\,$is the
function that equals $w(\psi _{1},\psi _{2},..,\psi _{i},..,\psi _{n})$ and
the Jacobian is given by%
\begin{equation}
||J(\alpha _{1},\alpha _{2},..,\alpha _{k},..\alpha _{n})||=\left\Vert 
\begin{array}{c}
\begin{tabular}{cccc}
$\frac{{\LARGE \partial \psi }_{1x}}{{\LARGE \partial \alpha }_{1x}}$ & $%
\frac{{\LARGE \partial \psi }_{1x}}{{\LARGE \partial \alpha }_{2x}}$ & $...$
& $\frac{{\LARGE \partial \psi }_{1x}}{{\LARGE \partial \alpha }_{nx}}$ \\ 
$\frac{{\LARGE \partial \psi }_{2x}}{{\LARGE \partial \alpha }_{1x}}$ & $%
\frac{{\LARGE \partial \psi }_{2x}}{{\LARGE \partial \alpha }_{2x}}$ & $...$
& $\frac{{\LARGE \partial \psi }_{2x}}{{\LARGE \partial \alpha }_{nx}}$ \\ 
$...$ & $...$ & $...$ & $...$ \\ 
$\frac{{\LARGE \partial \psi }_{nx}}{{\LARGE \partial \alpha }_{1x}}$ & $%
\frac{{\LARGE \partial \psi }_{nx}}{{\LARGE \partial \alpha }_{2x}}$ &  & $%
\frac{{\LARGE \partial \psi }_{nx}}{{\LARGE \partial \alpha }_{nx}}$%
\end{tabular}%
\begin{tabular}{cccc}
$\frac{{\LARGE \partial \psi }_{1x}}{{\LARGE \partial \alpha }_{1y}}$ & $%
\frac{{\LARGE \partial \psi }_{1x}}{{\LARGE \partial \alpha }_{2y}}$ & $...$
& $\frac{{\LARGE \partial \psi }_{1x}}{{\LARGE \partial \alpha }_{ny}}$ \\ 
$\frac{{\LARGE \partial \psi }_{2x}}{{\LARGE \partial \alpha }_{1y}}$ & $%
\frac{{\LARGE \partial \psi }_{2x}}{{\LARGE \partial \alpha }_{2y}}$ & $...$
& $\frac{{\LARGE \partial \psi }_{2x}}{{\LARGE \partial \alpha }_{ny}}$ \\ 
$...$ & $...$ & $...$ & $...$ \\ 
$\frac{{\LARGE \partial \psi }_{nx}}{{\LARGE \partial \alpha }_{1y}}$ & $%
\frac{{\LARGE \partial \psi }_{nx}}{{\LARGE \partial \alpha }_{2y}}$ &  & $%
\frac{{\LARGE \partial \psi }_{nx}}{{\LARGE \partial \alpha }_{ny}}$%
\end{tabular}
\\ 
\begin{tabular}{cccc}
$\frac{{\LARGE \partial \psi }_{1y}}{{\LARGE \partial \alpha }_{1x}}$ & $%
\frac{{\LARGE \partial \psi }_{1y}}{{\LARGE \partial \alpha }_{2x}}$ & $...$
& $\frac{{\LARGE \partial \psi }_{1y}}{{\LARGE \partial \alpha }_{nx}}$ \\ 
$\frac{{\LARGE \partial \psi }_{2y}}{{\LARGE \partial \alpha }_{1x}}$ & $%
\frac{{\LARGE \partial \psi }_{2y}}{{\LARGE \partial \alpha }_{2x}}$ & $...$
& $\frac{{\LARGE \partial \psi }_{2y}}{{\LARGE \partial \alpha }_{nx}}$ \\ 
$...$ & $...$ & $...$ & $...$ \\ 
$\frac{{\LARGE \partial \psi }_{ny}}{{\LARGE \partial \alpha }_{1x}}$ & $%
\frac{{\LARGE \partial \psi }_{ny}}{{\LARGE \partial \alpha }_{2x}}$ &  & $%
\frac{{\LARGE \partial \psi }_{ny}}{{\LARGE \partial \alpha }_{nx}}$%
\end{tabular}%
\begin{tabular}{cccc}
$\frac{{\LARGE \partial \psi }_{1y}}{{\LARGE \partial \alpha }_{1y}}$ & $%
\frac{{\LARGE \partial \psi }_{1y}}{{\LARGE \partial \alpha }_{2y}}$ & $...$
& $\frac{{\LARGE \partial \psi }_{1y}}{{\LARGE \partial \alpha }_{ny}}$ \\ 
$\frac{{\LARGE \partial \psi }_{2y}}{{\LARGE \partial \alpha }_{1y}}$ & $%
\frac{{\LARGE \partial \psi }_{2y}}{{\LARGE \partial \alpha }_{2y}}$ & $...$
& $\frac{{\LARGE \partial \psi }_{2y}}{{\LARGE \partial \alpha }_{ny}}$ \\ 
$...$ & $...$ & $...$ & $...$ \\ 
$\frac{{\LARGE \partial \psi }_{ny}}{{\LARGE \partial \alpha }_{1y}}$ & $%
\frac{{\LARGE \partial \psi }_{ny}}{{\LARGE \partial \alpha }_{2y}}$ &  & $%
\frac{{\LARGE \partial \psi }_{ny}}{{\LARGE \partial \alpha }_{ny}}$%
\end{tabular}%
\end{array}%
\right\Vert   \label{Eq.Jacobian}
\end{equation}%
Now using Eq.(\ref{Eq.RelnModeField}) 
\begin{eqnarray}
\frac{\partial \psi _{ix}}{\partial \alpha _{kx}} &=&\phi _{kix}\qquad \frac{%
\partial \psi _{ix}}{\partial \alpha _{ky}}=-\phi _{kiy}  \notag \\
\frac{\partial \psi _{iy}}{\partial \alpha _{kx}} &=&\phi _{kiy}\qquad \frac{%
\partial \psi _{iy}}{\partial \alpha _{ky}}=\phi _{kix}
\label{Eq.JacobianMatrix}
\end{eqnarray}%
and evaluating the Jacobian after showing that $(JJ^{T})_{ia\,kb}=\delta
_{ik}\delta _{ab}/\Delta x_{i}$ using the completeness relationship in Eq.(%
\ref{Eq.Completeness2}) we find that 
\begin{equation}
||J(\alpha _{1},\alpha _{2},..,\alpha _{k},..\alpha _{n})||=\dprod\limits_{i}%
\frac{1}{(\Delta x_{i})}  \label{Eq.Jacobian2}
\end{equation}%
and thus 
\begin{eqnarray}
\dint D^{2}\psi \,F[\psi (x)] &=&\lim_{n\rightarrow \infty }\lim_{\epsilon
\rightarrow 0}\didotsint d^{2}\alpha _{1}d^{2}\alpha _{2}..d^{2}\alpha
_{k}..d^{2}\alpha _{n}\,\dprod\limits_{i}\frac{1}{(\Delta x_{i})}  \notag \\
&&\times v(\alpha _{1},\alpha _{2},..,\alpha _{k},..\alpha _{n})\,f(\alpha
_{1},\alpha _{2},..,\alpha _{k},..\alpha _{n})
\label{Eq.PhaseSpaceIntegral2}
\end{eqnarray}%
This key result expresses the original functional integral as a phase space
integral over the expansion coefficients $\alpha _{k}$ of the function $\psi
(x)$ in terms of the mode functions $\phi _{k}(x)$.

The general result can be simplified with a \emph{special choice} of the
weight function%
\begin{equation}
w(\psi _{1},\psi _{2},..,\psi _{i},..,\psi _{n})=\dprod\limits_{i}(\Delta
x_{i})  \label{Eq.SimpleWeightFn}
\end{equation}%
and we then get a simple expression for the functional integral 
\begin{equation}
\dint D^{2}\psi \,F[\psi (x)]=\lim_{n\rightarrow \infty }\lim_{\epsilon
\rightarrow 0}\didotsint d^{2}\alpha _{1}d^{2}\alpha _{2}..d^{2}\alpha
_{k}..d^{2}\alpha _{n}\,\,f(\alpha _{1},\alpha _{2},..,\alpha _{k},..\alpha
_{n})  \label{Eq.PhaseSpaceIntegral3}
\end{equation}%
In this form of the functional integral the original functional $F[\psi (x)]$
has been replaced by the equivalent function $f(\alpha _{1},\alpha
_{2},..,\alpha _{k},..\alpha _{n})$ of the expansion coefficients $\alpha
_{k}$, and the functional integration is now replaced by a phase space
integration over the expansion coefficients.\pagebreak

\section{Appendix - Equivalence of Separate Mode and Field Theory Treatments}

\label{Appendix Equivalence of Separate Mode and Field Theory}

\subsection{Fokker-Planck and Functional Fokker-Planck Equations}

The functional Fokker-Planck equation for the distribution functional $P[%
\underrightarrow{\psi },\underrightarrow{\psi }^{\ast }]\,$must be
equivalent to the ordinary Fokker-Planck equation Eq. (\ref%
{Eq.NewFokker-Planck}) for the distribution function $P(\alpha ,\alpha
^{+},\alpha ^{\ast },\alpha ^{+\ast })$. The drift and diffusion terms can
be inter-related because functional and ordinary differentiation are
related, as explained in Appendix \ref{Appendix - Functional Calculus}. We
use%
\begin{eqnarray}
\left( \frac{\delta }{\delta \psi _{A}^{\mu }(x,t)}\right) _{x}
&=&\sum\limits_{k}\phi _{Ak}^{-\mu }(x,t)\,\frac{\partial }{\partial \alpha
_{A\mu k}(t)}  \notag \\
\frac{\partial }{\partial \alpha _{A\mu k}(t)} &=&\int dx\,\phi _{Ak}^{\mu
}(x,t)\left( \frac{\delta }{\delta \psi _{A}^{\mu }(x,t)}\right) _{x}
\label{Eq.FnalDiffnOrdDiffn3}
\end{eqnarray}%
in the functional Fokker-Planck equation (\ref%
{Eq.FnalFokkerPlanckHybridDistnFnal2}) which becomes 
\begin{eqnarray}
&&\frac{\partial }{\partial t}P[\underrightarrow{\psi },\underrightarrow{%
\psi }^{\ast }]\,  \notag \\
&=&\frac{\partial }{\partial t}P(\alpha ,\alpha ^{+},\alpha ^{\ast },\alpha
^{+\ast })  \notag \\
&=&\left\{ -\dsum\limits_{\mu A}\dint dx\sum\limits_{k}\phi _{Ak}^{-\mu
}(x,t)\,\frac{\partial }{\partial \alpha _{\mu Ak}}A_{A}^{\mu }(x)\right\}
P(\alpha ,\alpha ^{+},\alpha ^{\ast },\alpha ^{+\ast })  \notag \\
&&+\left\{ \frac{1}{2}\dsum\limits_{\mu A}\dsum\limits_{\nu B}\dint \dint
dxdy\sum\limits_{k}\phi _{Ak}^{-\mu }(x,t)\,\frac{\partial }{\partial \alpha
_{\mu Ak}}\sum\limits_{l}\phi _{Bl}^{-\nu }(y,t)\,\frac{\partial }{\partial
\alpha _{\nu Bl}}E_{A\,B}^{\mu \,\nu }(x,y)\right\} P(\alpha ,\alpha
^{+},\alpha ^{\ast },\alpha ^{+\ast })  \notag \\
&=&\left\{ -\dsum\limits_{\mu Ak}\,\frac{\partial }{\partial \alpha _{\mu Ak}%
}A_{Ak}^{\mu }+\frac{1}{2}\dsum\limits_{\mu Ak}\dsum\limits_{\nu Bl}\,\frac{%
\partial }{\partial \alpha _{\mu Ak}}\,\frac{\partial }{\partial \alpha
_{\nu Bl}}E_{Ak\,Bl}^{\mu \,\nu }\right\} P(\alpha ,\alpha ^{+},\alpha
^{\ast },\alpha ^{+\ast })
\end{eqnarray}%
and gives the Fokker-Planck equation (\ref{Eq.NewFokker-Planck}) with 
\begin{eqnarray}
A_{Ak}^{\mu } &=&\dint dx\,\phi _{Ak}^{-\mu }(x,t)A_{A}^{\mu }(x)  \notag \\
A_{A}^{\mu }(x) &=&\dsum\limits_{k}\phi _{Ak}^{\mu }(x,t)A_{Ak}^{\mu } 
\notag \\
E_{Ak\,Bl}^{\mu \,\nu } &=&\dint \dint dxdy\,\phi _{Ak}^{-\mu
}(x,t)E_{A\,B}^{\mu \,\nu }(x,y)\phi _{Bl}^{-\nu }(y,t)  \notag \\
E_{A\,B}^{\mu \,\nu }(x,y) &=&\dsum\limits_{kl}\phi _{Ak}^{\mu
}(x,t)E_{Ak\,Bl}^{\mu \nu }\phi _{Bl}^{\nu }(y,t)
\label{Eq.DriftDiffnRelnsHybrid}
\end{eqnarray}%
which is the same as Eqs. (\ref{Eq.NewFokker-Planck}). Note that in
particular 
\begin{eqnarray}
&&\dint \dint dxdy\,\phi _{Ak}^{-\mu }(x,t)\;\frac{1}{2}\left\{ \delta
_{A\,C}\delta _{B\,NC}\,\delta _{\mu \,-\nu }\left(
\sum_{m}\dsum\limits_{n}\phi _{Am}^{\mu }(x,t\mathbf{)}C_{Am\,Bn}^{\mu
}(t)\phi _{Bn}^{\nu }(y,t\mathbf{)}\right) \right\} \phi _{Bl}^{-\nu }(y,t) 
\notag \\
&&+\dint \dint dxdy\,\phi _{Ak}^{-\mu }(x,t)\;\frac{1}{2}\left\{ \delta
_{B\,C}\delta _{A\,NC}\,\delta _{\nu \,-\mu }\left(
\sum_{m}\dsum\limits_{n}\phi _{Bn}^{\nu }(y,t\mathbf{)}C_{Bn\,Am}^{\nu
}(t)\phi _{Am}^{\mu }(x,t\mathbf{)}\right) \right\} \phi _{Bl}^{-\nu }(y,t) 
\notag \\
&=&\frac{1}{2}\left\{ \delta _{A\,C}\delta _{B\,NC}\,\delta _{\mu \,-\nu
}C_{Ak\,Bl}^{\mu }+\delta _{B\,C}\delta _{A\,NC}\,\delta _{\nu \,-\mu
}C_{Bl\,Ak}^{\nu }\right\}
\end{eqnarray}%
as in Eq. (\ref{Eq.NewDiffusionMatrix}). Thus the functional and ordinary
Fokker-Planck equations for the hybrid case are \emph{consistent} with each
other.

\subsection{Ito Stochastic Equations for Fields and Phase Variables}

To see how the stochastic fields $\psi _{A}^{\mu s}(x,t\mathbf{)}$ are
related to the stochastic phases $\alpha _{\mu Ak}^{s}$, we find that the
Ito stochastic field equation (\ref{Eq.ItoSDEFieldsHyb}) becomes on
substituting for $\mathcal{G}^{\mu }(x)$ and $\mathcal{N}_{a}^{\mu }(x)$
from (\ref{Eq.HybRelnFFPEStochFldDE}), (\ref{Eq.GeeHyb}) and (\ref%
{Eq.NetaHyb})%
\begin{eqnarray}
\delta \psi _{A}^{\mu s}(x,t\mathbf{)} &\mathbf{=}&\left(
\dsum\limits_{k}\phi _{Ak}^{\mu }(x,t)A_{Ak}^{\mu }+\dsum\limits_{k}\phi
_{Ak}^{\mu }(x,t)\,\dsum\limits_{Bl}C_{Ak\,Bl}^{\mu }\,\alpha _{\mu
Bl}^{s}+\dsum\limits_{k}\frac{\partial }{\partial t}\phi _{Ak}^{\mu
}(x,t)\,\alpha _{\mu Ak}^{s}\right) \mathcal{\,}\delta t  \notag \\
&&+\dsum\limits_{ka}K_{Aka}^{\mu }\phi _{Ak}^{\mu }(x,t\mathbf{)}%
\int_{t}^{t+\delta t}dt_{1}\Gamma _{a}(t_{1})  \notag \\
&=&\dsum\limits_{k}\phi _{Ak}^{\mu }(x,t)\left\{ A_{Ak}^{\mu }\mathcal{\,}%
\delta t+\dsum\limits_{Bl}C_{Ak\,Bl}^{\mu }\,\alpha _{\mu Bl}^{s}\mathcal{\,}%
\delta t+\dsum\limits_{a}K_{Aka}^{\mu }\int_{t}^{t+\delta t}dt_{1}\Gamma
_{a}(t_{1})\right\}   \notag \\
&&+\dsum\limits_{k}\frac{\partial }{\partial t}\phi _{Ak}^{\mu
}(x,t)\,\alpha _{\mu Ak}^{s}\mathcal{\,}\delta t  \notag \\
&=&\dsum\limits_{k}\phi _{Ak}^{\mu }(x,t)\,\delta \alpha _{\mu
Ak}^{s}+\dsum\limits_{k}\frac{\partial }{\partial t}\phi _{Ak}^{\mu
}(x,t)\,\alpha _{\mu Ak}^{s}\mathcal{\,}\delta t
\end{eqnarray}%
on substituting for $\delta \alpha _{\mu Ak}^{s}$ from Eq.(\ref%
{Eq.ItoSDEFinalHyb}) 
\begin{equation}
\frac{\partial }{\partial t}\alpha _{\mu Ak}^{s}=A_{Ak}^{\mu
}+\sum_{l}C_{Ak\,Bl}^{\mu }\,\alpha _{B\mu l}^{s}+\sum_{a}K_{Aka}^{\mu
}\,\Gamma _{a}(t_{+})
\end{equation}%
and substituting from (\ref{Eq.TimeDerivModes2}) for the time derivative of
the mode function. Hence we see that the Ito stochastic equation for $\psi
_{A}^{\mu s}(x,t\mathbf{)}$ is 
\begin{equation}
\frac{\partial }{\partial t}\psi _{A}^{\mu s}(x,t\mathbf{)=}\dsum\limits_{k}%
\frac{\partial }{\partial t}\alpha _{\mu Ak}^{s}(t\mathbf{)\,}\phi
_{Ak}^{\mu }(x,t)+\dsum\limits_{k}\alpha _{\mu Ak}^{s}\mathcal{\,}\frac{%
\partial }{\partial t}\phi _{Ak}^{\mu }(x,t)  \label{Eq.ItoSDEBose5Hyb}
\end{equation}%
This shows that the \emph{stochastic fields} are given by 
\begin{equation}
\psi _{A}^{\mu s}(x,t\mathbf{)=}\sum_{k}\alpha _{\mu Ak}^{s}(t)\phi
_{Ak}^{\mu }(x,t\mathbf{)}  \label{Eq.StochasticFldHyb}
\end{equation}%
where the stochastic feature results only from the phase variables being
replaced by stochastic variables. These equations show that the stochastic
field equation for $\psi _{\mu }^{s}(x,t\mathbf{)}$ is equivalent to the
stochastic phase variable equation for the $\alpha _{\mu k}^{s}(t)$%
.\pagebreak 

\section{Appendix - Derivation of Langevin Equations}

\label{Appendix - Derivation of Langevin Equations}

Details in carrying out the stochastic averaging for the derivation of the
Langevin equations in Section \ref{Section 2 Separate Modes Case} are set
out here.

For the \textit{first order derivative} terms 
\begin{eqnarray}
&&\overline{\left\{ \sum_{\mu Ak}\left[ \frac{\partial }{\partial \alpha
_{\mu Ak}}F(\alpha ,\alpha ^{+})\right] \delta \alpha _{\mu
Ak}^{s}(t)\right\} }  \notag \\
&=&\overline{\sum_{\mu Ak}\left[ \frac{\partial }{\partial \alpha _{\mu Ak}}%
F(\alpha ,\alpha ^{+})\right] \mathcal{A}_{Ak}^{\mu }(\alpha _{\xi
Dm}^{s}(t))}\,\delta t  \notag \\
&&+\overline{\sum_{\mu Ak}\left[ \frac{\partial }{\partial \alpha _{\mu Ak}}%
F(\alpha ,\alpha ^{+})\right] \sum_{a}\mathcal{B}_{Aka}^{\mu }(\alpha _{\xi
Dm}^{s}(t))}\;\overline{\int_{t}^{t+\delta t}dt_{1}\Gamma _{a}(t_{1})} 
\notag \\
&=&\overline{\sum_{\mu Ak}\left[ \frac{\partial }{\partial \alpha _{\mu Ak}}%
F(\alpha ,\alpha ^{+})\right] \mathcal{A}_{Ak}^{\mu }(\alpha _{\xi
Dm}^{s}(t))}\,\delta t
\end{eqnarray}%
where the stochastic average rules for sums and products have been used, the
non-correlation between the averages of functions of $\alpha _{\mu Ak}^{s}(t)
$ at time $t$ and the $\Gamma $ at later times between $t$ to $t+\delta t$
is applied and the term involving $\overline{\Gamma _{a}(t_{1})}$ is equal
to zero from (\ref{Eq.GaussianMarkov}).

For the \textit{second order derivative} terms, we have on expanding the
product $\delta \alpha _{\mu Ak}^{s}(t)\delta \alpha _{\nu Bl}^{s}(t)$ and
using the stochastic average of a sum being the same as the sum of
stochastic averages 
\begin{eqnarray}
&&\overline{\left\{ \frac{1}{2}\sum_{\mu Ak}\sum_{\nu Bl}\left[ \frac{%
\partial }{\partial \alpha _{\mu Ak}}\frac{\partial }{\partial \alpha _{\nu
Bl}}F(\alpha ,\alpha ^{+})\right] \delta \alpha _{\mu Ak}^{s}(t)\delta
\alpha _{\nu Bl}^{s}(t)\right\} }  \notag \\
&=&\overline{%
\begin{array}{c}
\frac{1}{2}\sum_{\mu Ak}\sum_{\nu Bl}\left[ \frac{{\LARGE \partial }}{%
{\LARGE \partial \alpha }_{\mu Ak}}\frac{{\LARGE \partial }}{{\LARGE %
\partial \alpha }_{\nu Bl}}F(\alpha ,\alpha ^{+})\right] \\ 
\times \left[ \mathcal{A}_{Ak}^{\mu }(\alpha _{\xi Dm}^{s}(t))\delta t\,%
\mathcal{A}_{Bl}^{\nu }(\alpha _{\xi Dm}^{s}(t))\delta t\right]%
\end{array}%
}  \notag \\
&&+\overline{%
\begin{array}{c}
\frac{1}{2}\sum_{\mu Ak}\sum_{\nu Bl}\left[ \frac{{\LARGE \partial }}{%
{\LARGE \partial \alpha }_{\mu Ak}}\frac{{\LARGE \partial }}{{\LARGE %
\partial \alpha }_{\nu Bl}}F(\alpha ,\alpha ^{+})\right] \\ 
\times \left[ \mathcal{A}_{Ak}^{\mu }(\alpha _{\xi Dm}^{s}(t))\delta
t\,\sum_{b}\mathcal{B}_{Blb}^{\nu }(\alpha _{\xi
Dm}^{s}(t))\int_{t}^{t+\delta t}dt_{2}\Gamma _{b}(t_{2})\right]%
\end{array}%
}  \notag \\
&&+\overline{%
\begin{array}{c}
\frac{1}{2}\sum_{\mu Ak}\sum_{\nu Bl}\left[ \frac{{\LARGE \partial }}{%
{\LARGE \partial \alpha }_{\mu Ak}}\frac{{\LARGE \partial }}{{\LARGE %
\partial \alpha }_{\nu Bl}}F(\alpha ,\alpha ^{+})\right] \\ 
\times \left[ \sum_{a}\mathcal{B}_{Aka}^{\mu }(\alpha _{\xi
Dm}^{s}(t))\int_{t}^{t+\delta t}dt_{1}\Gamma _{a}(t_{1})\;\mathcal{A}%
_{Bl}^{\nu }(\alpha _{\xi Dm}^{s}(t))\delta t\right]%
\end{array}%
}  \notag \\
&&+\overline{%
\begin{array}{c}
\frac{1}{2}\sum_{\mu Ak}\sum_{\nu Bl}\left[ \frac{{\LARGE \partial }}{%
{\LARGE \partial \alpha }_{\mu Ak}}\frac{{\LARGE \partial }}{{\LARGE %
\partial \alpha }_{\nu Bl}}F(\alpha ,\alpha ^{+})\right] \\ 
\times \left[ \sum_{a}\mathcal{B}_{Aka}^{\mu }(\alpha _{\xi
Dm}^{s}(t))\int_{t}^{t+\delta t}dt_{1}\Gamma _{a}(t_{1})\;\sum_{b}\mathcal{B}%
_{Blb}^{\nu }(\alpha _{\xi Dm}^{s}(t))\int_{t}^{t+\delta t}dt_{2}\Gamma
_{b}(t_{2})\right]%
\end{array}%
}  \notag \\
&&
\end{eqnarray}%
Using the result that the stochastic averages for the functions of the $%
\alpha _{\xi Dm}^{s}(t)$ and the $\Gamma _{a}(t_{+})$ are uncorrelated we
find that 
\begin{eqnarray}
&&\overline{\left\{ \frac{1}{2}\sum_{\mu Ak}\sum_{\nu Bl}\left[ \frac{%
\partial }{\partial \alpha _{\mu Ak}}\frac{\partial }{\partial \alpha _{\nu
Bl}}F(\alpha ,\alpha ^{+})\right] \delta \alpha _{\mu Ak}^{s}(t)\delta
\alpha _{\nu Bl}^{s}(t)\right\} }  \notag \\
&=&\overline{\frac{1}{2}\sum_{\mu Ak}\sum_{\nu Bl}\left[ \frac{\partial }{%
\partial \alpha _{\mu Ak}}\frac{\partial }{\partial \alpha _{\nu Bl}}%
F(\alpha ,\alpha ^{+})\right] \left[ \mathcal{A}_{Ak}^{\mu }(\alpha _{\xi
Dm}^{s}(t))\,\mathcal{A}_{Bl}^{\nu }(\alpha _{\xi Dm}^{s}(t))\right] }%
\;\delta t^{2}  \notag \\
&&+\overline{\frac{1}{2}\sum_{\mu Ak}\sum_{\nu Bl}\left[ \frac{\partial }{%
\partial \alpha _{\mu Ak}}\frac{\partial }{\partial \alpha _{\nu Bl}}%
F(\alpha ,\alpha ^{+})\right] \left[ \mathcal{A}_{Ak}^{\mu }(\alpha _{\xi
Dm}^{s}(t))\delta t\,\sum_{b}\mathcal{B}_{Blb}^{\nu }(\alpha _{\xi
Dm}^{s}(t))\right] }\;  \notag \\
&&\times \overline{\int_{t}^{t+\delta t}dt_{2}\Gamma _{b}(t_{2})}\;\delta t 
\notag \\
&&+\overline{\frac{1}{2}\sum_{\mu Ak}\sum_{\nu Bl}\left[ \frac{\partial }{%
\partial \alpha _{\mu Ak}}\frac{\partial }{\partial \alpha _{\nu Bl}}%
F(\alpha ,\alpha ^{+})\right] \left[ \sum_{a}\mathcal{B}_{Aka}^{\mu }(\alpha
_{\xi Dm}^{s}(t))\;\mathcal{A}_{Bl}^{\nu }(\alpha _{\xi Dm}^{s}(t))\right] }%
\;  \notag \\
&&\times \overline{\int_{t}^{t+\delta t}dt_{1}\Gamma _{a}(t_{1})}\;\delta t 
\notag \\
&&+\overline{\frac{1}{2}\sum_{\mu Ak}\sum_{\nu Bl}\left[ \frac{\partial }{%
\partial \alpha _{\mu Ak}}\frac{\partial }{\partial \alpha _{\nu Bl}}%
F(\alpha ,\alpha ^{+})\right] \left[ \sum_{a}\mathcal{B}_{Aka}^{\mu }(\alpha
_{\xi Dm}^{s}(t))\;\sum_{b}\mathcal{B}_{Blb}^{\nu }(\alpha _{\xi Dm}^{s}(t))%
\right] }\;  \notag \\
&&\times \overline{\int_{t}^{t+\delta t}dt_{1}\Gamma
_{a}(t_{1})\int_{t}^{t+\delta t}dt_{2}\Gamma _{b}(t_{2})}
\end{eqnarray}%
Now the terms involving a single $\Gamma $ have a zero stochastic average,
whilst the terms with two $\Gamma $ give a stochastic average proportional
to $\delta t$ 
\begin{eqnarray}
\overline{\int_{t}^{t+\delta t}dt_{1}\Gamma _{a}(t_{1})\int_{t}^{t+\delta
t}dt_{2}\Gamma _{b}(t_{2})} &=&\int_{t}^{t+\delta t}dt_{1}\int_{t}^{t+\delta
t}dt_{2}\;\overline{\Gamma _{a}(t_{1})\Gamma _{b}(t_{2})}  \notag \\
&=&\int_{t}^{t+\delta t}dt_{1}\int_{t}^{t+\delta t}dt_{2}\;\delta
_{ab}\delta (t_{1}-t_{2})  \notag \\
&=&\delta _{ab}\,\delta t
\end{eqnarray}%
so that correct to order $\delta t$ the second order derivative term is 
\begin{eqnarray}
&&\overline{\left\{ \frac{1}{2}\sum_{\mu Ak}\sum_{\nu Bl}\left[ \frac{%
\partial }{\partial \alpha _{\mu Ak}}\frac{\partial }{\partial \alpha _{\nu
Bl}}F(\alpha ,\alpha ^{+})\right] \delta \alpha _{\mu Ak}^{s}(t)\delta
\alpha _{\nu Bl}^{s}(t)\right\} }  \notag \\
&=&\overline{\frac{1}{2}\sum_{\mu Ak}\sum_{\nu Bl}\left[ \frac{\partial }{%
\partial \alpha _{\mu Ak}}\frac{\partial }{\partial \alpha _{\nu Bl}}%
F(\alpha ,\alpha ^{+})\right] \left[ \sum_{a}\mathcal{B}_{Aka}^{\mu }(\alpha
_{\xi Dm}^{s}(t))\;\mathcal{B}_{Bla}^{\nu }(\alpha _{\xi Dm}^{s}(t))\right] }%
\;\delta t  \notag \\
&=&\overline{\frac{1}{2}\sum_{\mu Ak}\sum_{\nu Bl}\left[ \frac{\partial }{%
\partial \alpha _{\mu Ak}}\frac{\partial }{\partial \alpha _{\nu Bl}}%
F(\alpha ,\alpha ^{+})\right] \left[ [\mathcal{B}(\alpha _{\xi Dm}^{s}(t))%
\mathcal{\ B}^{T}(\alpha _{\xi Dm}^{s}(t))]_{Ak,Bl}^{\mu ,\nu }\right] }%
\;\delta t  \notag \\
&&
\end{eqnarray}

The remaining terms give stochastic averages correct to order $\delta t^{2}$
or higher so that we have correct to first order in $\delta t$ 
\begin{eqnarray}
&&\overline{F(\alpha ^{s}(t+\delta t),\alpha ^{s+}(t+\delta t))}-\overline{%
F(\alpha ^{s}(t),\alpha ^{s+}(t))}  \notag \\
&=&\left\{ \overline{\sum_{\mu Ak}\left[ \frac{\partial }{\partial \alpha
_{\mu Ak}}F(\alpha ,\alpha ^{+})\right] \mathcal{A}_{Ak}^{\mu }(\alpha _{\xi
Dm}^{s}(t))}\,\right\} \;\delta t  \notag \\
&&+\left\{ \overline{\frac{1}{2}\sum_{\mu Ak}\sum_{\nu Bl}\left[ \frac{%
\partial }{\partial \alpha _{\mu Ak}}\frac{\partial }{\partial \alpha _{\nu
Bl}}F(\alpha ,\alpha ^{+})\right] \left[ [\mathcal{B}(\alpha _{\xi
Dm}^{s}(t))\mathcal{\ B}^{T}(\alpha _{\xi Dm}^{s}(t))]_{Ak,Bl}^{\mu ,\nu }%
\right] }\right\} \;\delta t  \notag \\
&&
\end{eqnarray}%
\bigskip \pagebreak

\section{Appendix - Derivation of Functional Fokker-Planck Equation}

\label{Appendix Derivation of Functional Fokker-Planck Equation}

This Appendix contains details in the derivation of the functional
Fokker-Planck equation, the final form of which is set out in Section \ref%
{Section 3 - Quantum Field Case}.

\subsection{Form of Characteristic Functional}

In deriving (\ref{Eq.AuxCharFnal}) we note that integrals involving cross
terms between condensate and non-condensate fields are zero via mode
orthogonality and hence 
\begin{eqnarray}
\dint dx\,\hat{\Psi}(x\mathbf{)\,}\Xi ^{+}(x\mathbf{)} &\mathbf{=}&\dint dx\,%
\hat{\Psi}_{C}(x,t\mathbf{)\,}\Xi _{C}^{+}(x,t\mathbf{)+}\dint dx\,\hat{\Psi}%
_{NC}(x,t\mathbf{)\,}\Xi _{NC}^{+}(x,t\mathbf{)}  \notag \\
&\mathbf{=}&\sum\limits_{k\epsilon C,NC}\hat{a}_{k}\xi _{k}^{+}  \notag \\
\dint dx\,\Xi (x\mathbf{)}\,\hat{\Psi}^{\dag }(x\mathbf{)} &\mathbf{=}&\dint
dx\,\Xi _{NC}(x,t\mathbf{)}\,\hat{\Psi}_{NC}^{\dag }(x,t\mathbf{)+\dint }%
dx\,\Xi _{NC}(x,t)\,\hat{\Psi}_{NC}^{\dag }(x,t)  \notag \\
&\mathbf{=}&\sum\limits_{k\epsilon C,NC}\xi _{k}\hat{a}_{k}^{\dag }\,  \notag
\\
\frac{1}{2}\dint dx\,\Xi _{C}(x\mathbf{)\,}\Xi _{C}^{+}(x\mathbf{))} &%
\mathbf{=}&\frac{1}{2}\sum\limits_{k\epsilon C}\xi _{k}\xi _{k}^{+}
\label{Eq.ModeSumConNonConIntegRelns}
\end{eqnarray}

In deriving (\ref{Eq.DistnFnal2}) we note that

\begin{eqnarray}
\dint dx\,(\psi _{C}(x,t\mathbf{)\,}\Xi _{C}^{+}(x,t\mathbf{)+}\Xi _{C}(x,t%
\mathbf{)}\,\psi _{C}^{+}(x,t\mathbf{))} &\mathbf{=}&\sum\limits_{k\epsilon
C}(\alpha _{k}\xi _{k}^{+}+\xi _{k}\alpha _{k}^{+}\,)  \notag \\
\dint dx\,\psi _{NC}(x,t\mathbf{)\,}\Xi _{NC}^{+}(x,t) &\mathbf{=}%
&\sum\limits_{k\epsilon NC}\alpha _{k}\xi _{k}^{+}\qquad \dint dx\,\Xi
_{NC}(x,t\mathbf{)\,}\psi _{NC}^{+}(x,t)\mathbf{=}\sum\limits_{k\epsilon
NC}\xi _{k}\alpha _{k}^{+}  \notag
\end{eqnarray}

\subsection{Time Derivative of Characteristic Functional}

Details for the derivation of (\ref{Eq.TimeDerivIntegCondCharFlds}) \ are as
follows.

The time derivative in Eq.(\ref{Eq.TimeDerivIntegCondCharFlds00}) can be
evaluated using Eqs. (\ref{Eq.ConNonCondCharFldFns}), (\ref%
{Eq.TimeDerivModes1}) and (\ref{Eq.TimeDepCharFnVar}) 
\begin{eqnarray}
&&-\frac{1}{2}\frac{\partial }{\partial t}\dint dx\,\Xi _{C}(x,t\mathbf{)\,}%
\Xi _{C}^{+}(x,t\mathbf{)}  \notag \\
&=&-\frac{1}{2}\dint dx\,(\frac{\partial }{\partial t}\Xi _{C}(x,t\mathbf{)\,%
}\Xi _{C}^{+}(x,t\mathbf{)+}\Xi _{C}(x,t\mathbf{)\,}\frac{\partial }{%
\partial t}\Xi _{C}^{+}(x,t\mathbf{))}  \notag \\
&=&-\frac{1}{2}\dint dx\,(\left\{ \sum_{k\epsilon C}\dsum\limits_{l\epsilon
NC}C_{kl}(t)\xi _{l}(t)\phi _{k}(x,t\mathbf{)+}\sum_{k\epsilon
C}\dsum\limits_{l\epsilon NC}\xi _{k}(t)C_{kl}^{\ast }\phi _{l}(x,t)\right\} 
\notag \\
&&\times \sum_{m\epsilon C}\xi _{_{m}}^{+}(t)\phi _{m}^{\ast }(x,t\mathbf{))}
\notag \\
&&-\frac{1}{2}\dint dx\,(\sum_{m\epsilon C}\xi _{m}(t)\phi _{m}(x,t\mathbf{%
)\,}  \notag \\
&&\times \left\{ \sum_{k\epsilon C}\dsum\limits_{l\epsilon NC}C_{kl}^{\ast
}(t)\xi _{l}^{+}(t)\phi _{k}^{\ast }(x,t\mathbf{)+}\sum_{k\epsilon
C}\dsum\limits_{l\epsilon NC}\xi _{_{k}}^{+}(t)C_{kl}\phi _{l}^{\ast
}(x,t)\right\} \mathbf{)}  \notag \\
&=&-\frac{1}{2}\left\{ \sum_{k\epsilon C}\dsum\limits_{l\epsilon
NC}C_{kl}(t)\xi _{l}(t)\xi _{_{k}}^{+}(t)\right\} -\frac{1}{2}\left\{
\sum_{k\epsilon C}\dsum\limits_{l\epsilon NC}C_{kl}^{\ast }(t)\xi _{k}(t)\xi
_{l}^{+}(t)\right\}   \label{Eq.TimeDerivIntegCondCharFlds0}
\end{eqnarray}%
where the $l\epsilon C$ terms have been eliminated using $%
C_{kl}(t)+C_{lk}^{\ast }(t)=0$ and orthogonality of the modes for $l\epsilon
NC$ and $m\epsilon C$ has been used to eliminate the second and fourth
contributions.

Since for $l\epsilon NC$ and $k\epsilon C$ we have from (\ref%
{Eq.ConNonCondCharFldFns}) 
\begin{eqnarray}
\xi _{l}(t) &=&\dint dx\,\phi _{l}^{\ast }(x,t\mathbf{)}\Xi _{NC}(x,t\mathbf{%
)\qquad }\xi _{l}^{+}(t)=\dint dx\,\phi _{l}(x,t\mathbf{)}\Xi _{NC}^{+}(x,t%
\mathbf{)}  \notag \\
\xi _{k}(t) &=&\dint dy\,\phi _{k}^{\ast }(y,t\mathbf{)}\Xi _{C}(y,t\mathbf{%
)\qquad }\xi _{k}^{+}(t)=\dint dy\,\phi _{k}(y,t\mathbf{)}\Xi _{C}^{+}(y,t%
\mathbf{)}  \label{Eq.CondNonCondCharVar}
\end{eqnarray}%
we see that 
\begin{eqnarray}
&&-\frac{1}{2}\frac{\partial }{\partial t}\dint dx\,\Xi _{C}(x,t\mathbf{)\,}%
\Xi _{C}^{+}(x,t\mathbf{)}  \notag \\
&=&\frac{1}{2}\left\{ \dint \int dx\,dy\,\left\{ \sum_{k\epsilon
C}\dsum\limits_{l\epsilon NC}\phi _{l}^{\ast }(x,t\mathbf{)}C_{kl}(t)\phi
_{k}(y,t\mathbf{)}\right\} \mathbf{(}i\Xi _{NC}(x,t\mathbf{))}(i\Xi
_{C}^{+}(y,t\mathbf{))}\right\}  \notag \\
&&+\frac{1}{2}\left\{ \dint \int dx\,dy\,\left\{ \sum_{l\epsilon
C}\dsum\limits_{k\epsilon NC}\phi _{l}^{\ast }(x,t\mathbf{)}C_{lk}^{\ast
}(t)\phi _{k}(y,t\mathbf{)}\right\} \,(i\Xi _{C}(x,t\mathbf{))}\,(i\Xi
_{NC}^{+}(y,t\mathbf{))}\right\}  \notag \\
&&
\end{eqnarray}

The exponential factor in Eq. (\ref{Eq.TimeDerivIntegCondCharFlds00}) can be
recombined via (\ref{Eq.CharFnReln2}) with the normally ordered
characteristic functional $Tr\left\{ \exp i\dint dx\,(\hat{\Psi}(x\mathbf{)\,%
}\Xi ^{+}(x\mathbf{))}\right\} \,\hat{\rho}\,\left\{ \exp i\dint dx(\Xi (x%
\mathbf{)}\,\hat{\Psi}^{\dag }(x\mathbf{)})\right\} $ to produce the
original characteristic functional $\chi \lbrack \Xi _{C},\Xi _{C}^{+},\Xi
_{NC},\Xi _{NC}^{+}]$. Writing the characteristic functional in the form in
Eq. (\ref{Eq.DistnFnal2}) and noting that multiplication by $\mathbf{(}i\Xi
_{NC}(x,t\mathbf{))}(i\Xi _{C}^{+}(y,t\mathbf{))}$ etc. can be replaced by
functional differentiations such as $\frac{{\LARGE \delta }}{{\LARGE \delta
\psi }_{NC}^{+}{\LARGE (x,t)}}\frac{{\LARGE \delta }}{{\LARGE \delta \psi }%
_{C}{\LARGE (y,t)}}$ we see that the second term in Eq. (\ref%
{Eq.TimeDerivCharFnalConNonCond}) is the sum of two contributions 
\begin{eqnarray}
&&\frac{1}{2}\left\{ \dint \int dx\,dy\,\left\{ \sum_{k\epsilon
C}\dsum\limits_{l\epsilon NC}\phi _{l}^{\ast }(x,t\mathbf{)}C_{kl}(t)\phi
_{k}(y,t\mathbf{)}\right\} \mathbf{(}i\Xi _{NC}(x,t\mathbf{))}(i\Xi
_{C}^{+}(y,t\mathbf{))}\right\}   \notag \\
&&\times \chi \lbrack \Xi _{C},\Xi _{C}^{+},\Xi _{NC},\Xi _{NC}^{+}]  \notag
\\
&=&\int D^{2}\underrightarrow{\psi }\;\frac{1}{2}\dint \int dx\,dy\,\left\{
\sum_{k\epsilon C}\dsum\limits_{l\epsilon NC}\phi _{l}^{\ast }(x,t\mathbf{)}%
C_{kl}(t)\phi _{k}(y,t\mathbf{)}\right\} \left\{ \frac{\delta }{\delta \psi
_{NC}^{+}(x,t)}\frac{\delta }{\delta \psi _{C}(y,t\mathbf{)}}\right\}  
\notag \\
&&\times \exp (i\dint dx\,(\psi _{C}(x,t\mathbf{)\,}\Xi _{C}^{+}(x,t\mathbf{%
)+}\psi _{NC}(x,t)\,\Xi _{NC}^{+}(x,t)\mathbf{+}\,\Xi _{C}(x,t\mathbf{)}%
\,\psi _{C}^{+}(x,t\mathbf{)+\,}\Xi _{NC}(x,t)\,\psi _{NC}^{+}(x,t)\mathbf{)}
\notag \\
&&\times P[\underrightarrow{\psi },\underrightarrow{\psi }^{\ast }]  \notag
\\
&=&\int D^{2}\underrightarrow{\psi }\;\exp (i\dint dx\,\sum_{\mu }\psi ^{\mu
}(x,t\mathbf{)\,}\Xi ^{-\mu }(x,t\mathbf{))}  \label{Eq.FirstCouplingTerm} \\
&&\times \dint \int dx\,dy\,\left\{ \frac{1}{2}\sum_{k\epsilon
C}\dsum\limits_{l\epsilon NC}\phi _{NCl}^{+}(x,t\mathbf{)}%
C_{Ck\,NCl}^{-}(t)\phi _{Ck}^{-}(y,t\mathbf{)}\frac{\delta }{\delta \psi
_{NC}^{+}(x,t)}\frac{\delta }{\delta \psi _{C}^{-}(y,t\mathbf{)}}P[%
\underrightarrow{\psi },\underrightarrow{\psi }^{\ast }]\right\}   \notag
\end{eqnarray}%
and%
\begin{eqnarray}
&&\frac{1}{2}\left\{ \dint \int dx\,dy\,\left\{ \sum_{l\epsilon
C}\dsum\limits_{k\epsilon NC}\phi _{l}^{\ast }(x,t\mathbf{)}C_{lk}^{\ast
}(t)\phi _{k}(y,t\mathbf{)}\right\} \,(i\Xi _{C}(x,t\mathbf{))}\,(i\Xi
_{NC}^{+}(y,t\mathbf{))}\right\}   \notag \\
&&\times \chi \lbrack \Xi _{C},\Xi _{C}^{+},\Xi _{NC},\Xi _{NC}^{+}]  \notag
\\
&=&\int D^{2}\underrightarrow{\psi }\;\frac{1}{2}\dint \int dx\,dy\,\left\{
\sum_{l\epsilon C}\dsum\limits_{k\epsilon NC}\phi _{l}^{\ast }(x,t\mathbf{)}%
C_{lk}^{\ast }(t)\phi _{k}(y,t\mathbf{)}\right\} \left\{ \frac{\delta }{%
\delta \psi _{C}^{+}(x,t)}\frac{\delta }{\delta \psi _{NC}(y,t\mathbf{)}}%
\right\}   \notag \\
&&\times \exp (i\dint dx\,(\psi _{C}(x,t\mathbf{)\,}\Xi _{C}^{+}(x,t\mathbf{%
)+}\psi _{NC}(x,t)\,\Xi _{NC}^{+}(x,t)\mathbf{+}\,\Xi _{C}(x,t\mathbf{)}%
\,\psi _{C}^{+}(x,t\mathbf{)+\,}\Xi _{NC}(x,t)\,\psi _{NC}^{+}(x,t)\mathbf{)}
\notag \\
&&\times P[\underrightarrow{\psi },\underrightarrow{\psi }^{\ast }]  \notag
\\
&=&\int D^{2}\underrightarrow{\psi }\;\exp (i\dint dx\,\sum_{\mu }\psi ^{\mu
}(x,t\mathbf{)\,}\Xi ^{-\mu }(x,t\mathbf{))}  \label{Eq.SecondCouplingTerm}
\\
&&\times \dint \int dx\,dy\,\left\{ \frac{1}{2}\sum_{k\epsilon
NC}\dsum\limits_{l\epsilon C}\phi _{Cl}^{+}(x,t\mathbf{)}C_{Cl\,NCk}^{+}(t)%
\phi _{NCk}^{-}(y,t\mathbf{)}\frac{\delta }{\delta \psi _{C}^{+}(x,t)}\frac{%
\delta }{\delta \psi _{NC}^{-}(y,t\mathbf{)}}P[\underrightarrow{\psi },%
\underrightarrow{\psi }^{\ast }]\right\}   \notag
\end{eqnarray}%
where functional integration by parts has been applied twice.

Equating both sides of $\frac{{\LARGE \partial }}{{\LARGE \partial t}}\chi
\lbrack \Xi _{C},\Xi _{C}^{+},\Xi _{NC},\Xi _{NC}^{+}]$ via Eqs. (\ref%
{Eq.TimeDerivCharFnalSecondTerm}), (\ref{Eq.TimeDerivDensOprTerm}) and (\ref%
{Eq.TimeDerivCharFnalCondNonCond2}) gives the \emph{functional Fokker-Planck
equation }for the hybrid distribution function in the form

\begin{eqnarray}
&&\frac{\partial }{\partial t}P[\underrightarrow{\psi },\underrightarrow{%
\psi }^{\ast }]\,  \notag \\
&=&\left\{ -\dsum\limits_{\mu A}\dint dx\frac{\delta }{\delta \psi _{A}^{\mu
}(x,t)}A_{A}^{\mu }(x)+\frac{1}{2}\dsum\limits_{\mu A}\dsum\limits_{\nu
B}\dint \dint dxdy\frac{\delta }{\delta \psi _{A}^{\mu }(x,t)}\frac{\delta }{%
\delta \psi _{B}^{\nu }(y,t)}D_{A\,B}^{\mu \,\nu }(x,y)\right\} P[%
\underrightarrow{\psi },\underrightarrow{\psi }^{\ast }]  \notag \\
&&+\frac{1}{2}\left\{ \sum_{k\epsilon C}\dsum\limits_{l\epsilon NC}\dint
\int dx\,dy\,\phi _{NCl}^{+}(x,t\mathbf{)}C_{Ck\,NCl}^{-}(t)\phi
_{Ck}^{-}(y,t\mathbf{)}\frac{\delta }{\delta \psi _{NC}^{+}(x,t)}\frac{%
\delta }{\delta \psi _{C}^{-}(y,t\mathbf{)}}\right\} P[\underrightarrow{\psi 
},\underrightarrow{\psi }^{\ast }]  \notag \\
&&+\frac{1}{2}\left\{ \sum_{k\epsilon NC}\dsum\limits_{l\epsilon C}\dint
\int dx\,dy\,\phi _{Cl}^{+}(x,t\mathbf{)}C_{Cl\,NCk}^{+}(t)\phi
_{NCk}^{-}(y,t\mathbf{)}\frac{\delta }{\delta \psi _{C}^{+}(x,t)}\frac{%
\delta }{\delta \psi _{NC}^{-}(y,t\mathbf{)}}\right\} P[\underrightarrow{%
\psi },\underrightarrow{\psi }^{\ast }]  \notag \\
&=&\left\{ -\dsum\limits_{\mu A}\dint dx\frac{\delta }{\delta \psi _{A}^{\mu
}(x,t)}A_{A}^{\mu }(x)+\frac{1}{2}\dsum\limits_{\mu A}\dsum\limits_{\nu
B}\dint \dint dxdy\frac{\delta }{\delta \psi _{A}^{\mu }(x,t)}\frac{\delta }{%
\delta \psi _{B}^{\nu }(y,t)}D_{A\,B}^{\mu \,\nu }(x,y)\right\}  \notag \\
&&\times P[\underrightarrow{\psi },\underrightarrow{\psi }^{\ast }]  \notag
\\
&&+\frac{1}{2}\left\{ \dsum\limits_{\mu A}\dsum\limits_{\nu B}\dint \int
dx\,dy\,\frac{\delta }{\delta \psi _{A}^{\mu }(x,t\mathbf{)}}\frac{\delta }{%
\delta \psi _{B}^{\nu }(y,t)}\left( \sum_{k}\dsum\limits_{l}\phi _{Ak}^{\mu
}(x,t\mathbf{)}C_{Ak\,Bl}^{\mu }(t)\phi _{Bl}^{\nu }(y,t\mathbf{)}\right)
\delta _{A\,C}\delta _{B\,NC}\,\delta _{\mu \,-\nu }\right\}  \notag \\
&&\times P[\underrightarrow{\psi },\underrightarrow{\psi }^{\ast }]
\label{Eq.FnalFokkerPlanckHybridDistnFnal}
\end{eqnarray}%
\pagebreak

\section{Appendix - Derivation of Ito Stochastic Field Equations}

\label{Appendix Derivation of Ito Stochastic Field Equations}

In this Appendix details for the derivation of the Ito stochastic field
equations in Section \ref{Section 3 - Quantum Field Case} are set out.

\subsection{Derivative of Phase Space Functional Integral}

To evaluate the first term involving $\frac{{\LARGE \partial }}{{\LARGE %
\partial t}}F[\underrightarrow{\psi }]$ we first consider changes $\delta
\psi _{A}^{\mu }(x,t)$ in $\psi _{A}^{\mu }(x,t)$ during an interval $\delta
t$ that results in a change $\delta F[\underrightarrow{\psi }]$ to the
functional $F[\underrightarrow{\psi }]$. From Eq.(\ref{Eq.FnalDeriv}) we
have 
\begin{eqnarray}
F[\underrightarrow{\psi }+\delta \underrightarrow{\psi }]-F[\underrightarrow{%
\psi }] &\doteqdot &\dsum\limits_{\mu A}\int dx\,\left( \frac{\delta }{%
\delta \psi _{A}^{\mu }(x,t)}F[\underrightarrow{\psi }]\right) \delta \psi
_{A}^{\mu }(x,t)  \notag \\
\delta F[\underrightarrow{\psi }] &\doteqdot &\dsum\limits_{\mu A}\int
dx\,\left( \frac{\delta }{\delta \psi _{A}^{\mu }(x,t)}F[\underrightarrow{%
\psi }]\right) \frac{\partial \psi _{A}^{\mu }(x,t)}{\partial t}\delta t 
\notag \\
\frac{\partial }{\partial t}F[\underrightarrow{\psi }] &=&\dsum\limits_{\mu
A}\int dx\,\left( \frac{\delta }{\delta \psi _{A}^{\mu }(x,t)}F[%
\underrightarrow{\psi }]\right) \frac{\partial \psi _{A}^{\mu }(x,t)}{%
\partial t}
\end{eqnarray}%
Hence the first term in (\ref{Eq.TimeDerivFnalPhaseAverHybrid}) is given by%
\begin{eqnarray}
&&\int D^{2}\underrightarrow{\psi }\,\frac{\partial }{\partial t}F[%
\underrightarrow{\psi }]\,P[\underrightarrow{\psi },\underrightarrow{\psi }%
^{\ast }]  \notag \\
&=&\int D^{2}\underrightarrow{\psi }\,\sum_{\mu A}\dint dx\left( \frac{%
\delta }{\delta \psi _{A}^{\mu }(x,t)}F[\underrightarrow{\psi }]\right) 
\notag \\
&&\times \left\{ \int dy\,\dsum\limits_{B}\left( \dsum\limits_{kl}\phi
_{Ak}^{\mu }(x,t)C_{Ak\,Bl}^{\mu }\phi _{Bl}^{-\mu }(y,t)\right) \;\psi
_{B}^{\mu }(y,t)\right\} \,P[\underrightarrow{\psi },\underrightarrow{\psi }%
^{\ast }]  \notag \\
&&+\int D^{2}\underrightarrow{\psi }\,\sum_{\mu A}\dint dx\left( \frac{%
\delta }{\delta \psi _{A}^{\mu }(x,t)}F[\underrightarrow{\psi }]\right) 
\notag \\
&&\times \left\{ \dint dy\,\dsum\limits_{B}\left( \dsum\limits_{kl}\phi
_{Bl}^{\mu }(x,t)C_{Ak\,Bl}^{-\mu }\phi _{Ak}^{-\mu }(y,t)\right) \;\psi
_{A}^{\mu }(y,t)\right\} \,P[\underrightarrow{\psi },\underrightarrow{\psi }%
^{\ast }]  \notag \\
&=&\left\langle \sum_{\mu A}\dint dx\left( \frac{\delta }{\delta \psi
_{A}^{\mu }(x,t)}F[\underrightarrow{\psi }]\right) \left\{ \int
dy\,\dsum\limits_{B}\left( \dsum\limits_{kl}\phi _{Ak}^{\mu
}(x,t)C_{Ak\,Bl}^{\mu }\phi _{Bl}^{-\mu }(y,t)\right) \;\psi _{B}^{\mu
}(y,t)\right\} \right\rangle  \notag \\
&&+\left\langle \sum_{\mu A}\dint dx\left( \frac{\delta }{\delta \psi
_{A}^{\mu }(x,t)}F[\underrightarrow{\psi }]\right) \left\{ \dint
dy\,\dsum\limits_{B}\left( \dsum\limits_{kl}\phi _{Bl}^{\mu
}(x,t)C_{Ak\,Bl}^{-\mu }\phi _{Ak}^{-\mu }(y,t)\right) \;\psi _{A}^{\mu
}(y,t)\right\} \right\rangle  \notag \\
&&
\end{eqnarray}

Using the functional Fokker-Planck equation (\ref%
{Eq.FnalFokkerPlanckHybridDistnFnal2})\ we find after using integration by
parts that the second term in (\ref{Eq.TimeDerivFnalPhaseAverHybrid}) is 
\begin{eqnarray}
&&\int D^{2}\underrightarrow{\psi }\,F[\underrightarrow{\psi }]\,\frac{%
\partial }{\partial t}P[\underrightarrow{\psi },\underrightarrow{\psi }%
^{\ast }]  \notag \\
&=&\int D^{2}\underrightarrow{\psi }\,\left\{ \dsum\limits_{\mu A}\dint
dx\left( \frac{\delta }{\delta \psi _{A}^{\mu }(x,t)}F[\underrightarrow{\psi 
}]\right) A_{A}^{\mu }(x)\right\} P[\underrightarrow{\psi },\underrightarrow{%
\psi }^{\ast }]  \notag \\
&&+\int D^{2}\underrightarrow{\psi }\,\left\{ \frac{1}{2}\dsum\limits_{\mu
A}\dsum\limits_{\nu B}\dint \dint dxdy\left( \frac{\delta }{\delta \psi
_{A}^{\mu }(x,t)}\frac{\delta }{\delta \psi _{B}^{\nu }(y,t)}F[%
\underrightarrow{\psi }]\right) E_{A\,B}^{\mu \,\nu }(x,y)\right\} P[%
\underrightarrow{\psi },\underrightarrow{\psi }^{\ast }]  \notag \\
&=&\left\langle \dsum\limits_{\mu A}\dint dx\left( \frac{\delta }{\delta
\psi _{A}^{\mu }(x,t)}F[\underrightarrow{\psi }]\right) A_{A}^{\mu
}(x)\right\rangle  \notag \\
&&+\left\langle \frac{1}{2}\dsum\limits_{\mu A}\dsum\limits_{\nu B}\dint
\dint dxdy\left( \frac{\delta }{\delta \psi _{A}^{\mu }(x,t)}\frac{\delta }{%
\delta \psi _{B}^{\nu }(y,t)}F[\underrightarrow{\psi }]\right) E_{A\,B}^{\mu
\,\nu }(x,y)\right\rangle  \notag \\
&&
\end{eqnarray}%
where functional integration by parts has been used.

\subsection{Derivative of Stochastic Field Averages}

For the first order derivative terms we have on substituting from (\ref%
{Eq.ItoSDEFieldsHyb}) and expanding%
\begin{eqnarray}
&&\overline{\dint dx\,\left\{ \dsum\limits_{\mu A}\delta \psi _{A}^{\mu
s}(x,t)\,\left( \frac{\delta F[\psi ^{s},\psi ^{s+}]}{\delta \psi _{A}^{\mu
s}(x,t)}\right) _{x}\right\} }  \notag \\
&=&\overline{\dint dx\,\sum_{\mu A}\left( \frac{\delta F[\psi ^{s},\psi
^{s+}]}{\delta \psi _{A}^{\mu s}(x,t)}\right) _{x}\mathcal{G}_{A}^{\mu }(x)}%
\,\delta t  \notag \\
&&+\overline{\dint dx\,\sum_{\mu A}\left( \frac{\delta F[\psi ^{s},\psi
^{s+}]}{\delta \psi _{A}^{\mu s}(x,t)}\right) _{x}\sum_{a}\mathcal{N}%
_{Aa}^{\mu }(x)}\;\overline{\int_{t}^{t+\delta t}dt_{1}\Gamma _{a}(t_{1})} 
\notag \\
&=&\overline{\dint dx\,\sum_{\mu A}\left( \frac{\delta F[\psi ^{s},\psi
^{s+}]}{\delta \psi _{A}^{\mu s}(x,t)}\right) _{x}\mathcal{G}_{A}^{\mu }(x)}%
\,\delta t
\end{eqnarray}%
where the stochastic average rules for sums and products have been used, the
non-correlation between the averages of functions of $\psi _{\mu }^{s}(x,t)$
at time $t$ and the $\Gamma $ at later times between $t$ to $t+\delta t$ is
applied and the term involving $\overline{\Gamma _{a}(t_{1})}$ is equal to
zero from (\ref{Eq.GaussianMarkov}). Note that this term is proportional to $%
\delta t$.

For the second order derivative terms we have on substituting from (\ref%
{Eq.ItoSDEFieldsHyb}) and expanding we find that 
\begin{eqnarray}
&&\overline{\left\{ \frac{1}{2}\dint \dint dxdy\,\dsum\limits_{\mu A,\nu
B}\delta \psi _{A}^{\mu s}(x,t)\,\delta \psi _{B}^{\nu s}(y,t)\,\left( \frac{%
\delta ^{2}F[\psi ^{s},\psi ^{s+}]}{\delta \psi _{A}^{\mu s}(x,t)\delta \psi
_{B}^{\nu s}(y,t)}\right) _{x,y}\right\} }  \notag \\
&=&\overline{\frac{1}{2}\dint \dint dxdy\,\dsum\limits_{\mu A,\nu B}\,\left( 
\frac{\delta ^{2}F[\psi ^{s},\psi ^{s+}]}{\delta \psi _{A}^{\mu
s}(x,t)\delta \psi _{B}^{\nu s}(y,t)}\right) _{x,y}\left[ \mathcal{G}%
_{A}^{\mu }(x)\,\mathcal{G}_{B}^{\nu }(y)\right] }\;\delta t^{2}  \notag \\
&&+\overline{\frac{1}{2}\dint \dint dxdy\,\dsum\limits_{\mu A,\nu B}\,\left( 
\frac{\delta ^{2}F[\psi ^{s},\psi ^{s+}]}{\delta \psi _{A}^{\mu
s}(x,t)\delta \psi _{B}^{\nu s}(y,t)}\right) _{x,y}\mathcal{G}_{A}^{\mu
}(x)\delta t\,\sum_{b}\mathcal{N}_{Bb}^{\nu }(y)}  \notag \\
&&\times \;\overline{\int_{t}^{t+\delta t}dt_{2}\Gamma _{b}(t_{2})}\;\delta t
\notag \\
&&+\overline{\frac{1}{2}\dint \dint dxdy\,\dsum\limits_{\mu A,\nu B}\left( 
\frac{\delta ^{2}F[\psi ^{s},\psi ^{s+}]}{\delta \psi _{A}^{\mu
s}(x,t)\delta \psi _{B}^{\nu s}(y,t)}\right) _{x,y}\left[ \sum_{a}\mathcal{N}%
_{Aa}^{\mu }(x)\;\mathcal{G}_{B}^{\nu }(y)\delta t\right] }\;  \notag \\
&&\times \overline{\int_{t}^{t+\delta t}dt_{1}\Gamma _{a}(t_{1})}\;\delta t 
\notag \\
&&+\overline{\frac{1}{2}\dint \dint dxdy\,\dsum\limits_{\mu A,\nu B}\left( 
\frac{\delta ^{2}F[\psi ^{s},\psi ^{s+}]}{\delta \psi _{A}^{\mu
s}(x,t)\delta \psi _{B}^{\nu s}(y,t)}\right) _{x,y}\left[ \sum_{a}\mathcal{N}%
_{Aa}^{\mu }(x)\;\sum_{b}\mathcal{N}_{Bb}^{\nu }(y)\right] }\;  \notag \\
&&\times \overline{\int_{t}^{t+\delta t}dt_{1}\Gamma
_{a}(t_{1})\int_{t}^{t+\delta t}dt_{2}\Gamma _{b}(t_{2})}  \notag \\
&&
\end{eqnarray}%
where the stochastic average rules for sums and products have been used, the
non-correlation between the averages of functions of $\psi _{\mu }^{s}(x,t)$
at time $t$ and the $\Gamma $ at later times between $t$ to $t+\delta t$ is
applied. The terms involving a single $\Gamma $ have a zero stochastic
average, whilst from Eq.(\ref{Eq.GaussianMarkov}) the terms with two $\Gamma 
$ give a stochastic average proportional to $\delta t$ 
\begin{eqnarray}
\overline{\int_{t}^{t+\delta t}dt_{1}\Gamma _{a}(t_{1})\int_{t}^{t+\delta
t}dt_{2}\Gamma _{b}(t_{2})} &=&\int_{t}^{t+\delta t}dt_{1}\int_{t}^{t+\delta
t}dt_{2}\;\overline{\Gamma _{a}(t_{1})\Gamma _{b}(t_{2})}  \notag \\
&=&\int_{t}^{t+\delta t}dt_{1}\int_{t}^{t+\delta t}dt_{2}\;\delta
_{ab}\delta (t_{1}-t_{2})  \notag \\
&=&\delta _{ab}\,\delta t
\end{eqnarray}%
so that correct to order $\delta t$ the second order derivative term is 
\begin{eqnarray}
&&\overline{\left\{ \frac{1}{2}\dint \dint dxdy\,\dsum\limits_{\mu A,\nu
B}\delta \psi _{A}^{\mu s}(x,t)\,\delta \psi _{B}^{\nu s}(y,t)\,\left( \frac{%
\delta ^{2}F[\psi ^{s},\psi ^{s+}]}{\delta \psi _{A}^{\mu s}(x,t)\delta \psi
_{B}^{\nu s}(y,t)}\right) _{x,y}\right\} }  \notag \\
&=&\overline{\frac{1}{2}\dint \dint dxdy\,\dsum\limits_{\mu A,\nu B}\left( 
\frac{\delta ^{2}F[\psi ^{s},\psi ^{s+}]}{\delta \psi _{A}^{\mu
s}(x,t)\delta \psi _{B}^{\nu s}(y,t)}\right) _{x,y}\left[ \sum_{a}\mathcal{N}%
_{Aa}^{\mu }(x)\;\mathcal{N}_{Ba}^{\nu }(y)\right] }\;\delta t  \notag \\
&=&\overline{\frac{1}{2}\dint \dint dxdy\,\dsum\limits_{\mu A,\nu B}\left( 
\frac{\delta ^{2}F[\psi ^{s},\psi ^{s+}]}{\delta \psi _{A}^{\mu
s}(x,t)\delta \psi _{B}^{\nu s}(y,t)}\right) _{x,y}\left[ [\mathcal{N}(x)%
\mathcal{\ N}^{T}(y)]_{A,B}^{\mu ,\nu }\right] }\;\delta t  \notag \\
&&
\end{eqnarray}%
\bigskip 

\pagebreak

\end{document}